\documentclass[usenatbib]{mnras}

\usepackage{mathptmx}
\usepackage{txfonts}

\usepackage[T1]{fontenc}

\DeclareRobustCommand{\VAN}[3]{#2}
\let\VANthebibliography\thebibliography
\def\thebibliography{\DeclareRobustCommand{\VAN}[3]{##3}\VANthebibliography}

\usepackage{graphicx}	
\usepackage{color}
\usepackage{subcaption}
\usepackage{lscape}
\usepackage{multirow}
\usepackage{longtable}
\usepackage{microtype}
\usepackage{ulem}
\usepackage{array}
\newcolumntype{C}[1]{>{\centering\arraybackslash}p{#1}}
\newcolumntype{L}[1]{>{\raggedright\let\newline\\\arraybackslash\hspace{0pt}}m{#1}}

\usepackage{txfonts}
\usepackage[]{hyperref}
\newcommand{\hii}{H\textsc{ii}}
\newcommand{\solmass}{$M_{\odot}$} 
\newcommand{\mewm}{$\mu$m} 
\newcommand{\kms}{kms$^{-1}$}
\newcommand{\asec}{$^{\prime\prime}$}
 
\newcommand{\meth}{CH$_3$OH}

\newcommand{\lj}{\texttt{LumberJack} }

\definecolor{junkcol}{rgb}{1.0,0.6,0.2}
\definecolor{coiCol}{rgb}{0.0,0.4,0.6}
\definecolor{chanCol}{rgb}{0.6,0.3,0.6}
\definecolor{gafCol}{rgb}{0.0,0.9,0.1}

\title[TEMPO: Fragmentation and emission properties]{Tracing Evolution in Massive Protostellar Objects (TEMPO) - \textsc{I}:\\Fragmentation and emission properties of massive star-forming clumps in a luminosity limited ALMA sample}

\author[A. Avison et al.]{A. Avison $^{1,2,3}$\thanks{E-mail: adam.avison@skao.int} 
G. A. Fuller $^{1,2,4,5}$,
N. Asabre Frimpong $^{1,6}$,
S. Etoka $^{1}$,
M. Hoare $^{7}$, 
\newauthor
B.M. Jones $^{1,5}$, 
N. Peretto $^{8}$, 
A. Traficante $^{9}$, 
F. van der Tak $^{10,11}$,
J.E. Pineda $^{12}$,
M. Beltr\'an $^{13}$,
\newauthor
F. Wyrowski $^{14}$,
M. Thompson $^{15}$, 
S. Lumsden $^{7}$,
Z. Nagy $^{16,17}$, 
T. Hill $^{18}$,
S. Viti $^{19,20}$, 
\newauthor
F. Fontani $^{13}$,
P. Schilke $^{5}$ 
\\
$^{1}$Jodrell Bank Centre for Astrophysics, Department of Physics and Astronomy, School Of Natural Science, The University of Manchester, Manchester, M13 9PL, UK\\
$^{2}$UK ALMA Regional Centre Node, M13 9PL, UK\\
$^{3}$SKA Observatory, Jodrell Bank, Lower Withington, Macclesfield, SK11 9FT\\
$^{4}$Intituto de Astrof\'isica de Andalucia (CSIC), Glorieta de al Astronomia s/n E-18008, Granada, Spain\\
$^{5}$I. Physikalisches Institut, University of Cologne, Z\"ulpicher Str. 77, 50937 K\"oln, Germany\\
$^{6}$Ghana Space Science and Technology Institute, Radio Astronomy and Astrophysics Centre, Atomic-Haatso Road, Kwabenya, Accra, Greater Accra, Ghana\\
$^{7}$School of Physics and Astronomy, University of Leeds, Leeds LS2 9JT, UK\\
$^{8}$School of Physics and Astronomy, Cardiff University, Queens Buildings, The Parade, Cardiff CF24 3AA, UK\\
$^{9}$IAPS-INAF, Via Fosso del Cavaliere, 100, 00133 Rome, Italy\\
$^{10}$SRON Netherlands Institute for Space Research, Landleven 12, 9747 AD Groningen, The Netherlands\\
$^{11}$Kapteyn Astronomical Intsitute, University of Groningen, The Netherlands\\
$^{12}$Max-Planck-Institut für extraterrestrische Physik, Giessenbachstrasse 1, 85748 Garching, Germany\\
$^{13}$INAF-Osservatorio Astrofisico di Arcetri, Largo E. Fermi 5, 50125 Firenze, Italy \\
$^{14}$Max-Planck-Institut für Radioastronomie, Auf dem Hügel 69, 53121 Bonn, Germany\\
$^{15}$Centre for Astrophysics Research, Science and Technology Research Institute, University of Hertfordshire, College Lane, Hatfield,
Hertfordshire AL10 9AB, UK\\
$^{16}$Konkoly Observatory, Research Centre for Astronomy and Earth Sciences, E\"otv\"os Lor\'and Research Network (ELKH), \\    Konkoly Thege Mikl\'os \'ut 15-17, 1121 Budapest, Hungary\\
$^{17}$CSFK, MTA Centre of Excellence, Konkoly Thege Mikl\'os \'ut 15-17, 1121 Budapest, Hungary\\
$^{18}$Western Sydney University, Kingswood, Australia\\
$^{19}$Leiden Observatory, Leiden University, PO Box 9513, NL-2300 RA, Leiden, the Netherlands\\
$^{20}$Department of Physics and Astronomy, University College London, Gower Street, London, WC1E 6BT, United Kingdom
}

\date{Accepted XXX. Received YYY; in original form ZZZ}

\pubyear{2023}

\begin{document}
\label{firstpage}
\pagerange{\pageref{firstpage}--\pageref{lastpage}}
\maketitle

\begin{abstract}
The role of massive ($\geq$ 8\solmass) stars in defining the energy budget and chemical enrichment of the interstellar medium in their host galaxy is significant. In this first paper from the \textit{Tracing Evolution in Massive Protostellar Objects} (TEMPO) project we introduce a colour-luminosity selected (L$_*$ $\sim$ 3$\times10^3$ to 1$\times10^5$ L$_{\odot}$) sample of 38 massive star forming regions observed with ALMA at 1.3mm and explore the fragmentation, clustering and flux density properties of the sample. The TEMPO sample fields are each found to contain multiple fragments (between 2-15 per field). The flux density budget is split evenly (53\%-47\%) between fields where emission is dominated by a single high flux density fragment and those in which the combined flux density of fainter objects dominates. The fragmentation scales observed in most fields are not comparable with the thermal Jeans length, $\lambda_J$, being larger in the majority of cases, suggestive of some non-thermal mechanism. A tentative evolutionary trend is seen between luminosity of the clump and the `spectral line richness' of the TEMPO fields; with 6.7GHz maser associated fields found to be lower luminosity and more line rich. This work also describes a method of line-free continuum channel selection within ALMA data and a generalised approach used to distinguishing sources which are potentially star-forming from those which are not, utilising interferometric visibility properties. 
\end{abstract}




\section{Introduction}

Despite the importance of high-mass stars (M>8\solmass) on the galactic scale, due to their prodigious chemical and energetic feedback, our understanding of their formation and early evolution remains poorly understood \citep[e.g.][]{Tan14}. Answering the unresolved issues of massive star formation is not only important for the study of our Galactic environment but also has implications for the modelling of star formation and the evolution of the interstellar medium in extra-galactic sources throughout the star forming life-time of the Universe \citep{Kennicutt12}. 


Current discussion within the literature centres around two scenarios under which protostars may acquire the necessary mass to form high-mass stars, these are commonly termed the \textit{clump-fed} and \textit{core-fed} scenarios \citep[following e.g.][]{Wang10}. The \textit{core-fed} scenario posits that a stars final mass is correlated with the mass in the core from which is formed \citep{McKeeTan03, Tan14} thus requiring the presence of both low and high mass protostellar cores to create the distribution of stellar masses seen on the main sequence, with some core to final mass efficiency relating initial core mass to final stellar mass. However, currently there is little evidence for cores of sufficient mass to create the most massive stars of 10\solmass\ and greater \citep[e.g.]{Nony18, Sanhueza19}. Conversely, under the \textit{clump-fed} scenario the final mass of a star is not determined purely by material available within its natal core, but instead on its position within and the material available to it from the larger scales of the host clump. Such multi scale, hierarchical collapse removes the need for any relation between the initial mass of a protostellar core and the final mass of the star it forms, as the final mass is instead determined by the dynamical properties of the material on much larger scales and interaction/competition with other protostars in the protoclusters \citep{BonnellBate06, Wang10, Peretto13, Williams18, VazquezSemadeni19}.

An important observational indicator which can allow the discrimination between proposed evolutionary scenarios are the fragmentation of star forming clumps at early times within their evolution and the distribution (both spatially and in terms of the mass) of fragments within them\footnote{Throughout this paper we combine the nomenclature seen commonly within the literature \citep[e.g.]{Zhang09, TraficanteSqualo} when referring to structures of differing size is used. As such, objects of several to hundreds of pc are referred to as clouds, objects of $\sim$1 pc as clumps and objects $\leq 0.1$ pc as fragments, unless the are known to be star-forming in which case they are termed cores.}. Specifically, thermal Jeans fragmentation is considered to be consistent with global hierarchical collapse and competitive accretion models \citep{Sanhueza19} (c.f. \textit{clump-fed} models) whereas the need for turbulence or other mechanisms to support massive protostellar cores under the \textit{core-fed} scenario may indicate the presence of fragmentation on non-thermal scales.

There is some evidence for fragmentation on the thermal Jeans length ($\lambda_{J}\sim$0.1 pc at $T=$25 K and $n=10^5$cm$^{-3}$) as opposed to turbulent or filamentary fragmentation scales in samples of infrared dark (at 70\mewm) star forming clumps, when observed at high sensitivity and angular resolution \citep{Pillai11, Sanhueza19, Svoboda19}. Conversely, a number of authors have found evidence for filamentary, turbulently or magnetically supported fragmentation scales \citep{Wang14, Beuther15, Henshaw16, Fontani16, Lu18, Sokolov18, TraficanteSqualo} when studying high mass star forming infrared dark clouds (IRDCs) with \citet{TraficanteSqualo} finding evidence for an evolutionary relation of Jean's length as a function of $L/M$. 

The discrepancies between observational results may be attributable to a combination of factors such as differing sensitivities within observations or evolutionary dfferences in the samples of sources observed. The latter issue will be resolved over time as larger samples with varying sample selection criteria are published. It may also be the case that there is no ‘one true’ model for high-mass star formation and that attributes of different models are represented in
different regions and at different times in their evolution depending on the environment and starting conditions.


This paper represents the first in a series from the \textit{Tracing Evolution in Massive Protostellar Objects} (TEMPO) project. TEMPO has undertaken a systematic high resolution and high sensitivity survey using the world leading capabilities of ALMA to simultaneously study the chemistry, structure and fragmentation of a luminosity and colour selected sample of young high mass embedded objects. 

The two initial key goals of TEMPO are:
\begin{itemize}
\item Investigating how the mass and fragmentation of material in high-mass star-forming regions changes with luminosity and temperature. 
\item Investigating how the observed molecular gas chemical composition evolves (e.g. number of complex organic molecules present, high gas density tracer abundance) as a function of luminosity and spectral energy distribution (SED) properties. Asabre Frimpong et al. \textit{in prep.} will provide the first detailed analysis of the molecular emission recovered from the TEMPO data. 
\end{itemize}

The current paper begins to address the first goal and presents the population, clustering, flux density budget and fragmentation properties of our high-mass protostellar cluster sample as well as introducing and characterising the observations of the TEMPO project. Section 2 introduces the sample, the ALMA observations undertaken and data processing. Section 3 provides an overview of the observation results for continuum emission and the characteristics of this emission. In Section 4 the clustering, fragmentation and flux density budgets of the sample of observed fragments are discussed. Section 4 also comments on two properties of the TEMPO sample which relate to evolutionary characteristics. Using visibility analysis section 4 also addresses whether the detected fragments are likely currently star-forming or simply a transient conglomerations of material, and association with other star forming tracers. Section 5 discusses our initial TEMPO findings and Section 6 provides a summarised conclusion.


\section{Observations}
\subsection{The Sample}
\label{Sample:sec}
\begin{figure}
\begin{center}
\includegraphics[scale=0.56]{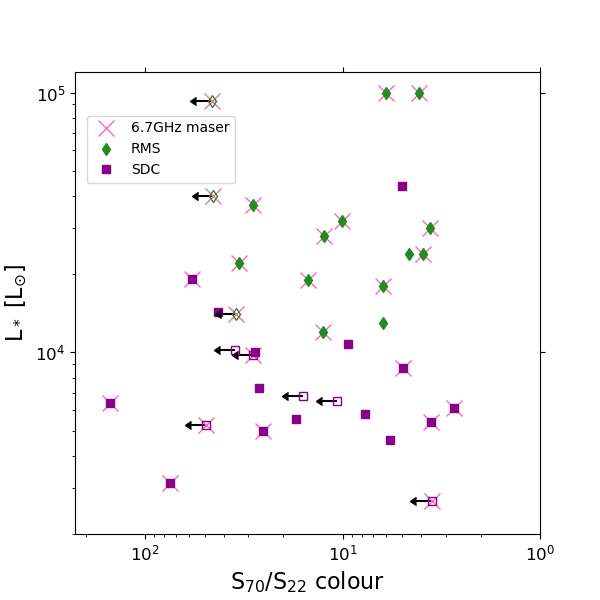}
\caption{S$_{70\mu m}$/S$_{22\mu m}$ colour $-$ luminosity plot for the fields in the sample, with RMS fields as green diamonds and SDC fields as purple squares. Luminosity derived from SED fitting to $Herschel$ data by \citet{Mottram11} for the RMS sources and \citet{Traficante15} for the SDC sources.  S$_{70\mu m}$/S$_{22\mu m}$ values are derived from Herschel (70\mewm) and WISE (22\mewm) measurements as presented in  \citet{Lumsden13} (for RMS sources) and \citet{Traficante15}. Unfilled markers denote those fields which are not detected at 22\mewm, listed with $\dagger$s in Table \ref{Selection:tab} and represent 22\mewm\ upper limits. Fields with a pink `$\times$' have an associated 6.7 GHz methanol maser detection from the Methanol MultiBeam (MMB) survey \citep{GreenTech}.}
\label{ColourLum:fig}
\end{center}
\end{figure}

The TEMPO sample comprises 38 luminosity and infrared colour selected fields known to host young high-mass embedded protostellar sources, selected from both the Red MSX Source (RMS) survey \citep{Lumsden13} and the Spitzer Dark Cloud (SDC) sample \citep{Peretto09} to cover a range of S$_{70\mu m}$/S$_{22\mu m}$ colours and exhibit luminosities above 3$\times 10^3L_{\odot}$, as seen in Figure \ref{ColourLum:fig}, a value which allows the sample to focus only on the most massive regions, i.e. those harbouring OB-type high-mass (proto)stars. The 70$\mu m$ data were taken from Herschel as part of the Hi-GAL survey \citep{Molinari10}. The selection criteria were used to ensure the presence of high-mass protostars (high $L_{\odot}$ values) and cover a range of evolutionary stages from mid-IR 22\mewm\ non-detections to S$_{70\mu m}$/S$_{22\mu m}\sim 1$. 

The choice of colour [22\mewm\ - 70\mewm] was made as the similar [24\mewm\ - 70\mewm] colour has been found to provide a good discrimination between sources with spectral energy distributions (SEDs) which are well fitted by embedded Zero Age Main Sequence (ZAMS) star models (and are thus relatively more evolved objects) and those which are best fit by a single optically thin greybody peaking at longer wavelengths than the ZAMS models (less evolved, relatively), \citep{Molinari08}, and bears a strong relation with source bolometric luminosity \citep{Molinari19}. Similarly, \citet{HughesMacLeod89} used the [60\mewm\ - 25\mewm] colour to define the colour space occupied by highly evolved infra-red sources which display \hii\ regions at optical wavelengths. The WISE 22\mewm\ data are used here rather than the \textit{Spitzer} MIPS 24\mewm\ data as the latter is saturated toward a number of the TEMPO fields.

The RMS Survey \citep{Lumsden13} was constructed using a subset of the v2.3 \textit{MSX} point source catalogue \citep{Egan03}, to generate a mid- and near-infrared colour selected sample of massive protostellar objects. The colour-selection criteria was complemented by additional higher resolution infra-red and radio observations to remove ultra compact \hii\ regions (UC\hii) and planetary nebulae, which exhibit similar colours, from the sample. As such the RMS is 90\% complete for massive protostellar objects within the survey's observed area $10^{\circ} <\ l\ < 350^{\circ}$, $b<5^{\circ}$.

The SDC sources from \citet{Peretto09} are drawn from an initial sample of $>$11,000 IRDCs seen in absorption at 8\mewm\ ($\tau_{8\mu m}>0.35$) in the GLIMPSE \citep{Churchwell09} data from the \textit{Spitzer} Space Telescope. Such 8\mewm\ opacities mean all the SDC IRDCs have column densities above 10$^{22}$cm$^{-2}$. The selected SDC sources as targets for the TEMPO sample are from the `starless and protostellar clumps embedded in the IRDCs' catalogue of \citet{Traficante15} and we use the  mass and luminosity properties for the selected sources from this work. All SDC sources selected for this current work have core masses $>$ 500 \solmass.   

Additionally, the TEMPO fields (both RMS and SDC) were chosen to be isolated across a range of IR wavelengths to avoid confusion and to have distances less than $\sim$6 kpc. The range of distances to our target fields covers 1.8 to 6.3 kpc (a factor of 3.5\footnote{For reference, at these distance 1\arcsec\ corresponds to a physical distance of 0.009 to 0.03 pc, respectively.}) which limits the lower range of observable spatial scales common within the data. There are 28 fields in the TEMPO sample (74\%) in which a 6.7\,GHz class-II methanol maser detected within the Methanol MultiBeam survey (MMB, \citealt{GreenTech}) is located with the observed ALMA primary beam. The 6.7\,GHz class-II methanol maser is known to be uniquely associated with high-mass protostellar objects \citep{Minier03, Xu08, Breen13}. The selection criteria properties for each field in the TEMPO sample are given in Table \ref{Selection:tab}.

Throughout this work fields drawn from the RMS survey are prefixed with `RMS-' (normally named simply after their Galactic coordinates i.e. Glll.lll$\pm$bb.bbb) to differentiate them from the sources from the SDC sample  (preceded with `SDC').

\subsection{ALMA Observations}
\label{obsprops:sec}
The observations were conducted in ALMA band 6 during Cycle 3 under project code 2015.1.01312.S. The project consisted of six separate scheduling blocks  each requiring a single execution to meet the requested sensitivity. The observations were made on the dates 7, 12 and 21 March 2016. The telescope was setup to observe 4 $\times$ 1.875GHz spectral windows (SPWs) with central frequencies of 225.2, 227.1, 239.8 and 241.9~GHz (equivalent to wavelengths of 1.33, 1.32, 1.25 and 1.24mm, respectively). Each SPW consisted of 1920 channels giving a frequency resolution of 976.562 kHz, equivalent to a velocity resolution of $\sim$1.25 \kms. During each observation the array was configured with minimum and maximum baseline lengths of 15.1m and 460.0m respectively. These values give an average resolution of $\sim$0.7-0.8\asec\ and maximum recoverable scale\footnote{The maximum recoverable scale (MRS) for an interferometer is the scale at which an interferometer can reliably recover all emission from a coherent object. The MRS does not relate to the scale over which an interferometer can recover any emission. Objects observed with an interferometer above this size scale are likely to have missing flux, and any associated images suffering from imaging artefacts, e.g. negative bowling, due to this. All recovered fragements in the TEMPO sample are below the MRS and no imaging artefacts are seen in the TEMPO image data.} within the data of 10.5\asec. At the average distance to the TEMPO fields, the average angular resolution gives a physical scale of 0.01 pc and the maximum recoverable scale is 0.2 pc. Table \ref{OBSPROPS:TAB} gives the observing properties of the data set. The data used within this work was extracted from the ALMA Archive and calibrated using scripts provided in the CASA \citep{McMullin07} data reduction software (versions 4.7 for calibration and 5.4 for analysis).

\onecolumn
\begin{landscape}
\begin{longtable}{l c c c c c c | c c C{0.5cm} C{0.5cm} c c C{0.5cm} | c c c}

\caption[]{Selection criteria properties of fields in the sample. Fields which have $\dagger$ in the $S_{70\mu m}/S_{22\mu m}$ column indicate a non-detection at 22\mewm. Distances and luminosities are taken from \citet[and online material from the RMS survey]{Lumsden13} for RMS sources and \citet{Traficante15} for SDC sources. A 6.7~GHz \meth\ maser is deemed associated with the field if it lays within one ALMA primary beam (at our observing frequency) of the field pointing position, maser postions were taken from the MMB survey catalogues \citep{MMB020to060,MMB186to330,MMB330to345,MMB345to006}. The value `\%-age BW' indicates the percentage of the total observed bandwidth (7.5 GHz) which was used in imaging once spectral line emission was removed (see \S \ref{processing:sec}). N, gives the number of sources detected in each field. FOV is the field of view in parsec, at the field distance. X$_{mean}$ is the mean edge length of the minimum spanning tree of sources in each field and $\lambda_{J}$ is the thermal Jeans length at T$_{clump}$. The values M$_{clump}$ and R$_{clump}$ are the clump mass, radii and Temperature as derived by \citet{Elia21} from \textit{Herschel} data.}
\\
\hline
  & Pointing & Pointing & & & & \meth\ & & & & & & & & \\
Field &  RA &  Dec  & L$_*$ & $S_{70\mu m}/S_{22\mu m}$ & D & Maser? & \textit{rms} & \%-age  & N & R$_{cl}$ & FOV & X$_{mean}$ & $\lambda_{J}$ & M$_{clump}$ & R$_{clump}$ & T$_{clump}$ \\
& [h:m:s] & [$^{\circ}$:$^{\prime}$:$^{\prime\prime}$] & [10$^4$L$_{\odot}$] & & [kpc] & [Y/N] & [mJy] & BW & - & [pc] & [pc] & [pc] & [pc] & [\solmass] & [pc] & [K]\\
\hline
RMS-G013.6562-00.5997 & 18:17:24.40 & -17:22:15.000 & 1.4 & 19.33 & 4.1 & Y & 0.15 & 34.2 & 6.0 & 0.2 & 0.44 & 0.13 & 0.05 & 7250.1 & 0.38 & 26.3\\
RMS-G017.6380+00.1566 & 18:22:26.40 & -13:30:12.000 & 10.0 & 4.09 & 2.2 & Y & 0.63 & 17.7 & 9.0 & 0.08 & 0.24 & 0.05 & 0.02 & 143.6 & 0.05 & 40.0\\
SDC18.816-0.447$\_1$ & 18:26:59.00 & -12:44:45.000 & 0.5 & 5.72 & 4.29 & N & 0.19 & 23.3 & 2.0 & 0.06 & 0.46 & 0.2 & 0.04 & 373.5 & 0.14 & 19.6\\
SDC20.775-0.076$\_1$ & 18:29:16.30 & -10:52:09.000 & 0.6 & $\dagger$ & 3.95 & N & 0.23 & 18.8 & 14.0 & 0.22 & 0.42 & 0.08 & 0.03 & 880.6 & 0.15 & 20.3\\
SDC20.775-0.076$\_3$ & 18:29:12.20 & -10:50:35.000 & 0.6 & 7.71 & 3.95 & N & 0.11 & 53.6 & 4.0 & 0.06 & 0.42 & 0.09 & 0.08 & 2006.1 & 0.35 & 28.1\\
SDC22.985-0.412$\_1$ & 18:34:40.10 & -09:00:39.000 & 0.3 & 74.97 & 4.59 & Y & 0.33 & 14.8 & 7.0 & 0.23 & 0.49 & 0.16 & 0.02 & 622.2 & 0.08 & 36.3\\
SDC23.21-0.371$\_1$ & 18:34:55.20 & -08:49:15.000 & 1.0 & $\dagger$ & 3.84 & Y & 0.26 & 34.2 & 9.0 & 0.14 & 0.41 & 0.11 & 0.04 & 11135.0 & 0.46 & 22.1\\
RMS-G023.3891+00.1851 & 18:33:14.30 & -08:23:57.000 & 2.4 & 3.93 & 4.5 & Y & 0.14 & 23.9 & 8.0 & 0.13 & 0.48 & 0.1 & 0.04 & 2138.3 & 0.24 & 23.5\\
SDC24.381-0.21$\_3$ & 18:36:40.60 & -07:39:14.000 & 0.6 & 17.19 & 3.61 & N & 0.16 & 21.2 & 11.0 & 0.27 & 0.39 & 0.1 & 0.02 & 1606.3 & 0.16 & 18.0\\
SDC24.462+0.219$\_2$ & 18:35:11.60 & -07:26:23.000 & 0.7 & 26.53 & 6.27 & N & 0.12 & 41.7 & 5.0 & 0.11 & 0.67 & 0.13 & 0.02 & 1098.0 & 0.12 & 21.3\\
SDC25.426-0.175$\_6$ & 18:37:30.20 & -06:41:16.000 & 1.0 & $\dagger$ & 3.98 & N & 0.14 & 40.2 & 3.0 & 0.05 & 0.43 & 0.13 & 0.07 & 663.8 & 0.2 & 35.5\\
SDC28.147-0.006$\_1$ & 18:42:42.50 & -04:15:34.000 & 0.5 & 25.23 & 4.49 & Y & 0.14 & 7.8 & 6.0 & 0.17 & 0.48 & 0.11 & 0.03 & 1078.4 & 0.16 & 19.9\\
SDC28.277-0.352$\_1$ & 18:44:21.90 & -04:17:39.000 & 0.5 & 3.55 & 3.12 & Y & 0.12 & 33.1 & 4.0 & 0.26 & 0.33 & 0.26 & 0.22 & 280.9 & 0.44 & 15.6\\
SDC29.844-0.009$\_4$ & 18:46:13.00 & -02:39:01.000 & 0.3 & $\dagger$ & 5.38 & Y & 0.69 & 7.4 & 8.0 & 0.18 & 0.58 & 0.08 & 0.03 & 7092.3 & 0.36 & 15.2\\
RMS-G029.8620-00.0444 & 18:45:59.60 & -02:45:07.000 & 2.8 & 12.4 & 4.9 & Y & 0.23 & 27.3 & 6.0 & 0.1 & 0.52 & 0.1 & 0.04 & 1055.5 & 0.18 & 22.6\\
SDC30.172-0.157$\_2$ & 18:47:08.20 & -02:29:58.000 & 0.7 & $\dagger$ & 4.16 & N & 0.24 & 4.6 & 2.0 & 0.04 & 0.45 & 0.11 & 0.33 & 79.8 & 0.28 & 40.0\\
RMS-G030.1981-00.1691 & 18:47:03.10 & -02:30:36.000 & 3.0 & 3.62 & 4.9 & Y & 0.15 & 6.1 & 3.0 & 0.13 & 0.52 & 0.2 & 0.06 & 372.8 & 0.18 & 25.1\\
SDC33.107-0.065$\_2$ & 18:52:08.20 & +00:08:13.000 & 1.9 & 58.3 & 4.54 & Y & 0.21 & 50.2 & 13.0 & 0.19 & 0.49 & 0.08 & 0.03 & 4405.5 & 0.26 & 26.4\\
RMS-G034.7569+00.0247 & 18:54:40.70 & +01:38:07.000 & 1.2 & 12.56 & 4.6 & Y & 0.14 & 19.7 & 6.0 & 0.19 & 0.49 & 0.12 & 0.04 & 348.4 & 0.12 & 22.6\\
RMS-G034.8211+00.3519 & 18:53:37.90 & +01:50:31.000 & 2.4 & 4.58 & 3.5 & N & 0.15 & 46.4 & 9.0 & 0.23 & 0.37 & 0.11 & 0.04 & 616.6 & 0.16 & 22.3\\
SDC35.063-0.726$\_1$ & 18:58:06.00 & +01:37:07.000 & 0.5 & $\dagger$ & 2.32 & Y & 0.29 & 39.3 & 9.0 & 0.13 & 0.25 & 0.06 & 0.01 & 379.8 & 0.06 & 25.5\\
SDC37.846-0.392$\_1$ & 19:01:53.50 & +04:12:51.000 & 4.4 & 5.03 & 4.08 & N & 0.66 & 36.0 & 9.0 & 0.18 & 0.44 & 0.09 & 0.06 & 4386.0 & 0.38 & 29.1\\
SDC42.401-0.309$\_2$ & 19:09:49.90 & +08:19:47.000 & 0.6 & 2.72 & 4.48 & Y & 0.09 & 63.0 & 4.0 & 0.1 & 0.48 & 0.14 & 0.06 & 210.2 & 0.12 & 40.0\\
SDC43.186-0.549$\_2$ & 19:12:08.90 & +08:52:08.000 & 1.4 & 42.68 & 4.15 & N & 0.24 & 54.1 & 12.0 & 0.26 & 0.44 & 0.11 & 0.05 & 3911.3 & 0.32 & 25.3\\
SDC43.311-0.21$\_1$ & 19:11:17.00 & +09:07:30.000 & 1.1 & 9.39 & 4.25 & N & 0.23 & 51.4 & 9.0 & 0.25 & 0.45 & 0.12 & 0.03 & 611.5 & 0.12 & 25.6\\
SDC43.877-0.755$\_1$ & 19:14:26.20 & +09:22:35.000 & 0.9 & 4.92 & 3.22 & Y & 0.25 & 70.2 & 11.0 & 0.17 & 0.34 & 0.08 & 0.04 & 6239.8 & 0.34 & 19.6\\
SDC45.787-0.335$\_1$ & 19:16:31.20 & +11:16:12.000 & 0.6 & 151.39 & 4.54 & Y & 0.36 & 2.1 & 5.0 & 0.14 & 0.49 & 0.14 & 0.02 & 237.6 & 0.08 & 25.9\\
SDC45.927-0.375$\_2$ & 19:16:56.20 & +11:21:54.000 & 1.0 & 27.69 & 4.21 & N & 0.11 & 39.7 & 5.0 & 0.16 & 0.45 & 0.13 & 0.06 & 885.0 & 0.24 & 23.4\\
RMS-G050.2213-00.6063 & 19:25:57.80 & +15:03:00.000 & 1.3 & 6.22 & 3.3 & N & 0.15 & 24.1 & 11.0 & 0.13 & 0.35 & 0.06 & 0.03 & 221.9 & 0.09 & 22.2\\
RMS-G326.6618+00.5207 & 15:45:02.80 & -54:09:03.000 & 1.4 & $\dagger$ & 1.8 & Y & 0.24 & 30.6 & 8.0 & 0.09 & 0.19 & 0.07 & 0.02 & 334.4 & 0.09 & 23.7\\
RMS-G327.1192+00.5103 & 15:47:32.80 & -53:52:39.000 & 3.7 & 28.55 & 4.9 & Y & 0.24 & 24.2 & 7.0 & 0.48 & 0.52 & 0.27 & 0.03 & 443.7 & 0.1 & 36.8\\
RMS-G332.0939-00.4206 & 16:16:16.50 & -51:18:25.000 & 9.3 & $\dagger$ & 3.6 & Y & 0.27 & 43.3 & 10.0 & 0.16 & 0.39 & 0.1 & 0.02 & 812.8 & 0.09 & 27.2\\
RMS-G332.9636-00.6800 & 16:21:22.90 & -50:52:59.000 & 2.2 & 33.34 & 3.2 & Y & 0.44 & 6.6 & 10.0 & 0.26 & 0.34 & 0.12 & 0.02 & 1723.6 & 0.16 & 23.2\\
RMS-G332.9868-00.4871 & 16:20:37.80 & -50:43:50.000 & 1.8 & 6.27 & 3.6 & Y & 0.21 & 16.2 & 4.0 & 0.1 & 0.39 & 0.1 & 0.04 & 602.1 & 0.16 & 22.4\\
RMS-G333.0682-00.4461 & 16:20:49.00 & -50:38:40.000 & 4.0 & $\dagger$ & 3.6 & Y & 0.4 & 21.7 & 15.0 & 0.29 & 0.39 & 0.1 & 0.01 & 2291.5 & 0.1 & 23.7\\
RMS-G338.9196+00.5495 & 16:40:34.00 & -45:42:08.000 & 3.2 & 10.09 & 4.2 & Y & 0.44 & 35.7 & 6.0 & 0.23 & 0.45 & 0.15 & 0.12 & 109.5 & 0.16 & 39.1\\
RMS-G339.6221-00.1209 & 16:46:06.00 & -45:36:44.000 & 1.9 & 15.07 & 2.8 & Y & 0.25 & 29.4 & 9.0 & 0.22 & 0.3 & 0.12 & 0.03 & 321.0 & 0.1 & 24.1\\
RMS-G345.5043+00.3480 & 17:04:22.90 & -40:44:24.000 & 10.0 & 6.05 & 2.0 & Y & 0.58 & 35.9 & 8.0 & 0.11 & 0.21 & 0.05 & 0.01 & 404.7 & 0.05 & 32.3\\
\hline
\label{Selection:tab}
\end{longtable}
\end{landscape}
\twocolumn
\begin{table*}
\caption[]{Observing properties of the ALMA data. $^a$ Average value of the synthesised beam across all fields. $^b$ Maximum recoverable scale in data, defined as $MRS=\frac{0.6\lambda}{b_{min}}$ where $b_{min}$ is the minimum baseline in the array.}
\begin{center}
\small
\begin{tabular}{c c c c c c c }
\hline
\hline
SPW & Central Freq. & Freq Range& Channel width & Synthesised beam$^{a}$ & P.A.$^{a}$ & MRS$^{b}$ \\
 & [GHz] & [GHz] & [\kms] & [\asec$\times$\asec] & [$^{\circ}$] & [\asec] \\
\hline
 0 & 239.8 & 238.86 $-$ 240.74 & 1.22 & 0.77 $\times$ 0.64 & 58 & 10.2 \\ 
 1 & 241.9 & 240.96 $-$ 242.84 & 1.21 & 0.77 $\times$ 0.64 & 46 & 10.2 \\ 
 2 & 227.1 & 226.16 $-$ 228.04 & 1.29 & 0.81 $\times$ 0.67 & 55 & 10.8 \\ 
 3 & 225.2 & 224.26 $-$ 226.14 & 1.30 & 0.82 $\times$ 0.68 & 56 & 10.9 \\ 
\hline
\end{tabular}
\end{center}
\label{OBSPROPS:TAB}
\end{table*}

\subsection{Continuum determination and imaging}
\label{processing:sec}

\subsubsection{Line Emission}
\begin{figure}
\begin{center}
    \includegraphics[scale=0.45]{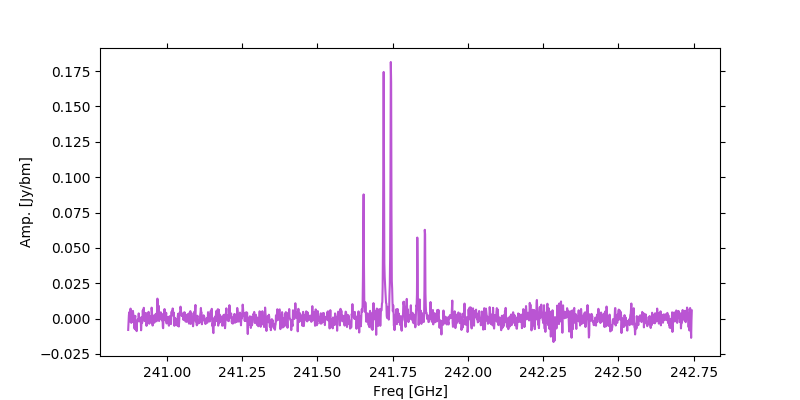}
\\
    \includegraphics[scale=0.45]{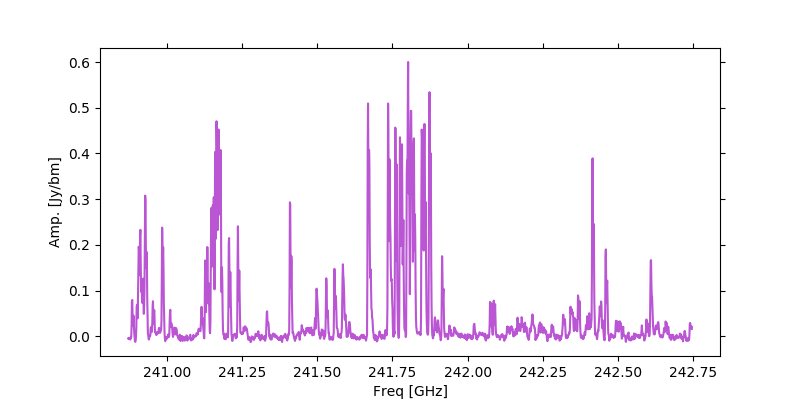}
\caption{Examples of whole spectral window (1.875GHz bandwidth) spectra from the ALMA data. These examples present spectral window 1 from a line `quiet' field (\textit{top:} SDC20.775$-$0.076\_1, $L_*$=6.5$\times10^3$L$_{\odot}$) and line dominated source (\textit{bottom: } SDC35.063$-$0.726\_1, $L_*$=5.2$\times10^3$L$_{\odot}$). The spectra are taken at the position of the strongest continuum detection in the respective fields. The two brightest lines in the top panel are \meth\ (5$_K$-4$_K$) transitions. }
\label{LineExamples:fig}
\end{center}
\end{figure}

The TEMPO target fields are young high-mass embedded protostellar objects meaning that all fields show some level of molecular line emission within the observations. Figure \ref{LineExamples:fig} shows sample spectra from SPW 1 for a molecular line `quiet' field and a line-dominated field. 


To extract continuum emission information about the sample we must remove channels containing molecular line emission from the spectra. To do this, a new CASA based task, \lj\footnote{See https://github.com/adam-avison/LumberJack for more information.}, was developed and used to process these data. \lj was used to process each field in the following way. 

\begin{enumerate}
\item The user selects the required ALMA measurement set and the target field within the measurement set to process.
\item \lj then generates an image cube of the whole target field at full spectral resolution in each SPW. 
\item The position of peak emission within each cube is located. This position is a single voxel (i.e. a position with a Right Ascension, Declination and velocity value. The spectrum along the velocity axis at this position (in RA and Dec) is extracted.
\item The returned spectrum is analysed to locate spectral lines using two complementary methods.

To analyse the spectrum, firstly, a sigma clipping analysis is used. This analysis derives the median and standard deviation values within the spectrum. Next, all channels with values which are either greater than the median value \textit{plus} the spectrum standard deviation multiplied by a clip factor, or less than the median value \textit{minus} the spectrum standard deviation multiplied by a clip factor, are excluded in iterative steps. The iterative analysis stops when either \textit{(a)} the signal-to-noise of the current spectrum is greater than in the previous iteration (here the signal-to-noise is defined as the maximum value in the current spectra divided by the spectrum median value) or \textit{(b)} the percentage change in the standard deviation of the spectrum between iterations is greater than a user defined tolerance. For the TEMPO sample the clip level was set to twice the standard deviation and the tolerance set to a percentage of 95.5\%.

Secondly, a gradient analysis is used to calculate the channel to channel gradient, $G$. $G$ is calculated as:

\begin{equation}
G=\frac{S_{ch}-S_{ch-1}}{\Delta ch}
\end{equation}

\noindent where $ch$ represents a channel number, $S$, the flux density in that channel, and $\Delta ch$ the channel width in units of channel (which here has a value of 1). Channels with $G > 3\sigma$, where $\sigma$ is the theoretical \textit{rms}-noise per channel of the data calculated by the \lj algorithm from the measurement set metadata\footnote{The theoretical $rms$-noise is calculated by extracting the time on-source, $\Delta t$, the median system temperature, $T_{sys}$, channel width in Hertz, $\Delta \nu$ and number of antennas, $N$ used during observation from the measurement set metadata. These values are then combined as:
\begin{equation}
\Delta S =\frac{2kT_{sys}}{A_{eff}\eta\sqrt{N(N-1)\Delta \nu \Delta t}}
\end{equation}

\noindent where $k$ is the Boltzmann constant, $A_{eff}$ the effective area of an ALMA antenna at the observing frequency and $\eta$ the aperture efficiency parameter ($\sim$0.7, \citealt{ALMATECHBOOK})
} are rejected as line contaminated. The combination of the line contaminated channels found using the sigma-clip and gradient analysis are combined to give a conservative first pass at the line free channels in the data set.

\item Following these steps a first pass `line-free' continuum image is made for the combined (e.g. all SPW) data. 

The user then defines continuum sources within this field for a second pass of line-free channel extraction. In the current work this was done using dendrogram analysis from the \texttt{astrodendro} python package \footnote{http://www.dendrograms.org/} to find all the candidate continuum sources in the field. The parameters used during the continuum determination are the same as used during the final source extraction and discussed fully in \S \ref{sourceExtr:sec}.

\item Using the position of these candidate continuum sources, additional spectra are extracted (for the fitted source sizes) and then step (iv) repeated for all spectra, with the line-free channels from each source in each SPW concatenated to created a final list of line-free channels for the target field. The final channel list comprises only channels determined as  line free for all sources in that field which ensures, as far as possible, no line contamination remains within the final images.  

\item The final line-free channel lists for each SPW are created as the output product of the \lj\ process.

\end{enumerate}

There are three potential limitations of note with the \lj analysis. First, typically the theoretical \textit{rms}-noise used in the gradient analysis will be smaller than the measured \textit{rms}-noise in an image as calibration errors are not accounted for when calculating the theoretical \textit{rms}-noise. The implication of this is that some low intensity spectral lines may be overlooked in the gradient analysis, however using a factor $> 3\sigma$ should tend to counteract this, as should the cross comparison with the sigma-clipping analysis. Second, using the positions of continuum sources within the field may lead to spectral line emission from e.g. molecular outflows not being fully excluded as this type of emission would tend to be offset from the position of the continuum sources. The use of a first pass continuum image and a second round of spectral line analysis acts to mitigate this. Visual inspection of the spectra, cubes and continuum images suggests that the effect of this latter limitation is minimal. The third limitation would occur in very line-rich objects within which there was a lot of velocity components or velocity gradients from the molecular material. This would give broad and potentially overlapping spectral line profiles across the observed spectrum and exclude possibly all channels within the observed frequency range. This case does not occur within the TEMPO sample.

To inspect the reliability of the \lj continuum extraction within the TEMPO sample, a sub-sample of eight ($\sim 20\%$) of the TEMPO fields were selected. The fields chosen were amongst the line richest of the RMS and SDC targets (four of each) and have been compared to the ARI-L continuum images available in the ALMA Archive \citep{Massardi21}. Considering all four spectral windows this gives a sample of 32 data points of comparison. The TEMPO and ARI-L continuum image peak flux density pixel values were used for the comparison as this tended to be toward the line richest source in a given field. The primary beam corrected images were used from both ARI-L and TEMPO (prior to self-calibration for TEMPO to ensure a fairer comparison). From this comparison we find all data points are within $\pm20\%$ of one another with the exception of three, showing a mean of 12$\%$ difference with a standard deviation of 18$\%$ (reducing to 8$\%$ and 4$\%$ when excluding the three outliers).

Given the absolute flux density calibration accuracy of ALMA being at the 10\% level in Band 6 \citep[e.g.][]{ALMATECHBOOK}, the amount of line emission removed, differences in CASA version used in calibrating and imaging the data and differences in imaging parameters (e.g. cell size, 0.13\asec ARI-L and 0.093\asec TEMPO) we believe that this constitutes a good matching between the TEMPO/LumberJack line extraction and that implemented by the ARI-L project. For the three data points beyond this range, one shows 25\% discrepancy between ARI-L and TEMPO which is considered marginal. The remaining two are for sources RMS-G013.6562$-$00.5997 in SPW0 (239.8GHz) at +42\% (ARI-L greater than TEMPO) and G326.6618+00.5207 in SPW1 (241.9GHz) at +82\% (again ARI-L greater than TEMPO). For these two objects the spectra are extremely line rich making continuum extract very difficult. We do note that in both cases, comparing the continuum values across all SPWs the TEMPO values are more consistent with a typical smoothly sloping spectral index than the ARI-L data.

The \lj derived line-free channel lists were used to create continuum images of each field in each SPW and as a single aggregate bandwidth (i.e. combined line free channels across all SPWs) continuum image using all line-free channels. The data were imaged in CASA using the task \texttt{tclean}, using `\textit{briggs}' weighting with the robust parameter set to 0.5. The \texttt{tclean} parameter \textit{deconvolver} was set to \textit{multiscale} as the data exhibit extended structure and this algorithm allows for the best quality images in such cases, \textit{scales} of 0, 6, 18, 26 and 43 pixels were used. These values correspond to a delta function, one and three times the beam size in pixels and approximately, 0.25 and 0.4 times the maximum recoverable scales of the data, respectively. The last two scales were found by manual inspection to produce the best images with the TEMPO data. The default \textit{smallscalebias} value of 0.6 was used throughout. 

\subsection{Self-calibration and Noise characteristics}

To ensure the highest dynamic range continuum maps for the TEMPO sample, an initial set of continuum images for the TEMPO fields (both combined continuum from all SPWs and continuum from each individual SPWs) were inspected to check if the respective signal-to-noise ratio was sufficient to undertake self-calibration of the data. For sources where self-calibration was possible (35/38 sources\footnote{The exceptions being SDC18.816$-$0.447$\_$1, SDC30.172$-$0.157$\_$2 and SDC45.927$-$0.375$\_$2}), up to three rounds of phase-only calibration were used to correct the phase solutions and produce the final maps used in our analysis. Amplitude self-calibration was not attempted as amplitude base calibration artifacts were not obvious within the dataset. The single SPW images were made with \textit{nterms} = 1 which assumes a flat spectrum due to fractional bandwidth considerations, whereas the combined SPW images used \textit{nterms} = 2. The cleaning masks for each source were created using CASA's auto-masking capabilities. Images of all fields were created both with and without primary beam correction. Figure \ref{Example_field_images:fig} give example images of the generated maps, with the rest of the sample shown in Appendix \ref{Maps:apdx} (available online).

Following the \lj processing described in Section \ref{processing:sec} the number of channels determined to be `line-free' and thus the total aggregate bandwidth in each SPW and each field is different. This results in the final continuum maps having a non-uniform sensitivity from field to field. An additional factor in the sensitivity achieved in each field is the spatial distribution of extended emission and any associated `missing' flux which is resolved out by the interferometer. Missing flux leads to artifacts such as negative `bowling' in the maps and has a significant effect on the determination of the noise characteristics of the images. 
 
 Table \ref{Noise_props:tab} gives characteristic values for the data set as a whole, with the final two rows giving the equivalent mass sensitivities for the combined spectral window images at $T$ = 15 K and $T$ = 30 K at the average distance to our target fields, D = 3.9~kpc. The sensitivity by field is listed in column 8 of Table \ref{Selection:tab} with column 9 giving the percentage of line free channels (across all four SPWs) found by the analysis described in the previous subsection as an indicator of the wealth of lines found in the sample.
 
\begin{table}
\caption[]{Characteristics \textit{rms}-noise and mass sensitivity properties by spectral window across the sample. `All' row gives the \textit{rms}-noise properties for the combined spectral window images. The mass sensitivities are calculated using the `All' noise values at temperatures of $T$=15K and $T$=30K, the average distance to our target fields D=3.9 kpc and using the dust opacities from \citet{OssenkopfHenning94} for protostellar cores. The model used was that including grains with ice mantels at a column density of 10$^6$cm$^{-3}$ (sixth column, including wavelength column), following e.g. \citet{vanderTak99}. Opacity value used was therefore, $\kappa$=8.99$\times10^{-1}$g$^{-1}$cm$^2$. }
\begin{center}
\small
\begin{tabular}{c c c c c}
\hline
\hline
SPW & \multicolumn{4}{c}{\textit{rms}-noise [mJy]}\\
 & \textit{mean} & \textit{median} & \textit{max.} & \textit{min.} \\
\hline
0 & 0.47 & 0.33 & 1.80 & 0.17 \\
1 & 0.56 & 0.37 & 3.65 & 0.17 \\
2 & 0.50 & 0.30 & 3.17 & 0.16 \\
3 & 0.44 & 0.30 & 2.28 & 0.15 \\
All & 0.26 & 0.23 & 0.69 & 0.09 \\
\hline
  \multicolumn{5}{c}{Mass sensitivity [\solmass]}\\
   & \textit{mean} & \textit{median} & \textit{max.} & \textit{min.} \\
\hline
$T$=15K & 2.5 & 2.2 & 6.5 & 0.9 \\
$T$=30K & 1.0 & 0.9 & 2.7 & 0.4 \\
\hline
\end{tabular}
\end{center}
\label{Noise_props:tab}
\end{table}

\section{Results}

The spectral line free ALMA continuum maps are given in Figures \ref{Example_field_images:fig} and A1 in Appendix \ref{Maps:apdx}. The observed and derived properties for each field as a whole can be found in Table \ref{Selection:tab}, which gives the \textit{rms}-noise value, percentage line free bandwidth, number of sources, and protocluster radius ($R_{cl}$), the field of view of the ALMA primary beam in parsecs at the used target distance, the mean edge length ($X_{mean}$) of a minimum spanning tree in each field and the thermal Jeans fragmentation length ($\lambda_{J}$), respectively. The derivation of $R_{cl}$ and  $\lambda_{J}$ are discussed in section \ref{ClustRadialFrag:sec} and $X_{mean}$ in section \ref{MST:sec}.

\begin{figure*}
    \centering
        \centering
        \includegraphics[width=.46\textwidth]{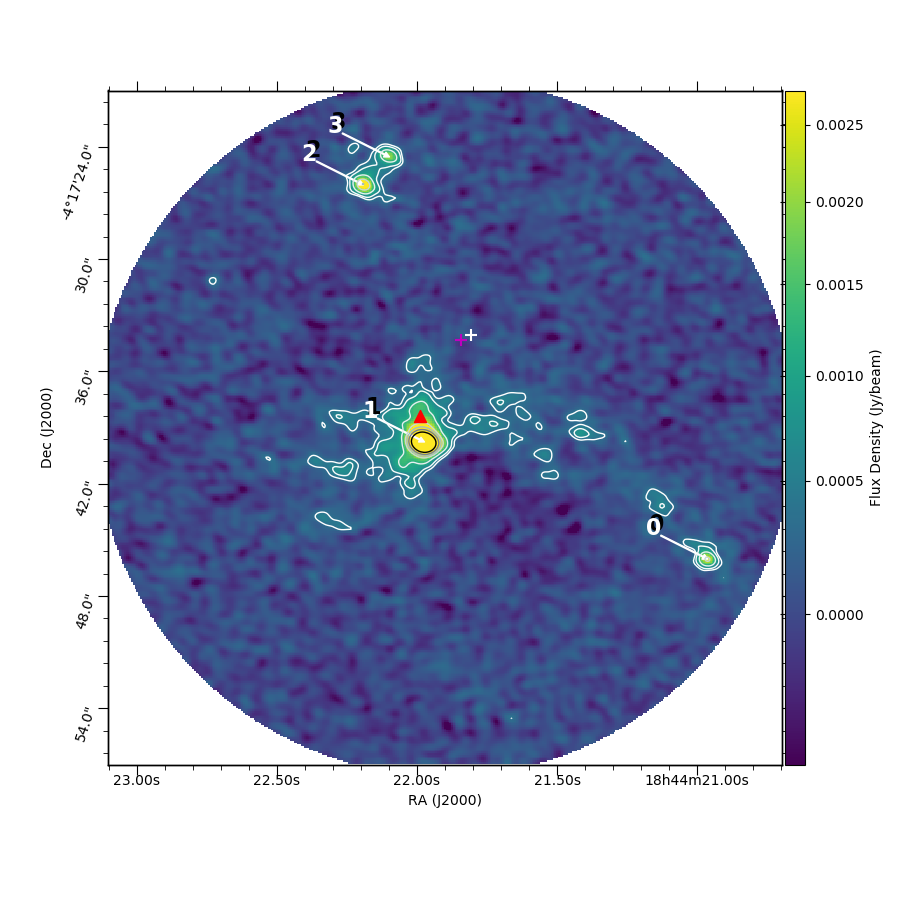}\qquad
        \includegraphics[width=.46\textwidth]{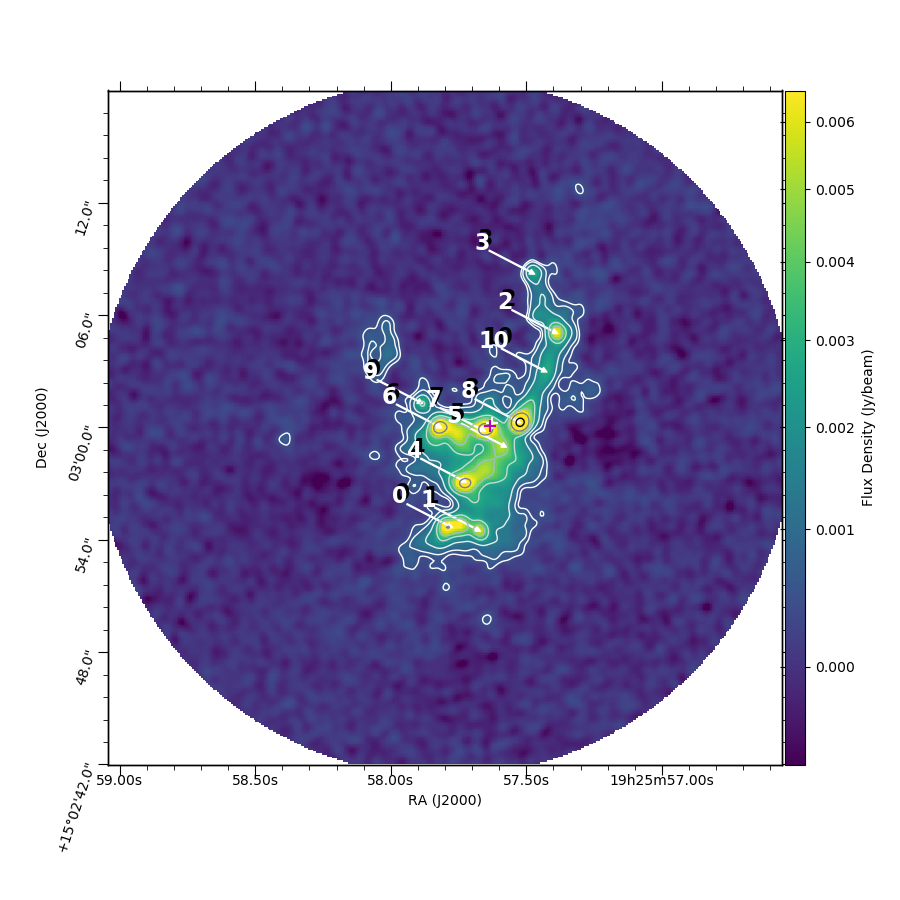}

    \caption{Example maps of the combined aggregate bandwidth continuum images for fields SDC28.277$-$0.352$\_$1 (left) and RMS-G050.2213$-$00.6063 (b, right). Contours are at 3, 5, 10, 20, 30, 50 and 100$\times$ the fields \textit{rms}-noise level, 0.19 and 0.46mJy respectively. The red triangle in (a) indicates the position of the 6.7GHz methanol maser in that field with position from \citet{MMB020to060}. The white and magenta `$+$' symbols give the average and normalised flux density weighted average position of sources in the field. Numbers and arrows indicate the detected sources in each field. Maps of all target fields can be found in Appendix \ref{Maps:apdx}, (available online).}
    \label{Example_field_images:fig}
\end{figure*}

The positions and properties of each detected source (hereafter refered to as a fragment) are given in Table \ref{BIGTABLE:tab}. Column 1 lists the target field (as found in Table \ref{Selection:tab}), column 2 the fragment ID in that field (from 0 to the $n^{th}$), columns 3 and 4 the Right Ascension and Declination of the source. Column 5 gives the measured continuum flux density in the map combining data from all SPWs. Column 6 gives an indication, the ASC$_{score}$, of the likelihood the source is actively star forming (as discussed in \S\ref{CentrallyCondensed:sec}), column 7 denotes which fragment is the brightest in the field, columns 8 and 9 indicate the most central fragment in the cluster for both an arithmetic and normalised flux density weighted average cluster centre, respectively. 

\subsection{Source extraction}
\label{sourceExtr:sec}
To generate the lists of fragments for each field a dendrogram analysis \citep{Rosolowsky08} was run on the final continuum maps for each SPW and on the combined SPW map using the \textsc{astrodendro} Python package. The dendrogram analysis used the following parameters \texttt{min$\_$value} = $5.0\times rms$, \texttt{min$\_$delta} = $1.0\times rms$ and a \texttt{min$\_$npix} equivalent to the number of pixels within the synthesised beam area (approximately 21 pixels). These parameters were selected after experimentation with the TEMPO data to yield realistic results and are consistent with those used by other authors on comparable data sets \citep[e.g][]{Henshaw16}. 

The resulting lists of fragments per image are cross matched in position, with fragments which have a matching peak position (within half the ALMA synthesised beam FWHM for a given field) in all individual images retained. As an independent additional check the \textit{GaussClumps} algorithm within the \textsc{StarLink} software package was run on the combined continuum image and the final dendrogram fragment list cross matched with the \textit{GaussClumps} list. The fragments retained from this cross comparison are our final fragment list for each field. The properties of these fragments are then extracted from each image.

During the dendrogram and \textit{GaussClumps} processing the non-primary beam corrected images were used, as primary beam correction increases the noise toward the edge of each map and leads to both algorithms including spurious noise features in their respective source lists. Using the final fragment lists the flux densities were extracted from the primary beam corrected maps. 

From the sample's 38 fields a total of 287 individual fragments were detected above 5-sigma (in the non-primary beam corrected maps). This gives an average of 7.6 fragments per observed field, with values ranging from 2 to 15 fragments in individual fields. 

\subsection{Protocluster radius and Jeans length}
\label{ClustRadialFrag:sec}
Using the extracted positions and flux densities of the fragments in each field, the protocluster radius and representative values of the Jeans length were derived. 

\begin{figure}
\begin{center}
 \includegraphics[scale=0.48]{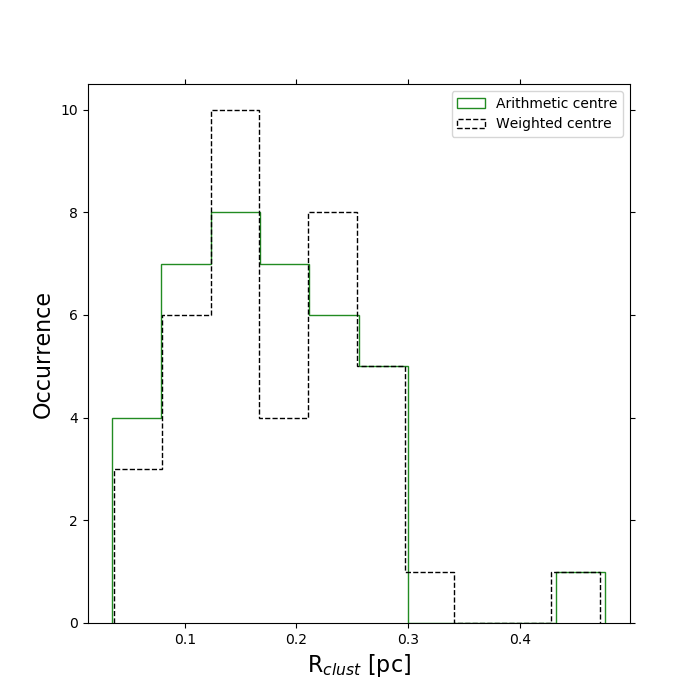}
\caption{Distribution of measured cluster radii across the 38 fields in the TEMPO sample, as measured from the arithmetic mean position and weighted mean position.}
\label{Rclust_hist:fig}
\end{center}
\end{figure}

The protocluster radius is defined here as the distance from the cluster centre to the furthest fragment position in that field, and makes the assumption that the whole cluster is observed within the ALMA primary beam of the TEMPO observations ($\sim$23\asec). The cluster centre is defined in two ways, first as the average position of all fragments in each cluster and second as the average of the flux density weighted fragment position (such that those with greater flux density are weighted more highly, this utilise the field normalised flux density, e.g. fragment flux density divided by the highest fragment flux density in the field). The distribution of cluster radii calculated using both methods can be seen in Figure \ref{Rclust_hist:fig} and values for each field are given in column 11 of Table \ref{Selection:tab}. Using either the arithmetic or weighted mean has little impact on the distribution of protocluster radii in this sample, both peaking between 0.1 and 0.2~pc, with a potential bimodality in the weighted case.

\begin{figure}
\begin{center}
\includegraphics[scale=0.56]{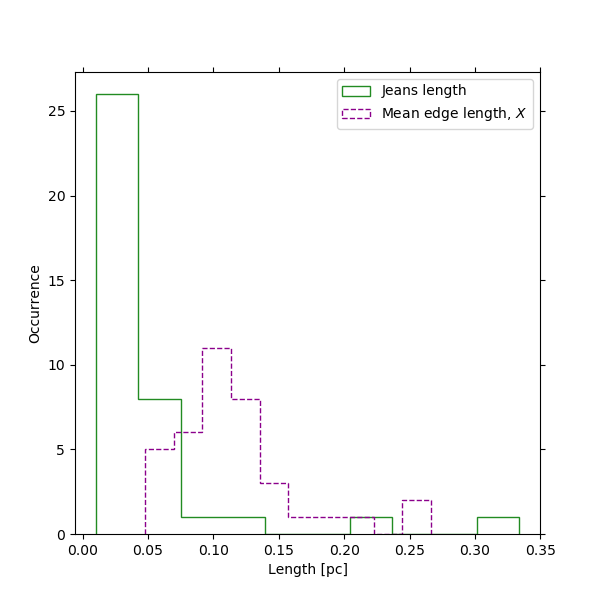}
\caption{Distribution of calculated Jeans lengths, $\lambda_J$, for 38 fields in the TEMPO sample (green solid lined histogram) and measured mean edge length, $X$, from the minimum spanning tree analysis of the sample (purple dashed histogram).}
\label{Jeans:fig}
\end{center}
\end{figure}

In the simplest case, i.e. with no magnetic or turbulent support against collapse, clump fragmentation is expected to occur on the scales of the Jeans length ($\lambda_J$). The $\lambda_J$ values for the TEMPO fields were calculated following the approach used by the ASHES survey \citep{Sanhueza19}:

\begin{equation}
\lambda_J = \sigma_{th}\sqrt{\frac{4\pi^2R_{clump}^3}{3GM_{clump}}}
\label{thermalJ:eqn}
\end{equation}

\noindent where M$_{clump}$ and R$_{clump}$ are the clump masses and radius respectively (columns 13 and 14 in Table \ref{Selection:tab}), for the TEMPO fields these  values were taken from \citet{Elia21}. $\sigma_{th}$ is the thermal velocity dispersion and is given by $\sigma_{th}=\sqrt{\frac{kT}{\mu m_H}}$, with $k$ the Boltzmann constant, $\mu$ the molecular weight (here =2.37) and $m_H$ the mass of the Hydrogen atom. The temperatures, T, used here is the T$_{clump}$ also from \citealt{Elia21}) (given in column 15  of Table \ref{Selection:tab}). Figure \ref{Jeans:fig} provides a histogram of $\lambda_J$ from the fields in the TEMPO sample, this value peaks at $\sim$0.025~pc, with a relatively narrow distribution throughout the sample excluding a few outliers at higher values.

\begin{figure}
\begin{center}
\includegraphics[scale=0.56]{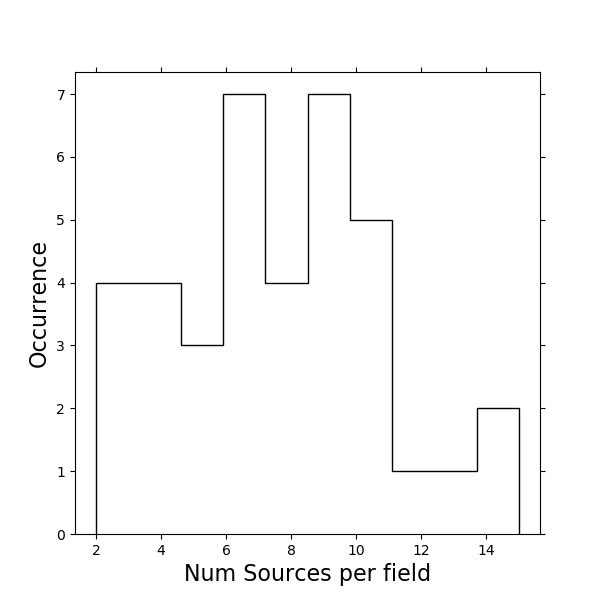}
\caption{Distribution of fragment counts across the 38 TEMPO fields.}
\label{Sources_per_field:fig}
\end{center}
\end{figure}

\subsection{Minimum Spanning Trees}
\label{MST:sec}

\begin{figure*}
    \centering
        \centering
        \includegraphics[width=.46\textwidth]{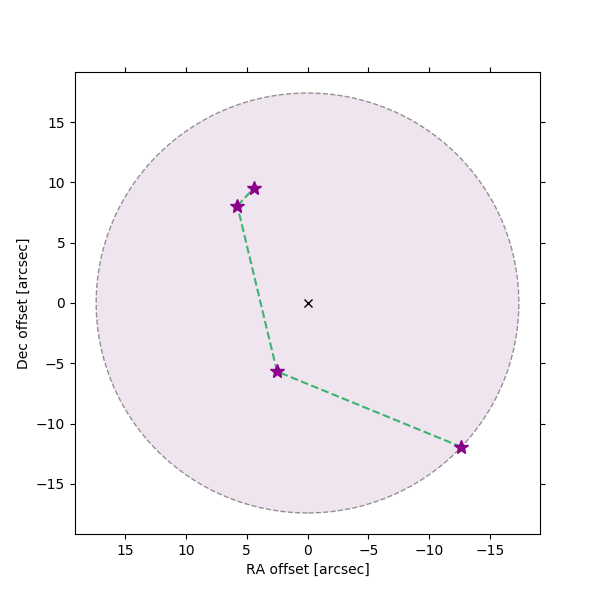}\qquad
        \includegraphics[width=.46\textwidth]{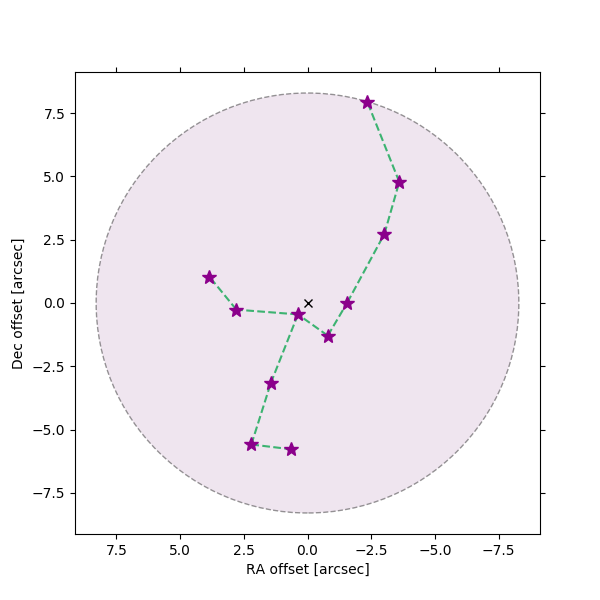}

    \caption{Example minimum spanning trees for fields SDC28.277$-$0.352$\_$1 (left) and RMS-G050.2213$-$00.6063 (right), the same fields as shown in the maps in Figure \ref{Example_field_images:fig}. Purple star icons denote source location, the green dashed lines are the edges of the minimum spanning tree. The purple shaded region is a circle of radius equal to the protocluster radius as defined in \S \ref{ClustRadialFrag:sec} centred at the average position of all sources in the field.}
    \label{Example_MST:fig}
\end{figure*}

Using the extracted fragment positions a set of minimum spanning trees (MST) were generated for each TEMPO field. The MSTs were created using the \texttt{minimum$\_$spanning$\_$tree} module within the Python \texttt{Scipy} module. MSTs provide a set of edges, which describe the minimised set of lines to connect points within a cluster of points. Within this analysis the MSTs are used to describe the mean edge length in the TEMPO clusters as part of the Fragmentation analysis \ref{Frag:sec} and in an investigation of the `Q'-value metric used to described source distributions in Appendix \ref{Qpara:appdx} (available online). Example MSTs are given in Figure \ref{Example_MST:fig}.

From the MSTs the average mean edge length is 0.12pc (not accounting for projection effects). This value is similar to the fiducial core scale of 0.1pc.The distribution of these values is shown in Figure \ref{Jeans:fig} as the purple dashed histogram. The implications of these measurements are discussed as part of the fragmentation analysis \S \ref{Frag:sec}.

\section{Analysis}
The initial focus of the TEMPO analysis is on the structure/fragmentation and the distribution of flux density detected in each of the sample fields. At this stage (\S \ref{ClustProp:sec} to \S \ref{EmissionProp:sec}) no attempt to categorise the detected fragments into star-forming cores and not star-forming fragments is made and, as such, all fragments are treated as potentially star forming. In \S \ref{CentrallyCondensed:sec} a potential interferometric classification into star forming core and non-star forming fragmented material is introduced.

\subsection{Clustering properties}
\label{ClustProp:sec}
\subsubsection{Nearest neighbours}
\label{Near_neighbours:sec}
Using the distance to each target field (given in Table \ref{Selection:tab}) the projected physical separation between each fragment in a given field was calculated from the observed angular separation\footnote{The effects of projection are accounted for when converting from observed angular separation to physical separation by dividing by a factor of $\frac{2}{\pi}$. This of course assumes the cluster is spherical in nature which may not be true in all cases.}. The number of neighbours per fragment within radial cut-offs of 0.03, 0.05 and 0.1 parsecs were inspected. These cut-offs were chosen to be representative as they are all within the fiducial protostellar core size scale (0.1~pc; \citealt[e.g][]{ZinneckerYorke07}) and above the lowest angular separation detectable within our data. This lower limit on detectable angular separation arises from the angular resolution of our data, objects separated by less than this scale would be observed as a single object. Taking the major axis of the average synthesised beam (0.82\asec) this lower limit would be 0.015~pc ($\sim$ 3200~au) at the average field distance of 3.9~kpc and covers a range from 0.007 to 0.025~pc over the TEMPO sample's distance range of 1.8 to 6.3~kpc. Below this it is not possible to distinguish between objects with the current data. 

\begin{figure}
\begin{center}
\includegraphics[scale=0.56]{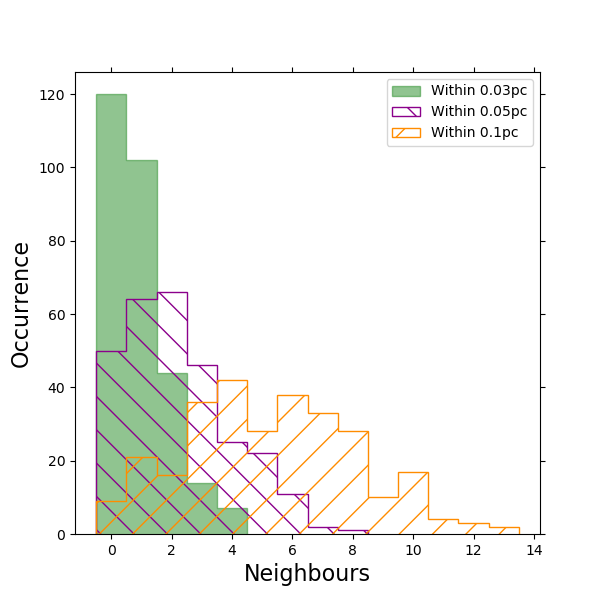}
\caption{Number of nearest neighbours for each fragment in the sample at cutoffs of 0.03~pc (filled green histogram), 0.05~pc (purple '\textbackslash' hatched histogram) and 0.1~pc (yellow '/' hatched histogram). Fragments in the Neighbours = 0 bin do not have a neighbour within that angular offset cut-off.}
\label{Near_neighbours:fig}
\end{center}
\end{figure}

Figure \ref{Near_neighbours:fig} shows the distribution of the number of nearest neighbours within each cut-off interval, including those which do not have a neighbour within that interval in the `Neighbours' equal to 0 bin. Figure \ref{Near_neighbours:fig} shows that very few of the sources within our sample are solitary. 

Over half of fragments (58.2\%) have a neighbour within 0.03~pc, increasing to 82.6\% of fragments with a neighbour within 0.05~pc and 96.9\% with a neighbour within our largest cut-off of 0.1~pc. Only 9 sources (3.1\% of the total sample) do not have a neighbour within the 0.1~pc cut-off. Coupling this with the number of fragments detected per field, ranging from 2 to 15, would suggest that our detected fragments are densely distributed within the target fields (cf. the observing field of view which is $\sim$23\asec, equivalent to $\sim$0.4~pc at the average field distance of 3.9~kpc). Together these values would seem to suggest that in most cases we are seeing in each field the fragmentation of a single star forming core (under e.g. the core accretion scenario) assuming the fiducial 0.1 pc size scale.

\subsubsection{Cluster radial profile properties}
\label{ClustRadial:sec}
\begin{figure}
\begin{center}
\includegraphics[scale=0.48]{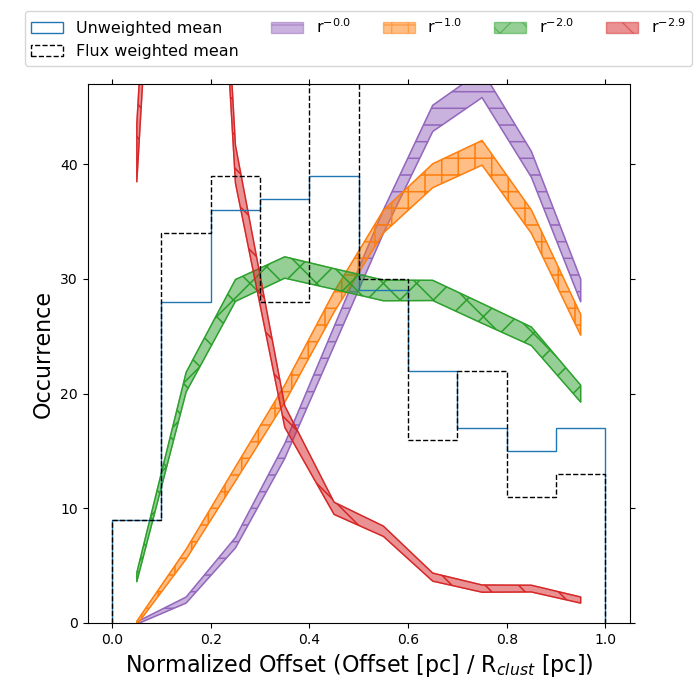}
\caption{Combined distribution of source position as offset from the mean position of all fragments in their respective fields (blue histogram) and the normalised flux density weighted mean position (black dashed histogram), normalised by the cluster radius $R_{clust}$ or weighted cluster radius, for each field. The filled regions show the expected normalized radial profiles, $r^{-\alpha}$ for values of $\alpha =$ 0.0, 1.0, 2.0 and 2.9 (purple with horizontal hatching, orange with `$+$' hatching, green with `$\times$' hatching and red with `\textbackslash'  hatching respectively). These profiles were drawn from 40,000 (10,000 per $\alpha$ value) randomly generated 3-dimension clusters with between 3 and 13 sources within them. See text for further details.} 
\label{norm_Radial_profile:fig}
\end{center}
\end{figure}

To examine the fragment density profiles of the protoclusters in the TEMPO fields, the positional offset for each fragment from their respective protocluster centre was calculated. Figure \ref{norm_Radial_profile:fig} gives the number of fragments at increasing radial offsets from both calculated cluster centres. We use the distance to each field from Table \ref{Selection:tab} to give a physical offset and normalised by the cluster radius. 

Figure \ref{norm_Radial_profile:fig} shows as filled lines the equivalent distribution of field normalised offsets from 40,000 randomly created 3-dimensional clusters. The randomly generated clusters have $N$ sources/fragments (for $N$ randomly selected between 3 and 13, to closely match the true field values without extremes c.f. 2 to 15 is the true range.) and radial profiles of $N(r)\propto r^{-\alpha}$, where $N(r)$ is the number of sources as a function of $r$ given the exponent $\alpha = $ 0.0, 1.0, 2.0 and 2.9 with 10,000 distributions per $\alpha$ value. To generate the cluster distributions the work of \citet{CartwrightWhitworth04} was followed, using their formulae:

\begin{equation}
r = \left(\frac{(3 - \alpha)R}{3}\right)^{\frac{1}{3-\alpha}}
\end{equation}

\begin{equation}
\theta = \cos^{-1}((2\Theta)-1)
\end{equation}

\begin{equation}
\phi = 2\pi\Phi
\end{equation}

\noindent where for each cluster $R$, $\Theta$ and $\Phi$ are randomly selected values between 0 and 1. The resulting $r$, $\theta$ and $\phi$ values are then converted to $x$, $y$, $z$ positions and projected into two dimensions. The projected 2D positions are used to calculate the offset from the cluster central position. The width of the filled lines in Figure \ref{norm_Radial_profile:fig} represent a $\pm$ 1- standard deviation at each histogram bin at a given normalised offset.

The observed data does not agree strongly with any of the plotted $r^{-\alpha}$ profiles, though visually both distributions appear closest to the $r^{-2.0}$ profile with exceptions of an excess between $\sim$0.2 and 0.5 for the normalized offset for both the averaged centre and normalised flux density weighted centre histograms.

As a more quantitative measure the observed data distributions were compared to the generated $r^{-\alpha}$ profiles using a two sample Kolmogorov-Smirnov test. With this method the null hypothesis, that the observed data are drawn from the same distribution as the generated profiles, is tested. Applying this test to the TEMPO data it is possible to reject the null hypothesis for TEMPO fields being drawn from an $r^{-2.9}$ profile with a $p$-value of 0.007 (0.031) (with comparisons to the weighted average values in brackets) these values indicate that the null hypothesis is rejected with only a $<0.7$\% ($<3.1$\%) probability of rejecting a true null, typically a $p$-value of less that 0.05 is considered sufficient to reject the null hypothesis. 

It is not possible to reject the null hypothesis for $\alpha$ values of 0, 1, or 2 with $p$-values of 0.68 (0.68), 0.97 (0.97) and 0.31 (0.11) respectively. This finding shows the TEMPO fields do not show a highly centrally condensed profile ($\alpha$=2.9) but beyond this it is not possible to not rule out that shallower radial profiles exist within our target fields. This may also suggest that different population distributions, e.g. fractal or broken power law, are present within the sample. The small source counts in the TEMPO sample limits the ability to conduct this analysis on a field by field basis. 

The $Q$-parameter, introduced by \citet{CartwrightWhitworth04}, has proven within the literature to be a useful diagnostic of stellar distributions within clusters. However, in testing this parameter for fields in the TEMPO sample it was found that the fragment counts were too small for $Q$ to be used robustly. A similar interpretation of the $Q$-parameter for small number clusters is seen in \citet{Parker18} in the case of L1622 for as many as 29 sources. Details of an investigation into the $Q$-value for small source/fragment counts conducted by the TEMPO team is presented in Appendix \ref{Qpara:appdx}.

\subsection{Fragmentation scales}
\label{Frag:sec}
In addition to the the cluster profile characteristics, the scales upon which the material in each field is fragmenting was investigated by comparing the source separations to the Jeans fragmentation length.

Table \ref{Selection:tab} lists in column 13 the calculated values of $\lambda_J$ for each observed field. The average $\lambda_J$ value is 0.05~pc. These values are compared to the mean edge length, $X$, which gives the distance between sources along the minimum spanning tree (this is the same $X$ as seen in Equation B1 corrected for projection effects by division of a factor $\frac{2}{\pi}$ \citep{Sanhueza19}. As can be see in Figure \ref{Jeans:fig}, $X$ peaks at $\sim$0.1~pc and covers a smaller range of values than the Jeans Lengths, but with typically higher values.

The ratio of $\lambda_J$ to $X$-values throughout the sample range from 0.33$\times$ to 9.1$\times$, with only one field (SDC30.172$-$0.157$\_$2\footnote{This field is one of the lowest SNR sources in the sample and contains only two fragments, which also may account for this result.}) having $\lambda_J$/$X$ less than 1. For the majority of TEMPO fields therefore the observed mean edge length between fragments is not consistent with thermal Jeans fragmentation and thus another, non-thermal, mechanism must be presented to account for the observed fragmentation. 

Filamentary or cylindrical fragmentation as seen in the works of \citet{Ostriker64,Henshaw16,Lu18} would tend to have length scales greater than those observed in the TEMPO fields. Using $T_{clump}$ from Table \ref{Selection:tab} and Equation 2 from \citet{Henshaw16} the $\lambda_{frag}$ for the TEMPO sample was calculated. As the hydrogen number density is unknown for the TEMPO sample values between $10^4$ and $10^6 $cm$^{-3}$ were input. Comparing of the $\lambda_{frag, f}$ to $X$ for each TEMPO source shows that $\lambda_{frag, f}$ is consistent with $X$ for 19 TEMPO fields at a density value of $5.0\times10^5$ cm$^{-3}$ and 32 TEMPO fields at value of $1.0\times10^6$ cm$^{-3}$, both densities appropriate for star forming regions. Meaning that filamentary fragmentation could account for the fragmentation scales seen some of the TEMPO fields. However it is noted that, morphologically the TEMPO sample do not appear particularly filamentary.

It is noted that the works of \citet[][]{Henshaw16,Lu18} have observed filamentary fragmentation in mosaic images of larger regions of sky than the present work and were targeted towards known filamentary objects, whereas the TEMPO sample had no such selection criteria. It is expected that the TEMPO fields observe the whole of the local star-forming core because to the physical scale of the ALMA field of view at the distances to the TEMPO sample is being greater than the fiducial star forming core size. However, it is not possible to rule out additional sources beyond the field of view limits without additional data to create mosaics covering a region of the sky.

Additionally, turbulent fragmentation can cause a deviation away from the Jeans length, in either direction \citep{Pineda15} and could potentially also account for the fragmentation scales seen in the TEMPO sample in addition to some filamentary fragmentation.

\subsection{Emission properties}
\label{EmissionProp:sec}
Beyond the physical structure of the observed fields, an examination of the distribution of observed flux density within each region was conducted. This analysis aimed at resolving whether the protoclusters comprise several equally bright fragments or are dominated by a single high flux density fragment. Due to relatively small numbers of fragments in each field, the combination of data across all observed fields was used to assess the general trend of flux density distribution within the sample. 

\begin{figure}
\begin{center}
\includegraphics[scale=0.56]{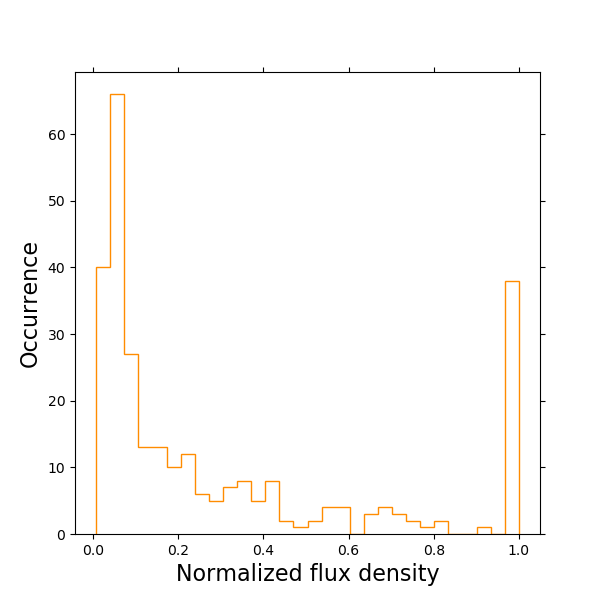}
\caption{Histogram of normalised flux density of each fragment in the sample. We normalise the flux density per field by the highest flux density fragment in the each field, $S_{frag}/S_{brightest}$.} 
\label{NormFlux:fig}
\end{center}
\end{figure}

Figure \ref{NormFlux:fig} gives the distribution of fragments, over all target fields, as a function of normalised flux density. The normalised flux density in  each field was defined as the division of each individual fragments observed flux density by that of the fragment with the highest flux density in its host field. As such the brightest fragment in each field will have a normalised flux density of 1 (and clearly seen in Figure \ref{NormFlux:fig} and all other sources values $<1$.

It is clear from Figure \ref{NormFlux:fig} that the TEMPO fields appear dominated by single (or infrequently a very small numbers) of bright fragment(s) with the remainder of the population being significantly fainter. Across the whole sample the majority (69.4\%) of fragments have $<20$\% of the flux density of the brightest fragment in their respective field.

To assess this, the ratio of the flux density of the brightest object to the sum of the flux density of all other fragments in a given field was calculated as, $\frac{S_{max}}{\Sigma S_{other}}$, hereafter termed $S_{budget}$. This value would be $\sim\leq1$ if the ``faint'' field fragments dominate the flux density budget or $>$1 if the brightest fragment dominates. Of the 38 TEMPO fields, 22 fields have an $S_{budget}$ $\leq$1 and as such the fainter fragments dominate the flux density budget, suggesting that the flux density is relatively evenly distributed amongst the fragments in these fields.

For the 16 fields with $S_{budget} >$1, indicating the flux density distribution is dominated by one (or a small number of) fragment(s), the ratio of the brightest fragment in that field to the second brightest was calculated. This allowed assessment of whether the flux density budget is dominated by a single source. Of these 16 fields, 14 contain a bright fragment which has a flux density at least 3$\times$ greater than that of the next brightest fragment in the field and as such these fields appear dominated by a single high flux density object. The remaining 2 fields (SDC18.816$-$0.447$\_$1 and SDC30.172$-$0.157$\_$2) contain a second fragment with between 0.83 and 0.91$\times$ the flux density of the brightest, with the remaining fragments in these fields not contributing significantly to the flux density budget. For these two fields, it is noted that both are found to contain only two fragments, and that these two fragments are separated by 0.12pc (5.9\arcsec\ at a distance of 4.3kpc for SDC18.816$-$0.447$\_$1) and 0.06pc (3.4 \arcsec\ at a distance of 4.2kpc for SDC30.172$-$0.157$\_$2). Not accounting for projection these separation are larger than $\lambda_J$ for SDC18.816$-$0.447$\_$1 and smaller than $\lambda_J$ for SDC30.172$-$0.157$\_$2 as calculated in Section \ref{Frag:sec} (c.f Table \ref{Selection:tab}). It is apparent from the TEMPO fields, that whilst the faint fragments dominate the number counts they do not typically dominate the flux density budget in a given field. 

Whilst is is possible to equate the measured flux density of a fragment to a mass for that fragment, this has not been attempted within the current work for the following reason. Given the expectation that each small scale fragment is internally heated by an evolving protostar, then to derive a meaningful masses would require knowledge of the temperatures of each fragment. This cannot be derived from the the continuum flux density alone and as such the analysis has been limited to discussion of flux density. Further investigation of the masses of the observed fragments will be conducted under a future work, when a more detailed analysis of the chemical properties of the TEMPO sample has been completed. Such an analysis should gives a reliable way to estimate temperatures and calculate meaningful masses.


\subsubsection{Brightest source properties}
\label{brightSource:sec}
Given the dominance, in terms of flux density, of single or small numbers of fragments within each TEMPO field, an analysis of the properties of these objects with respect to high-mass star-formation tracers, and their relative position in the TEMPO field was undertaken. Three samples were considered, in addition to the brightest fragment per field (sample size 38, one per field). Those being methanol maser associated TEMPO fragments (sample size 27, explained in next paragraph), IR object associated fragments (sample size 38) and the sample of the most central fragment in each TEMPO field (e.g. those fragments located closest to the non-intensity-weighted mean position in each TEMPO field, sample size 38).

There are 28 TEMPO fields with a known 6.7~GHz \meth\ maser source within the ALMA primary beam (see Table \ref{Selection:tab}), in each case there is only a single maser within the ALMA primary beam. A maximum offset limit between a TEMPO fragment peak position and the maser position of 2\arcsec\ (equivalent to a physical separation of $\sim$ 0.04~pc at the average source distance of 3.9~kpc) was applied to assign maser association with a TEMPO fragment. With this limit, the maximum offset retained is 1.4\arcsec\ (a physical separation of 0.03~pc at the assumed target distance). All other source-maser offsets are below this, with a minimum of 0.07\arcsec\, (0.8~milli parcsec at the source distance). This offset limit excludes the maser in field RMS-G034.8211$+$00.3519 for which the maser is offset by $\sim$ 13.8\arcsec\ from the nearest TEMPO source. It should be noted that the maser in this field only has a position recorded from the single dish Parkes Radio Telescope, rather than an interferometric position from ATCA in the MMB catalogues. Thus its positional accuracy is significantly lower.

Infrared sources were drawn from the Hi-GAL catalogues \citep{Elia21} at 70\mewm. Given the angular resolution of these Hi-GAL data, the maximum offset limit between the TEMPO fragment peak position and the IR source was limited to 5\arcsec\ (equivalent to a physical separation of $\sim$ 0.09~pc at the average source distance of 3.9~kpc) following the approach used by \citet{Jones20} for Hi-GAL - maser association. In cases where multiple TEMPO sources fell within this cutoff the source with the smallest offset was deemed the associated source. Using this limit, the maximum offset retained was 2.77\asec\ (a physical separation of 0.05~pc at the assumed target distance). 

\begin{figure}
\begin{center}
\includegraphics[scale=0.5,trim=50 100 100 0]{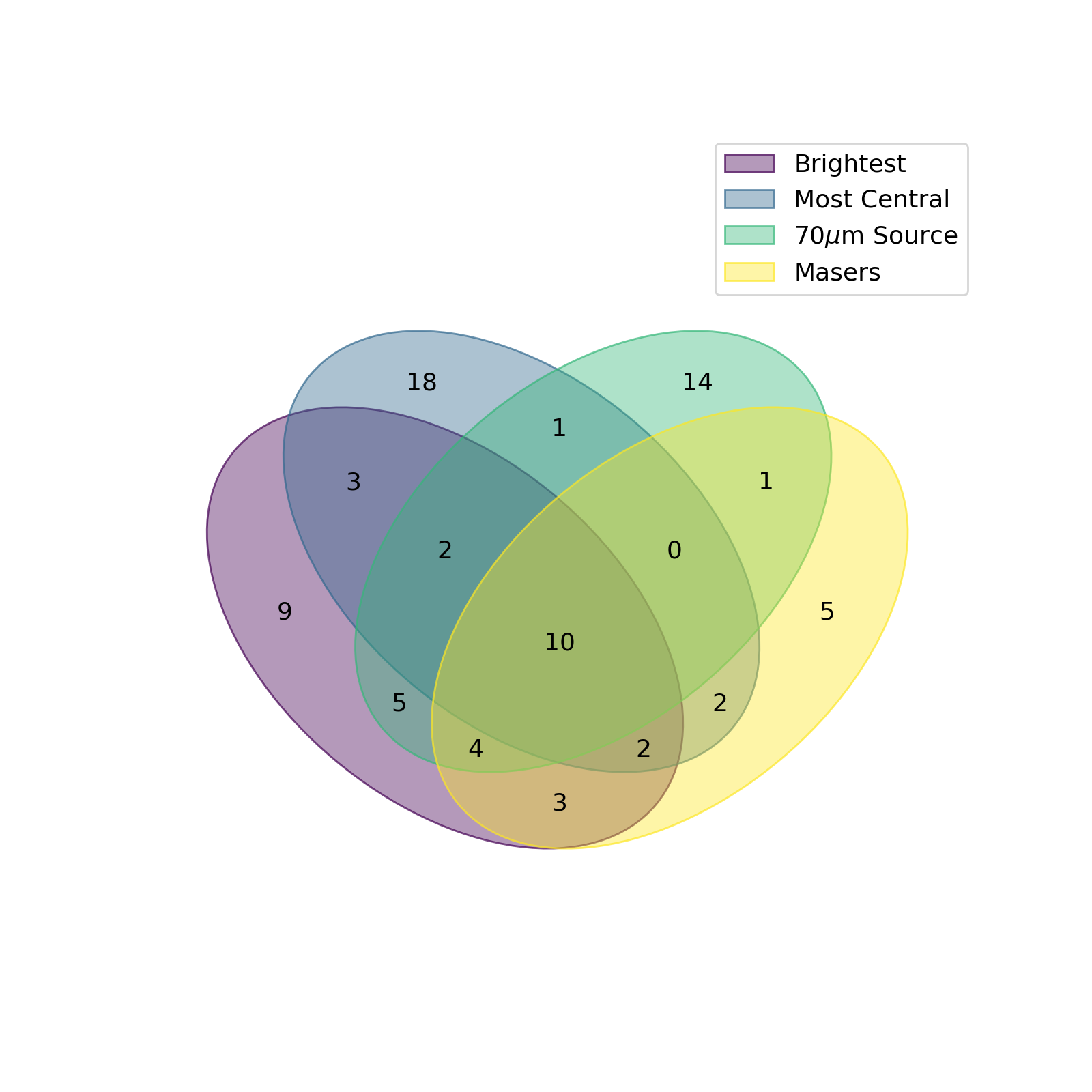}
\caption{Venn diagram showing the overlap of samples comprising, the brightest TEMPO field fragment (purple segment, 38 fragments total), the high-mass star forming tracing \meth\ masers associated fragments (yellow segment, 27 fragments total), infrared sources associated sources (detections at 70\mewm\ by \textit{Herschel} \citep{Elia21}, green segment, 38 fragments total) and the most central fragments in each TEMPO fields (arithmetic average, as the blue segment, 38 fragments total)}.
\label{BCIMVenn:fig}
\end{center}
\end{figure}

Figure \ref{BCIMVenn:fig} is a Venn diagram of the considered samples, with the given values indicating the number of fragments in each overlapping set. From this figure it can be seen that the brightest fragments in the TEMPO fields are commonly associated with the other sample types, with 76\% of the sample (29/38) being a member of at least one of the other sets. Looking at two sample comparisons the brightest TEMPO fragments are, perhaps unsurprisingly, most commonly associated with 70\mewm\ IR sources, (55\% of fields), followed by \meth\ masers (in 50\% of fields) and are also the most central source in their respective field in 45\% of cases. 

For the 50\% of fields which do not show a maser-brightest TEMPO source association, 11 (29\%) do not have a maser detection and of the remaining 8 sources, 6 have the second brightest source in the field associated with the maser. Viewed another way, in 70\% (19/27) of the TEMPO fields with a maser, the maser-associated fragments is also the brightest fragment. Such high overlap in membership of the brightest fragment and maser associated samples indicates that the brightest fragment in each field is a good proxy for the local high-mass star forming core candidate. All TEMPO fields were covered by the MMB survey at 6.7GHz meaning fields without a maser are due to a non-detection during that survey, not a lack of observational data. Thus the absence of \meth\ masers in 11 of the TEMPO fields may be indicative of a younger evolutionary stage in those fields, making the brightest fragments within these fields good candidates for follow-up maser observations to detect emergent masers or weak masers which were below the detection limit of the MMB survey. Alternatively, the absence of masers may simply be an inclination effect due to the beamed nature of maser emission.

Making the broad assumption that the brightest fragment in each field is also the most massive, it is interesting to note that the 55\% of TEMPO fields do not have the most massive fragment at their central position. High-mass Main Sequence stars are more commonly seen at the centre of stellar clusters and under the \textit{clump-fed} model are expected to spend at least part of their evolution there. This result is suggestive of either, some TEMPO fields being in early stages of evolution prior to the migration and settling of more massive cores at the cluster centre or the TEMPO observations are limited in either sensitivity or field of view meaning the sample are missing weaker (or out of field) sources thus skewing the true central position. Of course, a more robust investigation of the masses in the TEMPO sample requires a temperature measurement of each source (not just the clump temperature stated in Table \ref{Selection:tab}) which is beyond the scope of this work and will be addressed in a future paper.

\subsection{Looking for signatures of evolution}

A primary goal of the TEMPO survey was to look for evidence of evolution, or lack thereof, with the fields observed. Two of the properties derived from the continuum maps are worthy of note when inspected against the target clump luminosity\footnote{Here luminosity acts as a proxy of age, with lower luminosity indicating younger star forming clumps and vice versa.)} from \citet{Elia21}. These are namely the number of fragments (\S \ref{Near_neighbours:sec}) and percentage bandwidth which is spectral line-free within the data (column 9 in Table \ref{Selection:tab}). Beyond these two properties little indication of evolutionary trends are seen within the analysis conducted for this paper. Chemical and kinematic analysis of the TEMPO sample are to be published in future works (Asabre Frimpong et al. \textit{in prep.} and Wang et al. \textit{in prep.}).

\subsubsection{Number of fragments}
\begin{figure}
\begin{center}
\includegraphics[scale=0.5]{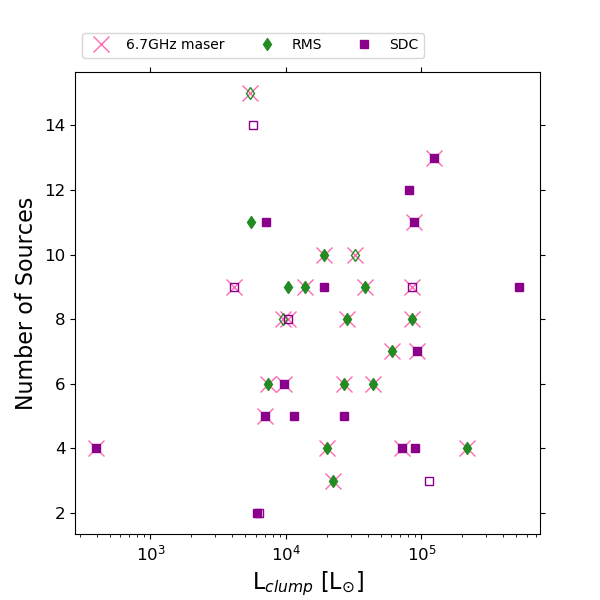}
\caption{Number of detected fragments as a function of clump luminosity \citep{Elia21}. Symbols are as per Figure \ref{ColourLum:fig}.}
\label{L_frag:fig}
\end{center}
\end{figure}

Figure \ref{L_frag:fig} gives the number of fragments extracted from the TEMPO fields as a function of clump luminosity from the work of \citet{Elia21}. Here we see no clear correlation between these two properties. This is note worthy as in a typical star-forming scenario as the source evolves the power output from the bipolar outflows will increase. Given this, one could expect greater disruption of the material in the field and thus a greater amount of fragmentation in more evolved clumps, something not seen in the TEMPO fields.

\subsubsection{Spectral line-free bandwidth}
A `by-product' of the \lj\ (\S \ref{processing:sec}) analysis conducted to find spectral line-free channels within the TEMPO data, the value of percentage bandwidth used in continuum, is also a measure of a fields line richness. The lower the available bandwidth for continuum imaging the higher the spectral line density within the target.  

\begin{figure}
\begin{center}
\includegraphics[scale=0.5]{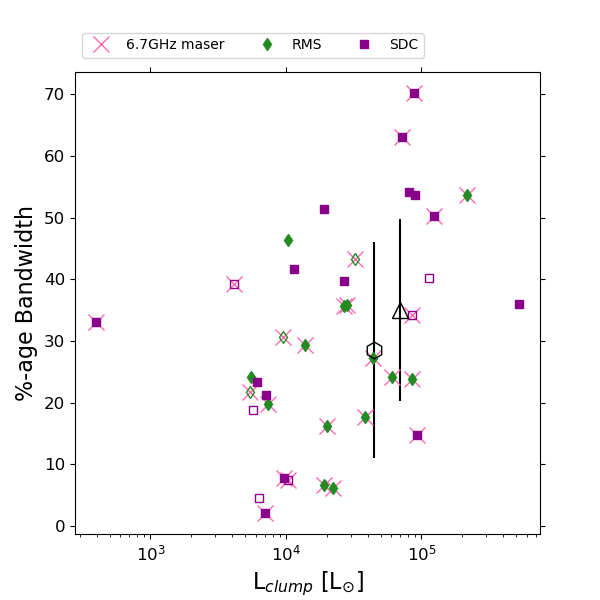}
\caption{Percentage spectral line-free bandwidth in the TEMPO ALMA data as a function of clump luminosity \citep{Elia21}. Symbols are as per Figure \ref{ColourLum:fig}, plus the black hexagon and black triangle mark the average values for the TEMPO field with and without an associated 6.7GHz \meth\ maser, respectively. And the associated error bars give the scale of 1 standard deviation of these samples.}
\label{L_BW:fig}
\end{center}
\end{figure}

Figure \ref{L_BW:fig} gives the percentage of the total observed ALMA bandwidth used in generating the continuum images as a function of clump luminosity \citep{Elia21}. Here there is evidence of a tentative correlation between $L_{clump}$ and percentage line-free bandwidth, (albeit with a large scatter at any given luminosity), with lower luminosity clumps having less line-free bandwidth (ergo more line rich) and higher luminosity clumps having a greater available bandwidth for continuum imaging (thus less spectral line emission). This could be explained in terms of evolution as the destruction of complex molecular species by the increasing radiation output of an evolving source as its luminosity increases. 

Also plotted in \ref{L_BW:fig} are the average values of two TEMPO field sub-samples, those which are 6.7GHz maser associated (hexagon marker, with average $L_{clump}$=4.5$x10^4$L$_{\odot}$ standard deviation 4.9$x10^4$L$_{\odot}$, and percentage bandwidth = 28.6 with standard deviation 17.6) and those which are not (triangle marker,  with average $L_{clump}$=6.9$x10^4$L$_{\odot}$ standard deviation 1.4$x10^5$L$_{\odot}$, and percentage bandwidth = 35.0 with standard deviation 14.8). A small offset is seen between these two samples which suggests that the lower luminosity, thus younger sample are preferentially the maser associated sources. Again this aligns with expected evolutionary traits of the methanol maser, which are thought to be destroyed as protostellar luminosity increases \citep{Breen10,Jones20}.

\subsection{Distinguishing between star-forming and non star-forming fragments}
\label{CentrallyCondensed:sec}

To conclude the discussion of the detected fragments within the TEMPO fields, an initial analysis into the nature of the detected fragments was conducted with the aim of distinguishing between fragments likely to be star-forming cores and those which are not star-forming, simply fragments (as they have hitherto been referred). This analysis compares the phase and amplitude properties of simulated interferometric visibility data of point sources, Gaussian profile sources and Gaussian plus point (hereafter Gaussian$+$Point) sources to the observed TEMPO visibility data. The three model profiles used were selected as a basis of comparison with the TEMPO fragments under the assumption that such profiles are likely to be present in actively star-forming cores. Particularly point-like, ergo unresolved, objects and point-like objects within extended envelopes. The use of a Gaussian profile as a comparison was a pragmatic choice as it provides a simple and quantifiable model of an centrally peaked, extended emission. A full description of the approach used is given in Appendix \ref{centCond:apdx}, whilst a summary is given in the following paragraphs.

A catalogue of point-like, Gaussian and Gaussian$+$Point source simulated datasets were created using the CASA task \texttt{simobserve}. The simulated data matched the TEMPO typical \textit{rms}-noise, FOV, synthesised beam shape, frequency tuning and bandwidth. The simulated datasets were created to cover a range of signal-to-noise ratios, differing source axis ratios (in the case of Gaussian \& Gaussian$+$Point models) and differing Gaussian peak emission to Point source peak emission ratios (for Gaussian$+$Point models)\footnote{The simulated data were generated with SNR values of 5, 10, 15, 20, 30, 40, 50, 60, 80, 100, 120, 150, 200, 300, 500, 750, and 1000. The SNR was defined as the peak pixel emission to off-source noise ratio. For Gaussian models axis ratios of 1:1, 2:1 and 3:1 were used. For Gaussian$+$Point models, Gaussian peak emission to point peak emission ratios of 1:1, 1:0.5 and 1:0.1 were used.}. 

For each SNR, axis ratio and peak emission ratios 100 simulated data sets were generated, each with a different thermal noise spatial distribution, controlled by a random seed value within \texttt{simobserve}. From the simulated dataset the amplitude and phase values were extracted at the position of the model source within them. Each simulated dataset contained a single source. The simulated amplitude and phase values were then used to generate empirical relations between SNR and amplitude and phase properties (c.f Appendix \ref{emprels:sec}).

\begin{figure}
\begin{center}
\includegraphics[scale=0.4]{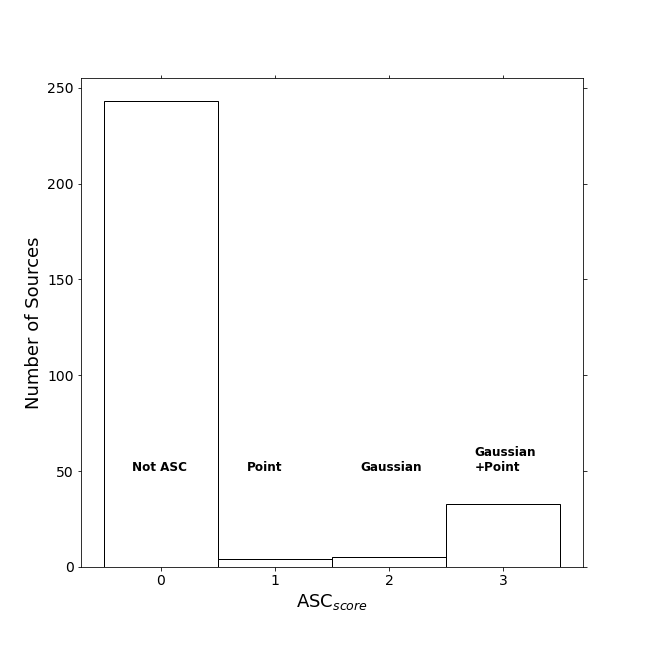}
\caption{Distribution of ASC$_{score}$ assigned to the TEMPO detected fragment sample. Individual scores per source are given in Table \ref{BIGTABLE:tab}}.
\label{ASCHist:fig}
\end{center}
\end{figure}

The same amplitude and phase properties were then extracted for each detected fragment in the real TEMPO data and compared to the relations generated from the simulated data. Based on the TEMPO fragment properties at its recorded SNR a decision tree (see Figure \ref{ASCflow:fig}), was followed to categorise each fragment into being either a point-like source (given a score of 1), Gaussian profile source (score of 2), Gaussian$+$point source (score of 3) or other morphology (score of 0). TEMPO fragments with scores of $\geq1$ are considered active star-formation candidates (ASCs) with this score hereafter referred to as the ASC$_{score}$. Figure \ref{ASCHist:fig} plots the breakdown of ASC$_{score}$ for the 287 sources detected in TEMPO and the ASC$_{score}$ of each source are given in column 13 of Table \ref{BIGTABLE:tab}.

\begin{figure}
\begin{center}
\includegraphics[scale=0.4]{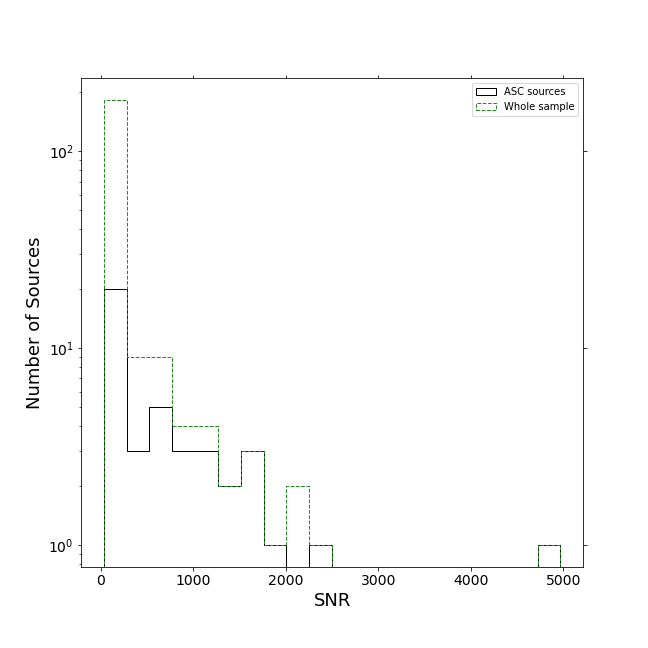}
\caption{Distribution of SNR for ASC sources (ASC$_{score}>0$). The whole sample are given in green-dashed and ASC sources as the black solid.}
\label{ASCScoreSNR:fig}
\end{center}
\end{figure}

Within the TEMPO sample, 42 fragments are found with an ASC$_{score} \geq 1$. Hereafter, these 42 fragments (14.6\% of the sample) are discussed together and labelled as actively star-forming candidates (ASC) sources. The remainder of fragments are considered not to be currently star-forming, as we do not see the required characteristics within our current data. This does not exclude the possibility that they are prestellar and in the future may coalesce further to go on to form protostars nor that they currently are star-forming but the recovered visibility data do not allow confirmation within these data. Alternately, fragments with ASC$_{score}$<1 are possibly clumps of material created by, for example, the disruptive effects of outflows from the protostellar sources \citep[e.g]{Arce07,Rosen20}. A full investigation of the gas kinematics and outflow properties of the sample will appear in a future works from the TEMPO project. Across the TEMPO sample 31 field have at least 1 ASC source, with only 7 fields having no detected ASC source. Specifically the fields without ASC are RMS-G017.6380$+$00.1566, SDC24.381-0.21$\_3$, SDC28.147-0.006$\_1$, SDC30.172-0.157$\_2$, RMS-G034.8211$+$00.3519, SDC45.787$-$0.335$\_$1, RMS-G332.9868$-$00.4871.

Figure \ref{ASCScoreSNR:fig} shows the distribution of SNR values for ASC sources overlaid on the SNR characteristics for the whole TEMPO sample. There is a fixed lower limit of SNR equal to 30 for ASC sources as specified in Appendix \ref{centCond:apdx}. It is clear that ASC sources are drawn from across the SNR parameter space and a large fraction is in the low SNR regime. This is to be expected for two reasons. Firstly, the low flux density (thus low SNR) fragments dominate the fragment counts in the TEMPO sample (c.f. \S \ref{EmissionProp:sec}) and secondly, the bounding conditions for ASC acceptance are broader at lower SNR (c.f. Equations \ref{pointfwhmupperbound:eqn} through \ref{gaussian2whmupperbound:eqn}). This latter point may also account for some of the high SNR fragments not being included in the ASC sample in that the stricter bounds at high SNR may exclude sources which are close to, but not within, those bounds. Though of course high flux density in the mm-wavelength regime does not automatically indicate a star-forming source.

\begin{figure}
\begin{center}
\includegraphics[scale=0.4]{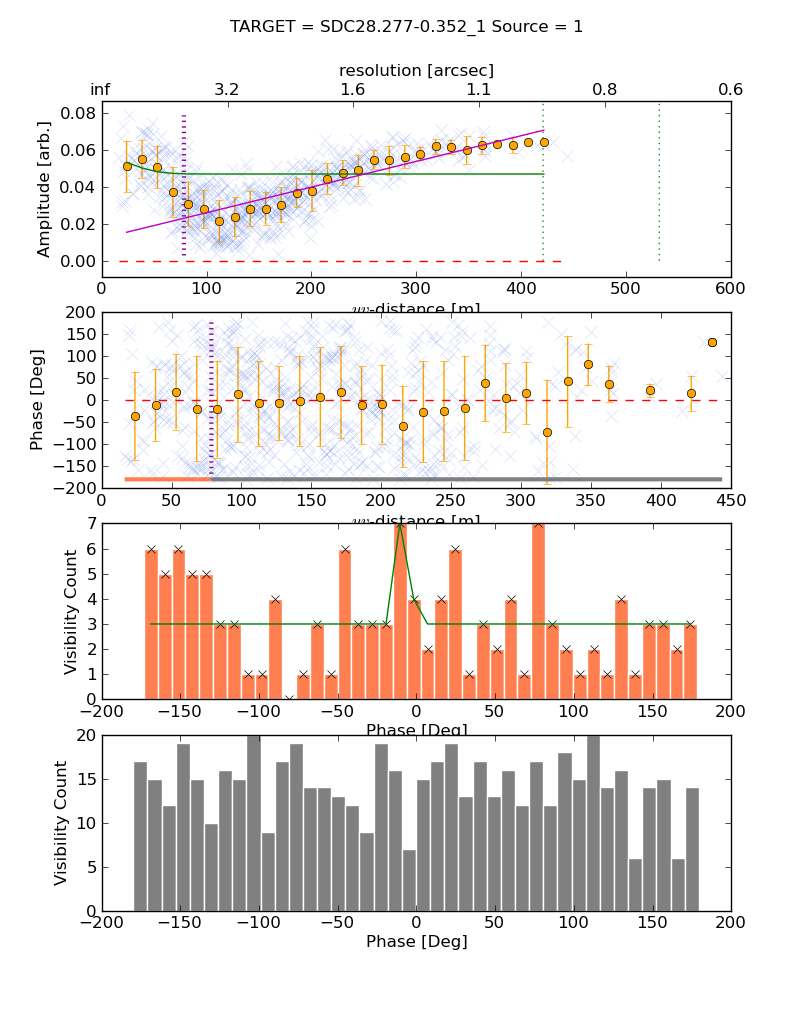}
\caption{Visibility analysis plot for SDC28.277$-$0.352$\_$1 fragment 1 c.f. Figure \ref{Example_field_images:fig}a. Showing signs of poor visibility subtraction using a single Gaussian component at the brightest field source position.}
\label{SDC28-277_AnP:fig}
\end{center}
\end{figure}

A third factor which maybe account for the exclusion of some high SNR fragments from the ASC sample is seen in the inspection of the post ASC analysis visibilities. The method used to extract the amplitude and phase data, uses a model subtraction of all other sources in the field to reduce their impact on the target sources visibility characteristics. However, inspecting the post-subtraction plots and images, some bright single sources do not appear well fit by a single simple Gaussian. Figure \ref{SDC28-277_AnP:fig}, presenting the visibility analysis plot for SDC28.277$-$0.352$\_$1 source 1 is a good example of such a problem, in which some residual Gaussian-like profile in amplitude and a large scatter in phase after source subtraction can be seen. A more robust modelling of the fragments, at the sub-resolution scales would be required to account for these kinds of source. Achieving this for the TEMPO sample size is beyond the scope of this current investigatory analysis.

\subsubsection{ASC characteristics} 

The ASC source sample was compared with the same three source samples as in \S \ref{brightSource:sec} to inspect for common characteristics within the ASC sample. The source samples used in the comparison were, 6.7GHz \meth\ maser associated sources, brightest TEMPO field sources, most central TEMPO field sources (using the arithmetic mean of field source positions). The latter two sample have a size of 38 (one per TEMPO field). 

\begin{figure}
\begin{center}
\includegraphics[scale=0.5,trim=90 100 100 0]{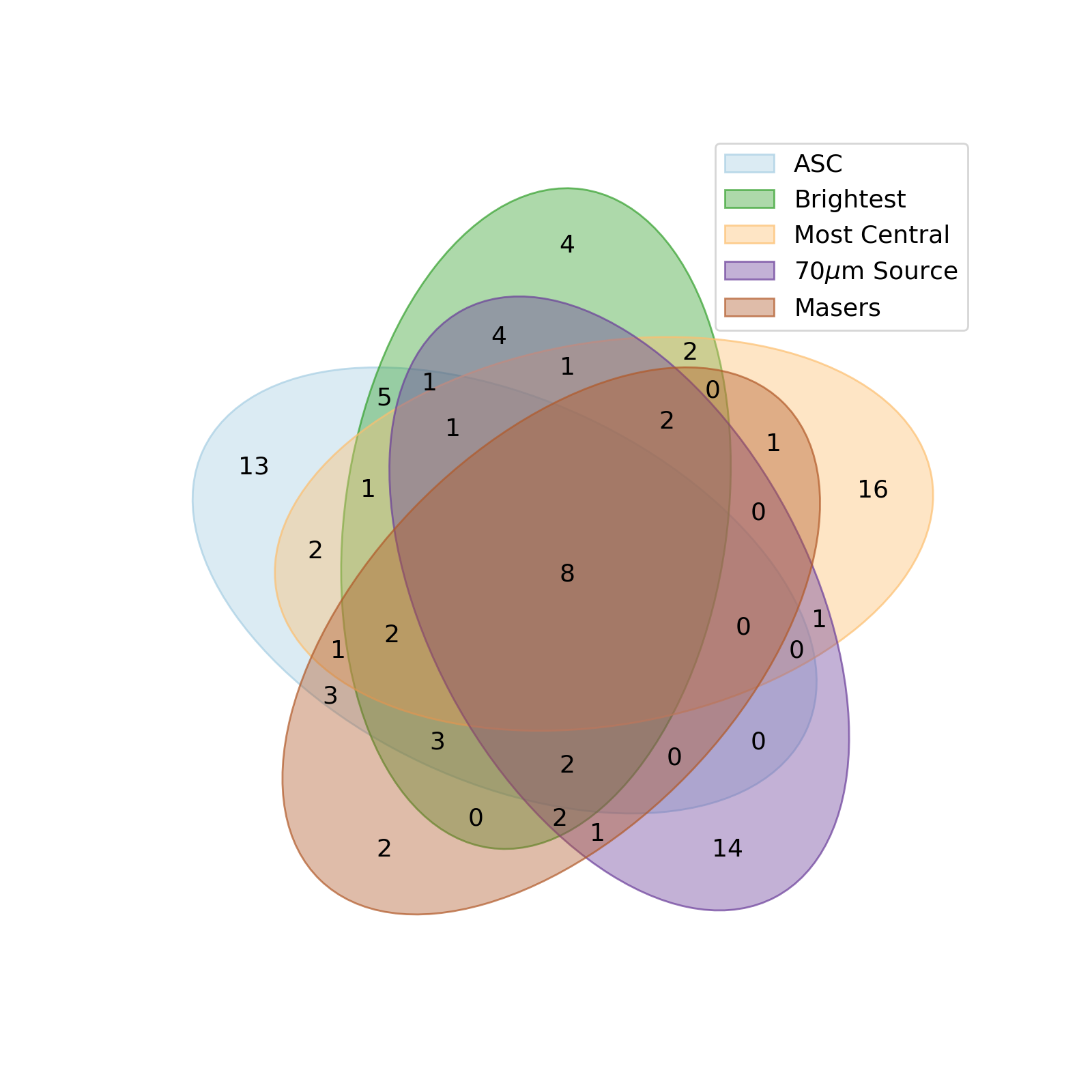}
\caption{Venn diagram showing the overlap of samples comprising, the Actively Star-forming Candidates (ASC, pale blue segment, sample size of 42), the \meth\ masers associated fragment sample (red segment, sample of 27 fragments), Infrared object associated fragments (detections at 70\mewm\ by \textit{Herschel} \citep{Elia21}, purple segment, sample of 38 fragments), the brightest fragments in each TEMPO fields (green segment, sample of 38 fragments) and the most central fragments in each TEMPO fields(arithmetic average, as the orange segment, sample of 38 fragments) }.
\label{ASCVenn:fig}
\end{center}
\end{figure}

Figure \ref{ASCVenn:fig} is a Venn diagram of the overlapping membership of the five samples. There is some observed overlap between members of the ASC sample and the brightest field fragment (22/38, 58\%), maser associated fragments (18/27, 67\%), 70\mewm\ IR source (16/38, 37\%) and most central field fragment (14/38, 37\%).

Such correspondence between the ASC candidates and star-forming core indicators (maser and brightest field source particularly), within the initial implementation of the described visibility analysis is a good indicator of the validity of the method. It provides additional constraints on those fragments within the TEMPO sample which are likely truly star-forming.

However, it is noted that in cases where the brightest fragment, \meth\ maser hosting fragment and/or IR counterpart fails to meet our ASC$_{score}$ criterion, visual inspection of the target field and post-analysis visibilities reveals indications that these sources are more complex than simple, single point-like or Gaussian sources and potentially maybe unresolved multiple systems. The technique also suffers from a requirement to know the exact position of an ASC source very accurately to recover the extracted visibility information without position errors affecting the recovered phase (see Appendix \ref{outstanding:sec} for more detail). The technique could be extended and modified to include a more general visibility parameters space analysis to determine better positions.

\section{Discussion}
The detected fragments in the TEMPO sample fields do not display a simple radial profile and may exhibit fractal or other distributions. This finding appears to agree with those found in other clusters, both observed and modelled, from within the literature. Though quantitative comparisons based on the $Q-$parameter cannot be made for the TEMPO fields, owing to the discussion given in Appendix \ref{Qpara:appdx}, qualitatively we find similarities to a number of young star forming clusters.  

The L1622, NGC2068/NGC2071 and NGC2023/NGC2024 star forming regions within Orion B are all found to be mildly substructured by \citet{Parker18} (all with $Q<$0.8 for source numbers of 29, 322, and 564 respectively), though they caution the use of the $Q$ value for the limited number of sources in the case of L1622 due to its low source numbers (for the same reasons discussed in Appendix \ref{Qpara:appdx}).

\citet{Sanhueza19} also use the $Q$-parameters on their IRDC derived sample finding in the majority of cases values indicative of substructure ($Q$<0.8). However, it is noted that the source numbers in the \citet{Sanhueza19} sample are between 13 to 37, so the validity of using $Q$ with these fields is unclear. Despite this, visual inspection of the reported fields, particularly when considering the published minimum spanning trees (their Figures 5 to 10) shows that most fields within their sample appear to contain some substructure. The region NGC 6334 I(N) was found by \citet{Hunter14} to be close to a $Q$ indicative of uniform density (0.82), though again for small source numbers. From the associated minimum spanning tree (MST) for this region whether or not the region is substructred or has a radial profile is unclear.

Using an alternative parameter, $\delta_{ADP,N}$ to gauge the level of substructure in the Orion Nebular Cluster (ONC) \citet{DaRio14}, (see their Equation 1), find that this more evolved stellar cluster has a low level of substructure \citep[see also][]{BateClarkeMcC98}. These authors note that the ONC appears somewhere between the substructured young Taurus molecular cloud \citep[see e.g.][]{CartwrightWhitworth04} and the radial distributions seen in globular clusters. 

Indeed, there is evidence within the literature, from both observed and modelled young stellar clusters, that cluster structure tends to evolve from an originally sub-structured formation toward a centrally concentrated final state \citep[e.g.][]{BonnellBateVine03,SchmejaKlessen06,Bate09,Maschberger10} as sufficient time for dynamical processing of the sources within the cluster elapses. The TEMPO sample was selected to give a range of ages of high-mass embedded protostars prior to the formation of an UC\hii\ region, as such some degree of substructure at these early times would be expected. This evolution of structure may also account for the distribution of source normalised offsets seen in Figure \ref{norm_Radial_profile:fig} with some fields beginning to show a more centrally concentrated profile than others. However, the TEMPO sample lacks sufficient source counts within individual fields to test this quantitatively as a function of e.g. IR colour. To further this analysis higher sensitivity and resolution images of the TEMPO fields is required to detected any fainter sources present and to resolve closely paired objects, which may currently appear as a single source within the TEMPO data.

The majority of the TEMPO fields show fragmentation on scales which are inconsistent with (with 87\% of fields having a mean edge length $X_{mean}$ $\geq 1.5\times$ up to 9.1$\times$ the $\lambda_J$) the thermal Jeans length when using the clump radii, mass and temperatures from \citet{Elia21} within the calculation. This is suggestive of a non-thermal fragmentation being present within the TEMPO fields. Similar results have been seen within other works. \citet{TraficanteSqualo} found in the SQUALO sample found a range of values of source separation to thermal Jeans length ratio (their $\lambda_{J,3D}$) of 1.06 $< \lambda_{J,3D}<$ 7.04, suggestive of some non-thermal fragmentation. SQUALO had similar observing characteristics to the TEMPO sample. Observations made over larger spatial scales, using mosaic rather than single pointing observations, also tend to find fragmentation scales which are better explained by turbulent or cylindrical fragmentation \citep{Henshaw16, Lu18}.

Conversely, the results seen by \citet{Svoboda19} targeted toward high-mass starless clump candidates, the ASHES sample \citep{Sanhueza19} toward 70\mewm\ dark high-mass clumps and in the CORE survey \citep{Beuther18} toward known high-mass star-forming regions, found fragmentation scales consistent with the thermal Jeans length scale. It is interesting to note that the calculation of the thermal Jeans length (Eqn. \ref{thermalJ:eqn}) is particularly sensitive to the value of $R_{clump}$ used, as it scales with $R_{clump}^{3/2}$. Using different $R_{clump}$ values for the TEMPO fields, for example those calculated by \citet{Traficante15} for the SDC fields and \citet{Urquhart14} for the RMS fields, brings the TEMPO field $X_{mean}$ values in to a more comparable range with thermal Jeans (in the range 0.5 - 1.5$\times\lambda_J$). Such sensitivity to changes in input is important to consider when it has such an impact on the findings. The use of \citet{Elia21} values has been retained within this work to allow use of a single consistently derived set of parameters from the literature.

\onecolumn
\begin{landscape}
\begin{longtable}{L{4.2cm}C{1.5cm} cccccccc C{1.2cm} C{1.2cm} C{1.2cm}}
\caption{Catalogue Table sources detected in the TEMPO sample. Column 1 gives the field Name as specified in Table \ref{Selection:tab}, Column 2 gives the source number of the TEMPO fragment in that field, Columns 3, 4, 5, 6 and 7 give the Right Ascension, Declination, fitted Major and Minor axes and source position angle for the detected sources from the Dendrogram analysis, respectively. Column 8 gives the measured source continuum flux density in mJy. Columns 9 gives the fragments ASC score as defined in Appendix \ref{centCond:apdx} and columns 10, 11 and 12 denote if the source is the Brightest in the field, the most central using the arithmetic mean (column 11) or weighted mean (column 12), respectively, with 1 indicating True and 0 indicating False.}\\
\hline
Field & Source & RA & Dec & Axis$_{maj}$ & Axis$_{min}$ & PA & S$_{combined}$ & ASC$_{score}$ & Brightest? & Central & Central\\
    & No.  & [h:m:s] & [$^{\circ}$:$^{\prime}$:$^{\prime\prime}$] & [$^{\prime\prime}$] & [$^{\prime\prime}$] & [Deg] & [mJy] & & & [mean]? & [weighted]?\\
\hline
\hline
RMS-G013.6562-00.5997\multirow{6}{*}{} & 0 & 18:17:24.374 & -17:22:14.720 & 0.703 & 0.484 & -177.99 & 4.707 & 0 & 0 & 0 & 0\\
 & 1 & 18:17:24.028 & -17:22:14.907 & 0.854 & 0.234 & -166.59 & 12.442 & 0 & 0 & 0 & 0\\
 & 2 & 18:17:23.878 & -17:22:14.346 & 0.662 & 0.26 & -171.96 & 6.785 & 0 & 0 & 0 & 0\\
 & 3 & 18:17:24.243 & -17:22:12.852 & 0.434 & 0.377 & -176.11 & 272.43 & 3 & 1 & 1 & 1\\
 & 4 & 18:17:24.348 & -17:22:11.824 & 0.252 & 0.166 & 148.05 & 9.851 & 0 & 0 & 0 & 0\\
 & 5 & 18:17:24.152 & -17:22:01.924 & 0.592 & 0.391 & 174.27 & 17.836 & 0 & 0 & 0 & 0\\
\hline
RMS-G017.6380+00.1566\multirow{9}{*}{} & 0 & 18:22:26.976 & -13:30:18.258 & 0.909 & 0.455 & 171.87 & 36.54 & 0 & 0 & 0 & 0\\
 & 1 & 18:22:26.778 & -13:30:17.978 & 1.241 & 0.375 & -179.81 & 38.38 & 0 & 0 & 0 & 0\\
 & 2 & 18:22:26.848 & -13:30:16.016 & 0.733 & 0.623 & -172.76 & 66.572 & 0 & 0 & 0 & 0\\
 & 3 & 18:22:26.573 & -13:30:16.016 & 1.45 & 0.439 & 178.98 & 31.153 & 0 & 0 & 1 & 1\\
 & 4 & 18:22:26.855 & -13:30:13.775 & 0.871 & 0.669 & 58.5 & 28.225 & 0 & 0 & 0 & 0\\
 & 5 & 18:22:26.253 & -13:30:12.654 & 0.644 & 0.453 & -170.14 & 37.276 & 0 & 0 & 0 & 0\\
 & 6 & 18:22:26.387 & -13:30:12.000 & 0.436 & 0.306 & 164.24 & 116.564 & 0 & 1 & 0 & 0\\
 & 7 & 18:22:26.432 & -13:30:10.879 & 1.211 & 0.597 & 175.27 & 20.292 & 0 & 0 & 0 & 0\\
 & 8 & 18:22:26.323 & -13:30:07.797 & 1.238 & 0.428 & -143.12 & 34.397 & 0 & 0 & 0 & 0\\
\hline
SDC18.816-0.447$\_1$\multirow{2}{*}{} & 0 & 18:26:58.872 & -12:44:51.912 & 0.957 & 0.665 & 125.48 & 21.696 & 3 & 0 & 0 & 1\\
 & 1 & 18:26:59.051 & -12:44:46.588 & 1.516 & 1.002 & -170.9 & 67.953 & 0 & 1 & 1 & 0\\
\hline
SDC20.775-0.076$\_1$\multirow{14}{*}{} & 0 & 18:29:16.617 & -10:52:20.115 & 1.287 & 0.519 & 177.02 & 14.244 & 0 & 0 & 0 & 0\\
 & 1 & 18:29:16.706 & -10:52:18.714 & 9.634 & 7.137 & 143.3 & 68.766 & 0 & 0 & 0 & 0\\
 & 2 & 18:29:16.528 & -10:52:06.291 & 1.6 & 1.117 & 117.18 & 12.391 & 0 & 0 & 0 & 1\\
 & 3 & 18:29:16.554 & -10:52:05.077 & 0.601 & 0.292 & 149.02 & 24.312 & 0 & 0 & 0 & 0\\
 & 4 & 18:29:16.541 & -10:52:03.676 & 0.592 & 0.353 & 152.42 & 27.997 & 0 & 0 & 0 & 0\\
 & 5 & 18:29:16.585 & -10:52:01.621 & 0.681 & 0.496 & 52.07 & 10.141 & 0 & 0 & 0 & 0\\
 & 6 & 18:29:16.319 & -10:51:59.660 & 1.585 & 1.316 & 174.22 & 11.975 & 0 & 0 & 0 & 0\\
 & 7 & 18:29:16.655 & -10:52:17.406 & 1.846 & 1.009 & 49.34 & 14.909 & 0 & 0 & 0 & 0\\
 & 8 & 18:29:16.414 & -10:52:11.428 & 0.616 & 0.464 & -140.84 & 95.474 & 2 & 1 & 0 & 0\\
 & 9 & 18:29:16.598 & -10:52:10.214 & 0.634 & 0.44 & 104.91 & 15.887 & 0 & 0 & 1 & 0\\
 & 10 & 18:29:16.351 & -10:52:09.374 & 0.565 & 0.406 & 50.88 & 40.176 & 0 & 0 & 0 & 0\\
 & 11 & 18:29:16.649 & -10:52:08.533 & 0.741 & 0.582 & -174.8 & 8.947 & 0 & 0 & 0 & 0\\
 & 12 & 18:29:16.211 & -10:52:08.346 & 0.906 & 0.407 & 141.98 & 19.575 & 0 & 0 & 0 & 0\\
 & 13 & 18:29:16.173 & -10:52:06.478 & 0.661 & 0.611 & -136.41 & 8.244 & 0 & 0 & 0 & 0\\
\hline
SDC20.775-0.076$\_3$\multirow{4}{*}{} & 0 & 18:29:12.080 & -10:50:35.934 & 1.017 & 0.424 & 179.85 & 45.017 & 0 & 0 & 0 & 0\\
 & 1 & 18:29:12.232 & -10:50:33.692 & 1.389 & 0.35 & 174.83 & 87.193 & 3 & 1 & 0 & 0\\
 & 2 & 18:29:11.908 & -10:50:33.599 & 0.777 & 0.353 & 162.19 & 11.237 & 3 & 0 & 1 & 1\\
 & 3 & 18:29:11.826 & -10:50:32.198 & 0.745 & 0.429 & 171.94 & 40.558 & 0 & 0 & 0 & 0\\
\hline
SDC22.985-0.412$\_1$\multirow{7}{*}{} & 0 & 18:34:39.842 & -09:00:46.285 & 1.084 & 0.452 & 134.38 & 17.605 & 0 & 0 & 0 & 0\\
 & 1 & 18:34:40.459 & -09:00:41.055 & 0.721 & 0.539 & 136.74 & 13.464 & 0 & 0 & 0 & 0\\
 & 2 & 18:34:40.100 & -09:00:40.401 & 1.768 & 0.775 & 123.03 & 14.743 & 0 & 0 & 0 & 0\\
 & 3 & 18:34:40.283 & -09:00:38.253 & 1.173 & 0.643 & 48.62 & 618.139 & 0 & 1 & 1 & 1\\
 & 4 & 18:34:40.220 & -09:00:32.836 & 1.037 & 0.518 & 80.01 & 38.889 & 0 & 0 & 0 & 0\\
 & 5 & 18:34:40.056 & -09:00:34.423 & 1.631 & 0.771 & 46.3 & 16.714 & 3 & 0 & 0 & 0\\
 & 6 & 18:34:40.403 & -09:00:29.193 & 1.309 & 0.622 & 108.4 & 33.663 & 0 & 0 & 0 & 0\\
\hline
SDC23.21-0.371$\_1$\multirow{9}{*}{} & 0 & 18:34:55.502 & -08:49:20.884 & 3.653 & 3.091 & 98.79 & 22.779 & 0 & 0 & 0 & 0\\
 & 1 & 18:34:55.654 & -08:49:20.604 & 0.871 & 0.846 & 179.4 & 10.403 & 0 & 0 & 0 & 0\\
 & 2 & 18:34:55.105 & -08:49:19.390 & 0.439 & 0.268 & 143.98 & 33.304 & 2 & 0 & 0 & 0\\
 & 3 & 18:34:55.206 & -08:49:14.813 & 0.936 & 0.855 & 69.75 & 1302.088 & 3 & 1 & 1 & 0\\
 & 4 & 18:34:55.899 & -08:49:15.467 & 1.658 & 1.086 & -173.62 & 15.711 & 3 & 0 & 0 & 0\\
 & 5 & 18:34:55.414 & -08:49:10.797 & 0.809 & 0.677 & 56.18 & 17.195 & 0 & 0 & 0 & 1\\
 & 6 & 18:34:55.257 & -08:49:09.956 & 0.601 & 0.35 & 99.88 & 27.666 & 0 & 0 & 0 & 0\\
 & 7 & 18:34:55.213 & -08:49:08.929 & 0.566 & 0.489 & 167.51 & 8.879 & 3 & 0 & 0 & 0\\
 & 8 & 18:34:55.282 & -08:49:07.435 & 1.249 & 0.613 & 88.62 & 12.061 & 0 & 0 & 0 & 0\\
\hline
RMS-G023.3891+00.1851\multirow{8}{*}{} & 0 & 18:33:14.117 & -08:23:59.989 & 0.767 & 0.483 & 174.48 & 6.82 & 0 & 0 & 0 & 0\\
 & 1 & 18:33:14.590 & -08:23:58.775 & 0.899 & 0.384 & -138.48 & 8.56 & 0 & 0 & 0 & 0\\
 & 2 & 18:33:14.325 & -08:23:57.467 & 0.795 & 0.632 & -143.39 & 159.398 & 3 & 1 & 1 & 1\\
 & 3 & 18:33:14.067 & -08:23:57.934 & 0.803 & 0.385 & 177.33 & 8.875 & 0 & 0 & 0 & 0\\
 & 4 & 18:33:13.891 & -08:23:55.879 & 0.842 & 0.52 & -135.14 & 10.897 & 0 & 0 & 0 & 0\\
 & 5 & 18:33:14.376 & -08:23:54.945 & 0.539 & 0.361 & 128.76 & 5.389 & 0 & 0 & 0 & 0\\
 & 6 & 18:33:14.306 & -08:23:52.330 & 0.744 & 0.375 & 133.9 & 17.409 & 0 & 0 & 0 & 0\\
 & 7 & 18:33:14.212 & -08:23:51.676 & 0.503 & 0.38 & 165.29 & 10.442 & 0 & 0 & 0 & 0\\
\hline
SDC24.381-0.21$\_3$\multirow{11}{*}{} & 0 & 18:36:41.165 & -07:39:24.087 & 0.497 & 0.443 & -151.73 & 28.952 & 0 & 0 & 0 & 0\\
 & 1 & 18:36:40.977 & -07:39:09.050 & 0.474 & 0.325 & 165.35 & 15.64 & 0 & 0 & 0 & 0\\
 & 2 & 18:36:40.870 & -07:39:09.237 & 1.035 & 0.667 & -172.1 & 5.897 & 0 & 0 & 0 & 0\\
 & 3 & 18:36:40.883 & -07:39:04.753 & 0.695 & 0.601 & 138.56 & 7.511 & 0 & 0 & 0 & 0\\
 & 4 & 18:36:40.732 & -07:38:57.842 & 0.645 & 0.511 & -164.26 & 30.456 & 0 & 0 & 0 & 0\\
 & 5 & 18:36:40.983 & -07:39:22.032 & 1.306 & 0.721 & 109.09 & 41.601 & 0 & 1 & 0 & 0\\
 & 6 & 18:36:41.040 & -07:39:19.511 & 0.953 & 0.596 & 86.38 & 11.725 & 0 & 0 & 0 & 0\\
 & 7 & 18:36:40.763 & -07:39:18.577 & 0.908 & 0.459 & 88.79 & 7.177 & 0 & 0 & 0 & 0\\
 & 8 & 18:36:40.788 & -07:39:15.588 & 0.537 & 0.401 & 138.12 & 23.985 & 0 & 0 & 1 & 0\\
 & 9 & 18:36:40.694 & -07:39:14.187 & 0.602 & 0.34 & 173.37 & 17.44 & 0 & 0 & 0 & 0\\
 & 10 & 18:36:40.688 & -07:39:12.039 & 0.702 & 0.351 & 90.41 & 26.65 & 0 & 0 & 0 & 1\\
\hline
SDC24.462+0.219$\_2$\multirow{5}{*}{} & 0 & 18:35:11.248 & -07:26:32.433 & 0.346 & 0.142 & 134.15 & 13.72 & 0 & 0 & 0 & 0\\
 & 1 & 18:35:11.317 & -07:26:31.313 & 0.511 & 0.39 & 107.99 & 129.213 & 2 & 1 & 1 & 0\\
 & 2 & 18:35:11.487 & -07:26:29.258 & 0.483 & 0.313 & 133.22 & 11.466 & 0 & 0 & 0 & 0\\
 & 3 & 18:35:11.104 & -07:26:28.137 & 0.814 & 0.359 & 80.33 & 6.239 & 0 & 0 & 0 & 1\\
 & 4 & 18:35:11.091 & -07:26:26.643 & 0.688 & 0.474 & 48.58 & 6.272 & 0 & 0 & 0 & 0\\
\hline
SDC25.426-0.175$\_6$\multirow{3}{*}{} & 0 & 18:37:30.564 & -06:41:21.884 & 0.591 & 0.409 & 159.39 & 7.3 & 0 & 0 & 0 & 1\\
 & 1 & 18:37:30.394 & -06:41:18.428 & 1.634 & 0.882 & -179.43 & 340.079 & 2 & 1 & 1 & 0\\
 & 2 & 18:37:30.664 & -06:41:17.775 & 0.455 & 0.367 & -154.14 & 11.221 & 0 & 0 & 0 & 0\\
\hline
SDC28.147-0.006$\_1$\multirow{6}{*}{} & 0 & 18:42:42.619 & -04:15:35.868 & 0.833 & 0.548 & 99.48 & 154.63 & 0 & 1 & 0 & 0\\
 & 1 & 18:42:42.800 & -04:15:35.588 & 0.526 & 0.294 & 175.6 & 22.367 & 0 & 0 & 0 & 0\\
 & 2 & 18:42:42.775 & -04:15:32.132 & 12.046 & 5.712 & 176.3 & 28.946 & 0 & 0 & 0 & 0\\
 & 3 & 18:42:42.675 & -04:15:29.797 & 1.086 & 0.707 & 67.01 & 6.312 & 0 & 0 & 1 & 0\\
 & 4 & 18:42:42.687 & -04:15:28.022 & 1.215 & 0.474 & -155.32 & 8.383 & 0 & 0 & 0 & 1\\
 & 5 & 18:42:42.494 & -04:15:23.352 & 1.089 & 0.66 & -163.7 & 8.122 & 0 & 0 & 0 & 0\\
\hline
SDC28.277-0.352$\_1$\multirow{4}{*}{} & 0 & 18:44:20.963 & -04:17:46.005 & 0.583 & 0.483 & 132.39 & 13.128 & 1 & 0 & 0 & 0\\
 & 1 & 18:44:21.975 & -04:17:39.747 & 0.606 & 0.55 & 88.11 & 107.87 & 3 & 1 & 1 & 1\\
 & 2 & 18:44:22.193 & -04:17:26.017 & 0.553 & 0.511 & -135.7 & 16.275 & 0 & 0 & 0 & 0\\
 & 3 & 18:44:22.100 & -04:17:24.523 & 0.652 & 0.524 & 174.13 & 10.954 & 0 & 0 & 0 & 0\\
\hline
SDC29.844-0.009$\_4$\multirow{8}{*}{} & 0 & 18:46:12.651 & -02:39:11.741 & 0.692 & 0.584 & 142.54 & 40.319 & 0 & 0 & 0 & 0\\
 & 1 & 18:46:12.682 & -02:39:10.433 & 1.76 & 1.288 & 121.64 & 42.937 & 0 & 0 & 0 & 0\\
 & 2 & 18:46:12.695 & -02:39:09.406 & 8.051 & 6.275 & 136.64 & 112.399 & 0 & 1 & 0 & 0\\
 & 3 & 18:46:12.863 & -02:39:06.978 & 1.14 & 0.511 & 134.27 & 36.827 & 0 & 0 & 1 & 1\\
 & 4 & 18:46:12.900 & -02:39:04.176 & 0.82 & 0.64 & 92.39 & 23.434 & 0 & 0 & 0 & 0\\
 & 5 & 18:46:12.938 & -02:39:02.868 & 0.504 & 0.355 & 135.78 & 41.906 & 3 & 0 & 0 & 0\\
 & 6 & 18:46:12.950 & -02:39:01.280 & 0.484 & 0.384 & 178.28 & 63.669 & 0 & 0 & 0 & 0\\
 & 7 & 18:46:13.012 & -02:38:59.225 & 3.717 & 2.276 & 53.28 & 55.205 & 0 & 0 & 0 & 0\\
\hline
RMS-G029.8620-00.0444\multirow{6}{*}{} & 0 & 18:45:59.500 & -02:45:08.681 & 0.703 & 0.367 & 114.94 & 50.949 & 0 & 0 & 0 & 0\\
 & 1 & 18:45:59.569 & -02:45:06.626 & 0.379 & 0.308 & 153.72 & 61.41 & 3 & 1 & 0 & 0\\
 & 2 & 18:45:59.899 & -02:45:06.253 & 0.943 & 0.503 & -163.15 & 8.992 & 0 & 0 & 0 & 0\\
 & 3 & 18:45:59.625 & -02:45:05.506 & 0.355 & 0.239 & 140.29 & 14.411 & 3 & 0 & 1 & 1\\
 & 4 & 18:45:59.644 & -02:45:02.050 & 8.562 & 6.583 & 175.17 & 34.426 & 0 & 0 & 0 & 0\\
 & 5 & 18:45:59.550 & -02:45:02.050 & 8.757 & 6.921 & 45.21 & 33.466 & 0 & 0 & 0 & 0\\
\hline
SDC30.172-0.157$\_2$\multirow{2}{*}{} & 0 & 18:47:07.863 & -02:30:04.631 & 0.92 & 0.686 & 132.38 & 18.971 & 0 & 0 & 1 & 1\\
 & 1 & 18:47:07.913 & -02:30:01.269 & 0.84 & 0.695 & -140.92 & 20.416 & 0 & 1 & 0 & 0\\
\hline
RMS-G030.1981-00.1691\multirow{3}{*}{} & 0 & 18:47:03.069 & -02:30:36.374 & 1.338 & 0.896 & 77.88 & 132.207 & 3 & 1 & 1 & 1\\
 & 1 & 18:47:03.362 & -02:30:37.868 & 0.719 & 0.648 & -141.49 & 5.754 & 0 & 0 & 0 & 0\\
 & 2 & 18:47:02.901 & -02:30:30.770 & 0.907 & 0.732 & 125.16 & 5.803 & 0 & 0 & 0 & 0\\
\hline
SDC33.107-0.065$\_2$\multirow{13}{*}{} & 0 & 18:52:08.032 & +00:08:08.890 & 1.601 & 0.561 & 55.14 & 14.62 & 0 & 0 & 0 & 0\\
 & 1 & 18:52:07.664 & +00:08:13.280 & 0.49 & 0.295 & -168.71 & 54.242 & 0 & 0 & 0 & 0\\
 & 2 & 18:52:08.138 & +00:08:13.560 & 0.376 & 0.268 & -176.06 & 19.249 & 0 & 0 & 0 & 0\\
 & 3 & 18:52:08.007 & +00:08:14.027 & 0.379 & 0.161 & 158.27 & 12.825 & 0 & 0 & 0 & 0\\
 & 4 & 18:52:08.437 & +00:08:14.775 & 0.615 & 0.495 & -166.67 & 14.717 & 0 & 0 & 0 & 0\\
 & 5 & 18:52:08.150 & +00:08:10.198 & 0.608 & 0.176 & -163.28 & 9.862 & 0 & 0 & 0 & 1\\
 & 6 & 18:52:08.075 & +00:08:10.385 & 0.346 & 0.293 & 96.05 & 7.458 & 0 & 0 & 1 & 0\\
 & 7 & 18:52:07.521 & +00:08:11.225 & 1.075 & 0.546 & 64.37 & 15.123 & 0 & 0 & 0 & 0\\
 & 8 & 18:52:07.963 & +00:08:11.786 & 0.43 & 0.37 & 47.8 & 255.478 & 3 & 1 & 0 & 0\\
 & 9 & 18:52:08.275 & +00:08:11.692 & 0.66 & 0.373 & -173.31 & 69.758 & 0 & 0 & 0 & 0\\
 & 10 & 18:52:08.530 & +00:08:11.412 & 3.138 & 1.587 & -178.47 & 18.553 & 0 & 0 & 0 & 0\\
 & 11 & 18:52:07.826 & +00:08:12.253 & 0.392 & 0.286 & -179.59 & 153.617 & 0 & 0 & 0 & 0\\
 & 12 & 18:52:08.555 & +00:08:12.720 & 0.981 & 0.73 & -177.96 & 7.765 & 0 & 0 & 0 & 0\\
\hline
RMS-G034.7569+00.0247\multirow{6}{*}{} & 0 & 18:54:40.719 & +01:38:00.649 & 1.325 & 0.557 & 90.87 & 7.116 & 0 & 0 & 0 & 0\\
 & 1 & 18:54:40.625 & +01:38:01.676 & 7.671 & 5.681 & 50.77 & 23.453 & 0 & 0 & 0 & 0\\
 & 2 & 18:54:40.812 & +01:38:03.918 & 0.726 & 0.624 & -163.24 & 5.231 & 0 & 0 & 0 & 0\\
 & 3 & 18:54:40.974 & +01:38:04.105 & 3.224 & 1.057 & 64.4 & 12.342 & 0 & 0 & 0 & 0\\
 & 4 & 18:54:40.737 & +01:38:06.346 & 0.782 & 0.475 & 61.61 & 114.044 & 3 & 1 & 1 & 1\\
 & 5 & 18:54:40.445 & +01:38:12.230 & 1.828 & 1.451 & -177.64 & 6.947 & 0 & 0 & 0 & 0\\
\hline
RMS-G034.8211+00.3519\multirow{9}{*}{} & 0 & 18:53:38.112 & +01:50:36.044 & 0.899 & 0.622 & -162.28 & 9.559 & 0 & 0 & 0 & 0\\
 & 1 & 18:53:38.268 & +01:50:24.929 & 0.766 & 0.598 & 89.9 & 8.244 & 0 & 0 & 0 & 0\\
 & 2 & 18:53:38.567 & +01:50:26.890 & 0.934 & 0.358 & 178.41 & 12.696 & 0 & 0 & 0 & 0\\
 & 3 & 18:53:37.694 & +01:50:28.478 & 0.879 & 0.526 & -179.82 & 39.686 & 0 & 0 & 0 & 0\\
 & 4 & 18:53:38.324 & +01:50:27.451 & 0.432 & 0.291 & 158.6 & 9.095 & 0 & 0 & 0 & 0\\
 & 5 & 18:53:38.212 & +01:50:27.824 & 0.521 & 0.307 & -144.0 & 26.717 & 0 & 0 & 0 & 1\\
 & 6 & 18:53:37.944 & +01:50:29.786 & 1.159 & 0.514 & -167.31 & 117.312 & 0 & 1 & 1 & 0\\
 & 7 & 18:53:38.043 & +01:50:32.308 & 6.674 & 4.256 & 78.21 & 23.465 & 0 & 0 & 0 & 0\\
 & 8 & 18:53:37.190 & +01:50:34.456 & 0.758 & 0.569 & -172.02 & 31.016 & 0 & 0 & 0 & 0\\
\hline
SDC35.063-0.726$\_1$\multirow{9}{*}{} & 0 & 18:58:05.664 & +01:36:59.154 & 1.401 & 0.593 & -137.23 & 18.854 & 0 & 0 & 0 & 0\\
 & 1 & 18:58:05.533 & +01:36:59.715 & 0.493 & 0.35 & 155.75 & 53.356 & 0 & 0 & 0 & 0\\
 & 2 & 18:58:06.143 & +01:37:07.560 & 0.505 & 0.319 & 149.3 & 225.376 & 3 & 1 & 0 & 0\\
 & 3 & 18:58:06.262 & +01:37:07.467 & 0.341 & 0.161 & -163.26 & 31.357 & 0 & 0 & 0 & 0\\
 & 4 & 18:58:06.392 & +01:37:10.082 & 1.128 & 0.79 & 120.31 & 11.651 & 0 & 0 & 0 & 0\\
 & 5 & 18:58:06.610 & +01:37:02.237 & 0.726 & 0.442 & 124.0 & 81.652 & 0 & 0 & 0 & 0\\
 & 6 & 18:58:06.592 & +01:37:05.039 & 0.436 & 0.38 & 175.36 & 52.275 & 0 & 0 & 0 & 0\\
 & 7 & 18:58:06.716 & +01:37:05.599 & 3.68 & 1.85 & -173.02 & 31.899 & 0 & 0 & 0 & 0\\
 & 8 & 18:58:06.106 & +01:37:05.692 & 0.391 & 0.241 & 152.56 & 11.555 & 0 & 0 & 1 & 1\\
\hline
SDC37.846-0.392$\_1$\multirow{9}{*}{} & 0 & 19:01:53.656 & +04:12:44.462 & 0.836 & 0.465 & -171.3 & 43.87 & 0 & 0 & 0 & 0\\
 & 1 & 19:01:53.537 & +04:12:48.852 & 1.302 & 0.979 & 48.21 & 1708.024 & 0 & 1 & 0 & 0\\
 & 2 & 19:01:53.650 & +04:12:53.242 & 0.663 & 0.409 & 172.21 & 33.47 & 3 & 0 & 0 & 0\\
 & 3 & 19:01:53.413 & +04:12:53.802 & 0.644 & 0.346 & 121.75 & 35.893 & 0 & 0 & 0 & 0\\
 & 4 & 19:01:53.556 & +04:12:53.802 & 1.149 & 0.51 & 168.49 & 24.494 & 0 & 0 & 1 & 1\\
 & 5 & 19:01:53.494 & +04:12:55.390 & 0.995 & 0.846 & -173.0 & 23.363 & 0 & 0 & 0 & 0\\
 & 6 & 19:01:53.444 & +04:12:56.791 & 1.175 & 0.538 & 147.35 & 74.85 & 0 & 0 & 0 & 0\\
 & 7 & 19:01:53.687 & +04:12:57.445 & 0.979 & 0.765 & 49.93 & 69.035 & 0 & 0 & 0 & 0\\
 & 8 & 19:01:53.281 & +04:12:56.511 & 1.077 & 0.574 & 109.24 & 33.079 & 0 & 0 & 0 & 0\\
\hline
SDC42.401-0.309$\_2$\multirow{4}{*}{} & 0 & 19:09:49.397 & +08:19:41.863 & 1.019 & 0.55 & 137.43 & 9.791 & 0 & 0 & 0 & 0\\
 & 1 & 19:09:49.931 & +08:19:44.105 & 0.319 & 0.21 & -148.74 & 8.054 & 0 & 0 & 0 & 0\\
 & 2 & 19:09:49.862 & +08:19:45.506 & 0.672 & 0.432 & 48.48 & 217.087 & 3 & 1 & 1 & 0\\
 & 3 & 19:09:49.365 & +08:19:44.385 & 1.56 & 0.906 & 154.42 & 5.252 & 0 & 0 & 0 & 1\\
\hline
SDC43.186-0.549$\_2$\multirow{12}{*}{} & 0 & 19:12:09.121 & +08:51:58.753 & 0.786 & 0.581 & 113.88 & 11.398 & 0 & 0 & 0 & 0\\
 & 1 & 19:12:08.692 & +08:52:08.000 & 0.959 & 0.634 & 152.82 & 15.05 & 0 & 0 & 0 & 0\\
 & 2 & 19:12:09.354 & +08:52:13.604 & 8.174 & 6.277 & 159.16 & 34.392 & 0 & 0 & 0 & 0\\
 & 3 & 19:12:09.032 & +08:52:14.164 & 0.563 & 0.325 & -162.64 & 103.479 & 3 & 0 & 0 & 0\\
 & 4 & 19:12:09.215 & +08:52:15.005 & 0.556 & 0.387 & 164.74 & 135.08 & 3 & 1 & 0 & 0\\
 & 5 & 19:12:08.755 & +08:52:10.802 & 0.466 & 0.43 & 85.17 & 22.247 & 0 & 0 & 0 & 1\\
 & 6 & 19:12:09.165 & +08:52:10.335 & 9.103 & 7.136 & 155.17 & 57.184 & 0 & 0 & 0 & 0\\
 & 7 & 19:12:08.566 & +08:52:10.709 & 1.359 & 1.337 & 139.34 & 10.35 & 0 & 0 & 0 & 0\\
 & 8 & 19:12:09.272 & +08:52:12.203 & 0.84 & 0.622 & 160.1 & 23.93 & 0 & 0 & 0 & 0\\
 & 9 & 19:12:08.490 & +08:52:12.857 & 1.151 & 0.856 & -175.18 & 13.921 & 0 & 0 & 0 & 0\\
 & 10 & 19:12:08.850 & +08:52:12.390 & 0.966 & 0.618 & 179.82 & 8.25 & 0 & 0 & 1 & 0\\
 & 11 & 19:12:08.711 & +08:52:13.324 & 0.817 & 0.453 & 168.17 & 13.881 & 0 & 0 & 0 & 0\\
\hline
SDC43.311-0.21$\_1$\multirow{9}{*}{} & 0 & 19:11:16.697 & +09:07:16.644 & 7.898 & 7.001 & 151.23 & 72.133 & 0 & 0 & 0 & 0\\
 & 1 & 19:11:17.076 & +09:07:24.116 & 1.037 & 0.623 & 113.1 & 18.259 & 0 & 0 & 0 & 0\\
 & 2 & 19:11:17.372 & +09:07:24.489 & 0.869 & 0.589 & 159.31 & 11.639 & 0 & 0 & 0 & 0\\
 & 3 & 19:11:17.448 & +09:07:26.077 & 1.14 & 0.733 & 161.61 & 9.796 & 0 & 0 & 0 & 0\\
 & 4 & 19:11:17.107 & +09:07:26.171 & 1.334 & 0.5 & 76.34 & 9.464 & 0 & 0 & 1 & 1\\
 & 5 & 19:11:16.968 & +09:07:28.599 & 0.679 & 0.479 & -140.48 & 23.726 & 0 & 0 & 0 & 0\\
 & 6 & 19:11:17.214 & +09:07:31.588 & 0.682 & 0.407 & 156.36 & 308.593 & 3 & 1 & 0 & 0\\
 & 7 & 19:11:16.975 & +09:07:32.055 & 0.76 & 0.466 & -171.13 & 7.457 & 0 & 0 & 0 & 0\\
 & 8 & 19:11:17.347 & +09:07:32.989 & 0.607 & 0.238 & 152.99 & 20.427 & 0 & 0 & 0 & 0\\
\hline
SDC43.877-0.755$\_1$\multirow{11}{*}{} & 0 & 19:14:26.768 & +09:22:23.979 & 1.189 & 0.914 & 138.1 & 20.488 & 0 & 0 & 0 & 0\\
 & 1 & 19:14:26.755 & +09:22:25.286 & 0.691 & 0.393 & -168.18 & 18.896 & 0 & 0 & 0 & 0\\
 & 2 & 19:14:26.143 & +09:22:41.445 & 1.163 & 0.936 & 79.06 & 11.541 & 0 & 0 & 0 & 0\\
 & 3 & 19:14:26.837 & +09:22:25.660 & 0.977 & 0.635 & 144.74 & 19.332 & 0 & 0 & 0 & 0\\
 & 4 & 19:14:26.402 & +09:22:27.715 & 9.276 & 8.274 & 100.5 & 47.371 & 0 & 0 & 0 & 0\\
 & 5 & 19:14:26.490 & +09:22:29.116 & 0.706 & 0.505 & 82.19 & 14.49 & 1 & 0 & 1 & 0\\
 & 6 & 19:14:26.288 & +09:22:29.209 & 0.964 & 0.455 & 89.19 & 11.06 & 0 & 0 & 0 & 0\\
 & 7 & 19:14:26.295 & +09:22:32.665 & 0.581 & 0.37 & 106.76 & 10.029 & 0 & 0 & 0 & 1\\
 & 8 & 19:14:26.181 & +09:22:34.066 & 0.916 & 0.441 & 135.7 & 234.998 & 0 & 1 & 0 & 0\\
 & 9 & 19:14:26.377 & +09:22:36.121 & 0.741 & 0.323 & 124.7 & 161.769 & 3 & 0 & 0 & 0\\
 & 10 & 19:14:26.724 & +09:22:40.511 & 0.717 & 0.563 & -146.1 & 11.228 & 0 & 0 & 0 & 0\\
\hline
SDC45.787-0.335$\_1$\multirow{5}{*}{} & 0 & 19:16:31.302 & +11:16:05.462 & 0.823 & 0.703 & 81.68 & 12.445 & 0 & 0 & 0 & 0\\
 & 1 & 19:16:31.067 & +11:16:08.171 & 1.053 & 0.934 & -146.14 & 13.492 & 0 & 0 & 1 & 1\\
 & 2 & 19:16:31.251 & +11:16:09.758 & 0.647 & 0.489 & 126.5 & 21.606 & 0 & 0 & 0 & 0\\
 & 3 & 19:16:31.086 & +11:16:11.907 & 0.893 & 0.764 & 55.78 & 241.117 & 0 & 1 & 0 & 0\\
 & 4 & 19:16:30.787 & +11:16:14.055 & 7.884 & 6.5 & 167.69 & 46.958 & 0 & 0 & 0 & 0\\
\hline
SDC45.927-0.375$\_2$\multirow{5}{*}{} & 0 & 19:16:55.984 & +11:21:42.792 & 1.006 & 0.867 & -149.71 & 5.99 & 0 & 0 & 0 & 0\\
 & 1 & 19:16:56.136 & +11:21:47.836 & 0.669 & 0.444 & 102.99 & 5.619 & 0 & 0 & 0 & 1\\
 & 2 & 19:16:56.175 & +11:21:50.077 & 0.831 & 0.468 & 76.57 & 33.926 & 3 & 1 & 1 & 0\\
 & 3 & 19:16:56.086 & +11:21:52.692 & 0.704 & 0.39 & 130.8 & 12.698 & 0 & 0 & 0 & 0\\
 & 4 & 19:16:56.232 & +11:21:58.016 & 0.836 & 0.507 & -177.86 & 25.674 & 3 & 0 & 0 & 0\\
\hline
RMS-G050.2213-00.6063\multirow{11}{*}{} & 0 & 19:25:57.774 & +15:02:54.676 & 0.557 & 0.29 & -168.9 & 18.868 & 0 & 0 & 0 & 0\\
 & 1 & 19:25:57.671 & +15:02:54.489 & 0.405 & 0.314 & 150.99 & 7.491 & 0 & 0 & 0 & 0\\
 & 2 & 19:25:57.387 & +15:03:05.044 & 0.519 & 0.371 & 98.65 & 15.144 & 0 & 0 & 0 & 0\\
 & 3 & 19:25:57.471 & +15:03:08.219 & 0.678 & 0.475 & 109.5 & 7.699 & 0 & 0 & 0 & 0\\
 & 4 & 19:25:57.723 & +15:02:57.105 & 0.817 & 0.303 & -146.63 & 35.796 & 0 & 1 & 0 & 0\\
 & 5 & 19:25:57.574 & +15:02:58.973 & 0.503 & 0.325 & 115.3 & 5.18 & 0 & 0 & 0 & 0\\
 & 6 & 19:25:57.813 & +15:03:00.000 & 0.681 & 0.311 & 162.24 & 28.359 & 0 & 0 & 0 & 0\\
 & 7 & 19:25:57.652 & +15:02:59.813 & 0.43 & 0.316 & 165.08 & 16.795 & 0 & 0 & 1 & 1\\
 & 8 & 19:25:57.523 & +15:03:00.280 & 0.46 & 0.316 & 52.02 & 28.785 & 0 & 0 & 0 & 0\\
 & 9 & 19:25:57.884 & +15:03:01.308 & 0.544 & 0.494 & 114.21 & 5.6 & 0 & 0 & 0 & 0\\
 & 10 & 19:25:57.426 & +15:03:02.989 & 0.913 & 0.285 & 71.63 & 10.811 & 1 & 0 & 0 & 0\\
\hline
RMS-G326.6618+00.5207\multirow{8}{*}{} & 0 & 15:45:02.768 & -54:09:16.823 & 1.002 & 0.648 & -167.1 & 69.391 & 0 & 0 & 0 & 0\\
 & 1 & 15:45:02.524 & -54:09:15.049 & 1.1 & 0.807 & 59.65 & 18.543 & 0 & 0 & 0 & 0\\
 & 2 & 15:45:02.609 & -54:09:13.367 & 31.705 & 24.132 & -152.67 & 197.379 & 0 & 1 & 0 & 0\\
 & 3 & 15:45:03.055 & -54:09:03.560 & 0.48 & 0.308 & 51.79 & 66.805 & 0 & 0 & 0 & 0\\
 & 4 & 15:45:02.864 & -54:09:03.093 & 0.37 & 0.331 & -165.04 & 85.377 & 0 & 0 & 1 & 0\\
 & 5 & 15:45:02.343 & -54:09:01.973 & 1.256 & 0.934 & 61.81 & 9.585 & 0 & 0 & 0 & 1\\
 & 6 & 15:45:02.417 & -54:09:00.198 & 0.897 & 0.632 & 146.89 & 9.996 & 0 & 0 & 0 & 0\\
 & 7 & 15:45:02.407 & -54:08:55.902 & 0.602 & 0.532 & -172.95 & 14.735 & 2 & 0 & 0 & 0\\
\hline
RMS-G327.1192+00.5103\multirow{7}{*}{} & 0 & 15:47:32.958 & -53:52:43.483 & 0.939 & 0.541 & 48.2 & 20.114 & 0 & 0 & 0 & 0\\
 & 1 & 15:47:32.536 & -53:52:43.203 & 1.088 & 0.386 & 137.84 & 15.642 & 0 & 0 & 0 & 0\\
 & 2 & 15:47:32.304 & -53:52:39.467 & 0.604 & 0.489 & -155.61 & 38.166 & 0 & 0 & 0 & 0\\
 & 3 & 15:47:32.726 & -53:52:38.720 & 0.91 & 0.634 & -174.47 & 373.72 & 3 & 1 & 1 & 1\\
 & 4 & 15:47:32.335 & -53:52:35.824 & 1.253 & 1.118 & 88.5 & 9.491 & 0 & 0 & 0 & 0\\
 & 5 & 15:47:34.015 & -53:52:34.703 & 1.476 & 1.162 & -170.16 & 14.635 & 0 & 0 & 0 & 0\\
 & 6 & 15:47:32.040 & -53:52:34.143 & 1.094 & 0.559 & 55.73 & 16.48 & 0 & 0 & 0 & 0\\
\hline
RMS-G332.0939-00.4206\multirow{10}{*}{} & 0 & 16:16:16.988 & -51:18:29.950 & 0.874 & 0.574 & 156.72 & 19.053 & 0 & 0 & 0 & 0\\
 & 1 & 16:16:16.550 & -51:18:19.956 & 0.786 & 0.447 & -163.03 & 80.685 & 0 & 0 & 0 & 0\\
 & 2 & 16:16:16.380 & -51:18:17.061 & 0.792 & 0.507 & 66.21 & 23.143 & 0 & 0 & 0 & 0\\
 & 3 & 16:16:16.699 & -51:18:17.061 & 0.724 & 0.503 & 68.51 & 15.57 & 0 & 0 & 0 & 0\\
 & 4 & 16:16:16.649 & -51:18:27.615 & 0.491 & 0.219 & -166.37 & 19.434 & 0 & 0 & 0 & 0\\
 & 5 & 16:16:17.207 & -51:18:26.868 & 1.249 & 0.56 & 133.39 & 14.813 & 0 & 0 & 0 & 0\\
 & 6 & 16:16:16.460 & -51:18:25.280 & 0.469 & 0.375 & 57.37 & 376.691 & 3 & 1 & 0 & 0\\
 & 7 & 16:16:16.550 & -51:18:24.159 & 0.333 & 0.242 & 59.58 & 14.485 & 0 & 0 & 1 & 0\\
 & 8 & 16:16:16.331 & -51:18:22.291 & 0.755 & 0.456 & 142.32 & 20.463 & 0 & 0 & 0 & 0\\
 & 9 & 16:16:16.759 & -51:18:21.824 & 0.363 & 0.265 & 137.53 & 17.8 & 1 & 0 & 0 & 1\\
\hline
RMS-G332.9636-00.6800\multirow{10}{*}{} & 0 & 16:21:22.594 & -50:53:05.164 & 1.098 & 0.624 & -171.51 & 24.495 & 3 & 0 & 0 & 0\\
 & 1 & 16:21:22.851 & -50:53:03.577 & 0.708 & 0.582 & 51.47 & 49.414 & 0 & 0 & 0 & 0\\
 & 2 & 16:21:22.564 & -50:53:01.335 & 0.612 & 0.319 & 152.54 & 17.997 & 0 & 0 & 0 & 0\\
 & 3 & 16:21:22.880 & -50:52:59.747 & 0.24 & 0.216 & -159.88 & 48.117 & 3 & 0 & 0 & 1\\
 & 4 & 16:21:22.959 & -50:52:58.626 & 0.248 & 0.209 & 165.18 & 52.006 & 0 & 0 & 0 & 0\\
 & 5 & 16:21:23.245 & -50:52:56.572 & 1.275 & 0.639 & 117.12 & 26.134 & 0 & 0 & 0 & 0\\
 & 6 & 16:21:22.811 & -50:52:56.291 & 0.58 & 0.375 & -155.35 & 73.564 & 0 & 1 & 1 & 0\\
 & 7 & 16:21:21.587 & -50:52:56.011 & 0.695 & 0.545 & 141.6 & 47.643 & 0 & 0 & 0 & 0\\
 & 8 & 16:21:23.384 & -50:52:54.517 & 1.372 & 0.479 & 135.17 & 30.728 & 0 & 0 & 0 & 0\\
 & 9 & 16:21:22.061 & -50:52:52.275 & 2.157 & 0.687 & 163.27 & 30.227 & 0 & 0 & 0 & 0\\
\hline
RMS-G332.9868-00.4871\multirow{4}{*}{} & 0 & 16:20:37.672 & -50:43:52.615 & 9.77 & 4.897 & 156.62 & 23.565 & 0 & 0 & 0 & 0\\
 & 1 & 16:20:37.839 & -50:43:51.588 & 0.945 & 0.464 & 174.74 & 14.518 & 0 & 0 & 0 & 0\\
 & 2 & 16:20:37.790 & -50:43:49.626 & 0.726 & 0.487 & -169.58 & 66.333 & 0 & 1 & 1 & 1\\
 & 3 & 16:20:37.633 & -50:43:44.022 & 1.013 & 0.816 & 46.16 & 22.637 & 0 & 0 & 0 & 0\\
\hline
RMS-G333.0682-00.4461\multirow{15}{*}{} & 0 & 16:20:48.813 & -50:38:47.192 & 1.205 & 0.706 & 83.08 & 16.72 & 0 & 0 & 0 & 0\\
 & 1 & 16:20:48.990 & -50:38:43.923 & 0.624 & 0.273 & 126.74 & 59.986 & 0 & 0 & 0 & 0\\
 & 2 & 16:20:48.951 & -50:38:38.412 & 0.327 & 0.229 & -149.57 & 26.04 & 0 & 0 & 0 & 0\\
 & 3 & 16:20:48.568 & -50:38:38.319 & 0.402 & 0.21 & -141.16 & 24.13 & 0 & 0 & 0 & 0\\
 & 4 & 16:20:49.236 & -50:38:37.665 & 1.089 & 0.502 & 76.89 & 23.394 & 0 & 0 & 0 & 0\\
 & 5 & 16:20:48.126 & -50:38:37.758 & 1.057 & 0.553 & -148.08 & 25.686 & 0 & 0 & 0 & 0\\
 & 6 & 16:20:48.028 & -50:38:35.984 & 1.187 & 0.774 & 55.63 & 26.818 & 0 & 0 & 0 & 0\\
 & 7 & 16:20:48.794 & -50:38:33.836 & 0.47 & 0.419 & -169.86 & 37.76 & 0 & 0 & 0 & 0\\
 & 8 & 16:20:47.832 & -50:38:28.698 & 10.477 & 6.758 & -145.25 & 186.908 & 0 & 0 & 0 & 0\\
 & 9 & 16:20:48.764 & -50:38:43.549 & 1.037 & 0.691 & 48.7 & 14.95 & 0 & 0 & 0 & 0\\
 & 10 & 16:20:48.980 & -50:38:42.335 & 0.439 & 0.31 & 74.91 & 42.44 & 0 & 0 & 0 & 0\\
 & 11 & 16:20:49.196 & -50:38:41.308 & 0.527 & 0.125 & -139.45 & 48.701 & 0 & 0 & 0 & 0\\
 & 12 & 16:20:48.980 & -50:38:40.467 & 0.631 & 0.353 & -164.52 & 480.716 & 3 & 1 & 0 & 0\\
 & 13 & 16:20:48.372 & -50:38:40.467 & 0.59 & 0.231 & 177.23 & 28.151 & 0 & 0 & 0 & 0\\
 & 14 & 16:20:48.676 & -50:38:39.626 & 0.405 & 0.207 & 45.88 & 43.333 & 0 & 0 & 1 & 1\\
\hline
RMS-G338.9196+00.5495\multirow{6}{*}{} & 0 & 16:40:33.082 & -45:42:14.538 & 0.598 & 0.442 & 116.9 & 75.288 & 0 & 0 & 0 & 0\\
 & 1 & 16:40:33.688 & -45:42:09.868 & 0.515 & 0.377 & 142.82 & 98.854 & 0 & 0 & 0 & 0\\
 & 2 & 16:40:34.036 & -45:42:08.747 & 0.487 & 0.179 & 117.8 & 174.486 & 0 & 1 & 0 & 0\\
 & 3 & 16:40:33.768 & -45:42:08.280 & 0.485 & 0.326 & 130.91 & 29.941 & 0 & 0 & 1 & 1\\
 & 4 & 16:40:34.018 & -45:42:07.346 & 0.245 & 0.226 & -137.9 & 100.37 & 3 & 0 & 0 & 0\\
 & 5 & 16:40:33.724 & -45:42:02.770 & 0.955 & 0.42 & 55.53 & 57.357 & 0 & 0 & 0 & 0\\
\hline
RMS-G339.6221-00.1209\multirow{9}{*}{} & 0 & 16:46:06.018 & -45:36:51.005 & 0.749 & 0.507 & 157.63 & 16.124 & 0 & 0 & 0 & 0\\
 & 1 & 16:46:06.819 & -45:36:49.137 & 0.856 & 0.51 & -135.63 & 19.075 & 0 & 0 & 0 & 0\\
 & 2 & 16:46:05.822 & -45:36:40.638 & 0.554 & 0.416 & 50.3 & 12.545 & 0 & 0 & 0 & 0\\
 & 3 & 16:46:05.555 & -45:36:40.731 & 1.183 & 0.428 & 170.09 & 14.089 & 0 & 0 & 0 & 0\\
 & 4 & 16:46:05.982 & -45:36:43.720 & 0.934 & 0.421 & 99.07 & 154.839 & 3 & 1 & 0 & 1\\
 & 5 & 16:46:05.475 & -45:36:42.879 & 0.799 & 0.575 & 53.24 & 9.796 & 0 & 0 & 0 & 0\\
 & 6 & 16:46:06.142 & -45:36:42.319 & 0.721 & 0.29 & 156.32 & 9.834 & 0 & 0 & 1 & 0\\
 & 7 & 16:46:07.246 & -45:36:40.731 & 0.814 & 0.655 & 144.74 & 104.558 & 0 & 0 & 0 & 0\\
 & 8 & 16:46:06.463 & -45:36:40.824 & 0.984 & 0.459 & 162.96 & 10.712 & 0 & 0 & 0 & 0\\
\hline
RMS-G345.5043+00.3480\multirow{8}{*}{} & 0 & 17:04:23.722 & -40:44:31.005 & 1.243 & 0.588 & 131.58 & 48.419 & 0 & 0 & 0 & 0\\
 & 1 & 17:04:23.451 & -40:44:27.269 & 0.644 & 0.337 & 51.02 & 58.091 & 0 & 0 & 0 & 0\\
 & 2 & 17:04:23.262 & -40:44:25.494 & 0.384 & 0.33 & 52.8 & 38.329 & 0 & 0 & 0 & 0\\
 & 3 & 17:04:23.155 & -40:44:24.654 & 0.463 & 0.352 & -165.92 & 38.053 & 0 & 0 & 1 & 1\\
 & 4 & 17:04:22.958 & -40:44:24.934 & 0.33 & 0.254 & 81.78 & 34.942 & 0 & 0 & 0 & 0\\
 & 5 & 17:04:22.900 & -40:44:22.786 & 0.47 & 0.38 & -153.98 & 850.958 & 3 & 1 & 0 & 0\\
 & 6 & 17:04:22.793 & -40:44:21.572 & 0.346 & 0.197 & 55.2 & 55.957 & 0 & 0 & 0 & 0\\
 & 7 & 17:04:22.580 & -40:44:19.143 & 1.154 & 0.648 & -147.87 & 27.117 & 0 & 0 & 0 & 0\\
\hline
\hline
\label{BIGTABLE:tab}
\end{longtable}
\end{landscape}
\twocolumn



The emission fraction values within the TEMPO sample appears to be largely consistent with other high-mass star forming regions which have been independently studied. For example, in the study of NGC6334 I(N) \citet{Hunter14} find $\sim$83\% of their sources have a flux density $<20\%$ of the bright source (derived from values in their Table 2). The \citet{Hunter14} data has a slightly lower sensitivity to TEMPO\footnote{c.f. TEMPO $\Delta S =$ 0.26 / 0.23 / 0.69 / 0.09 mJy (mean/median/max/min).} with $\Delta S = 2.2$mJy beam$^{-1}$. Both studies were conducted at 1.3mm. 

The work on G28.34+0.06 P1 also at 1.3mm by \citet{Zhang15}, however, finds lower values of 47\% of sources with $<20$\% of the highest flux density. The \citet{Zhang15} data is slightly higher sensitivity to the mean value of the TEMPO data at 0.075mJy beam$^{-1}$. Whilst not as high percentage as those reported by \citet{Hunter14} and the TEMPO result, \citet{Zhang15} note in their paper there is an under abundance of low mass cores in their target field, which would drive the low mass percentage down in this source. The authors suggest this may be caused by lower mass stars forming later and the trend seen across these various studies may be indicative of the relative evolutionary stages across the studied sources.

\section{Conclusion}

The TEMPO survey conducted a high resolution (0.8\asec), high sensitivity (mean \textit{rms}-noise $\sim$0.26mJy equivalent to $\sim$1.0-2.5\solmass\ for $T$ = 30/15K respectively) ALMA survey of 38 colour-luminosity selected high-mass star forming regions. The continuum emission from fragments within the survey sample fields has been imaged and the clustering, fragmentation, and distribution of emission has been assessed. Additionally we have undertaken analysis to gauge whether the observed sources are matter over-densities or centrally condensed (and therefore likely currently star-forming). 

Our key findings are given in the following bullet point list:

\begin{itemize}
    \item Each field has between 2 and 15 detected fragments (average 7.6).
    \item The observed clusters in our target fields do not show distributions consistent with a simple radial profile ($r^{-\alpha}$) for $\alpha$=0,1,2 but it is possible to exclude higher $\alpha$ values.
    \item The \citet{CartwrightWhitworth04} $Q$ parameter does not work to distinguish fractal from radial cluster distributions for small number ($N<15$) clusters. See Appendix \ref{Qpara:appdx}.
    \item The fragmentation scale, calculated as the mean edge length for the minimum spanning tree in each field, is not consistent with thermal Jeans fragmentation for the majority of fields in the TEMPO sample. With 33 (87\% of the sample) having a mean edge length, $X$, greater than or equal to 1.5$\times$ the thermal Jeans fragmentation scale suggesting that some other mode of fragmentation may be in effect in these fields. The remaining 5 fields have fragmentation scales comparable with (or in one case) smaller than the thermal Jeans length.
    \item Across the whole sample the majority ($\sim$69\%) of detected fragments have a low flux density compared to the brightest source in that field, where low is defined as $<$20\% of the flux of the brightest fragment in their respective fields. The flux budget within the TEMPO fields is divided approximately evenly, 47\%:53\% between fields where the sum of low flux density fragments is greater than that of the brightest field fragment, and where the highest flux density fragment dominates.  For the latter fields, predominantly the brightest fragment has greater than $3\times$ the flux density of the next brightest object, indicating that these fields are truly dominated by a single high flux density object.
    \item The brightest fragment in each TEMPO field is commonly associated with high-mass star formation activity as traced by class II 6.7~GHz \meth\ maser (70\% of fields with a maser present) and with the local 70\mewm\ source (55\% of fields). This suggests a good correlation between the brightest 1.3mm TEMPO fragment and the high-mass star forming core candidate in each field.
    \item Two noteworthy trends are seen when comparing derived properties from the TEMPO continuum maps to clump luminosity. Firstly, the number of fragments detected shows no correlation with increasing luminosity. Given outflow power from the evolving protostar(s) in each field could be expected to increase with age, the disruption of nearby material and thus number of fragments could be expected to increase. This is not seen. Secondly, the amount of spectral line-free bandwidth for each source shows a weak positive correlation with increasing luminosity, suggesting the younger (lower luminosity) fields are more line rich than their more evolved (higher luminosity) counterparts. Splitting the TEMPO sample between those with an associated 6.7GHz\meth\ maser and those without, there is some indication that the maser associated sub-sample tends toward the younger, lower luminosity fields. Something which is expected from maser lifetime and pumping mechanism literature.
    \item The interferometric visibilities properties of the TEMPO fragments were investigated and compared to those of point-like, Gaussian and Gaussian$+$Point profiles, to provide an indication of the centrally condensed nature of the fragments and thus whether they are actively star-forming or not. This implemented technique recovered 42 fragments ($\sim$ 15\% of the sample) which match the empirically derived criteria to be considered actively star-forming at their respective SNR. These actively star-forming candidates show a high correspondence with the class II \meth\ masers sample (67\%) and 70 \mewm\ IR sources (37\% of sample). It is noted (c.f Appendix \ref{centCond:apdx}) that the visibility analysis applied suffers some limitation, for complex and potentially unresolved objects requiring further analysis beyond the scope of this paper. However, it may be instructive to apply this technique to a wider sample of star-forming regions observed by ALMA to further establish an `active star forming core' criteria in the ALMA-era, over reliance on classic clump scale tracers.
 
\end{itemize}

\section*{Acknowledgements}

A.A. acknowledges support from STFC grants ST/T001488/1 and ST/P000827/1.

G.A.F acknowledges support from the Collaborative Research Centre 956, funded by the Deutsche Forschungsgemeinschaft (DFG) project ID 184018867.
G.A.F also acknowledges financial support from the State Agency for Research of the Spanish MCIU through the AYA 2017-84390-C2-1-R grant (co-funded by FEDER) 
from grant PID2020-114461GB-I00, funded by MCIN/AEI/10.13039/5011000110
and through the "Center of Excellence Severo Ochoa" award for the Instituto de Astrof\'isica de Andalucia
(SEV-2017-0709). 

Z.N. acknowledges funding from the European Research Council (ERC) under the European Union's Horizon 2020 research and innovation programme under grant agreement No 716155 (SACCRED), from the ESA PRODEX contract nr. 4000132054, and from the J\'anos Bolyai Research Scholarship of the Hungarian Academy of Sciences.

J.E.P. was supported by the Max-Planck Society.

This paper makes use of the following ALMA data: ADS/JAO.ALMA\#2015.1.01312.S. ALMA is a partnership of ESO (representing its member states), NSF (USA) and NINS (Japan), together with NRC (Canada) and NSC and ASIAA (Taiwan), in cooperation with the Republic of Chile. This research made use of astrodendro, a Python package to compute dendrograms of Astronomical data (http://www.dendrograms.org/) and Astropy, a community-developed core Python package for Astronomy \citep{AstropyPaper}.


\section*{Data Availability}
The observational ALMA data is publicly available via the ALMA Archive under project code 2015.1.01312.S. The data underlying the analysis within this article are available in the article and in its online supplementary material and will be shared on reasonable request to the corresponding author.


\bibliographystyle{mnras}
\bibliography{proj_1312.bib} 



\appendix
\section{Continuum Maps of all target fields}
\label{Maps:apdx}
\renewcommand\thesubfigure{\roman{subfigure}}
\begin{figure*}
    \centering
    \begin{subfigure}[t]{0.5\textwidth}
        \centering
        \includegraphics[scale=0.3]{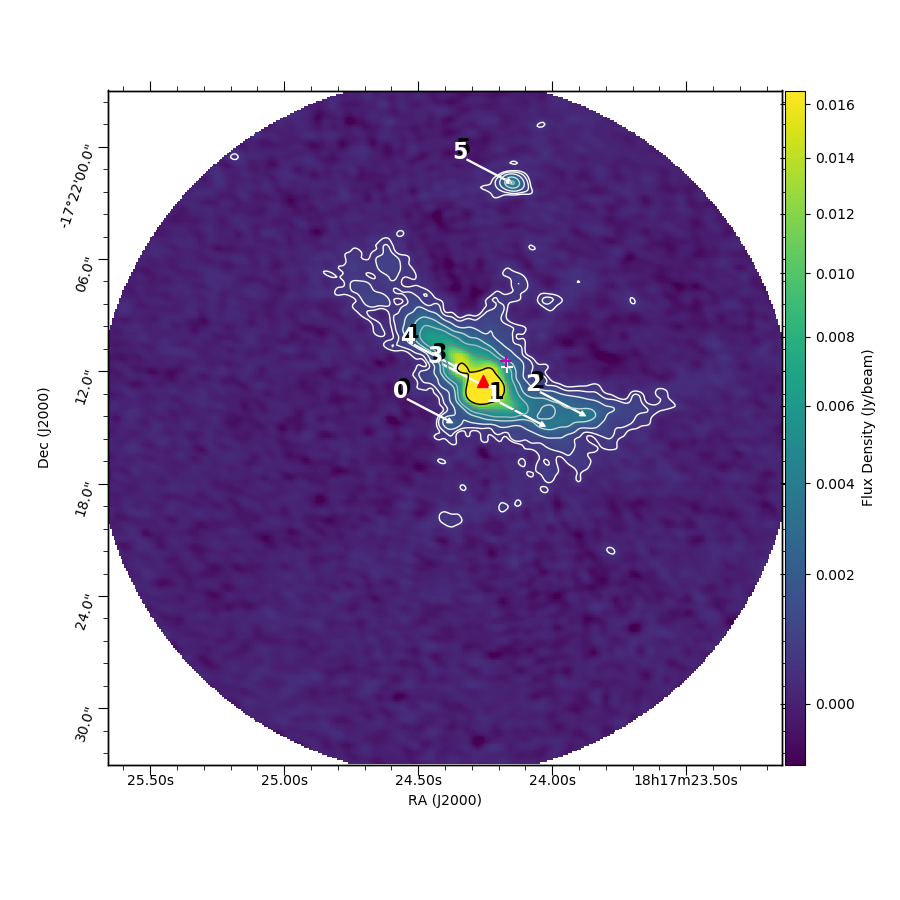}
        \caption{G013.6562$-$00.5997}
        \label{G013p6562-00p5997:fig}
    \end{subfigure}%
    ~ 
    \begin{subfigure}[t]{0.5\textwidth}
        \centering
	\includegraphics[scale=0.3]{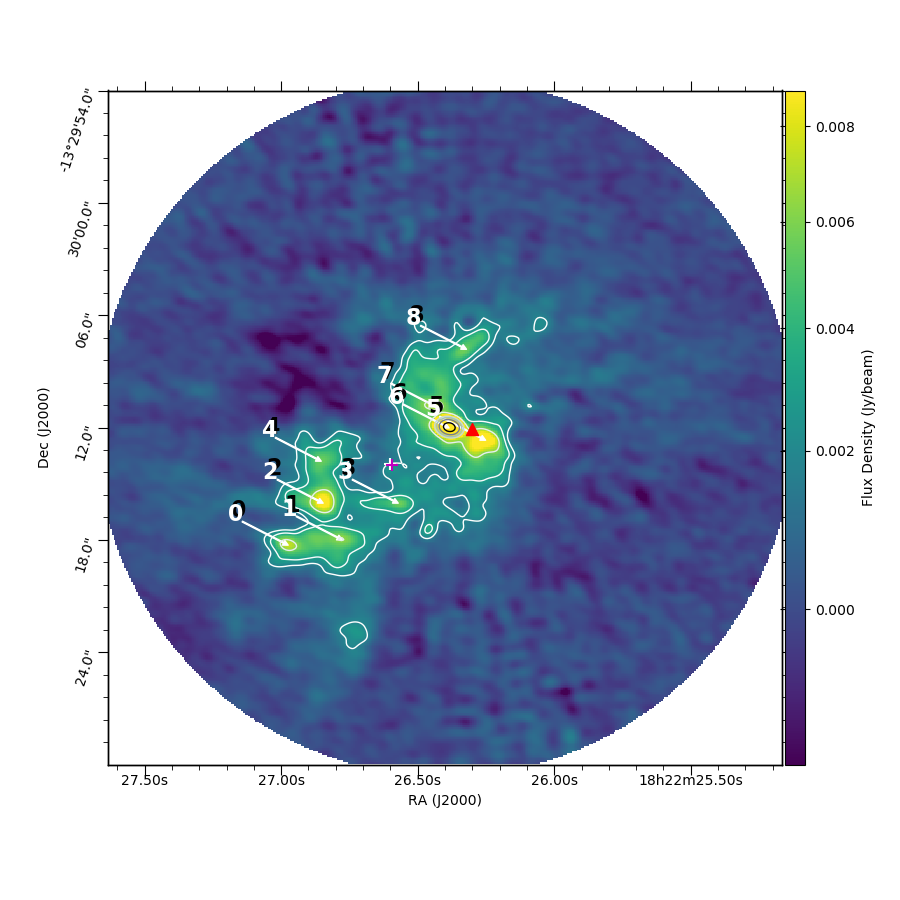}
        \caption{G017.6380$+$00.1566}
        \label{G017.6380+00p1566:fig}
    \end{subfigure}
    
    \begin{subfigure}[t]{0.5\textwidth}
        \centering
        \includegraphics[scale=0.3]{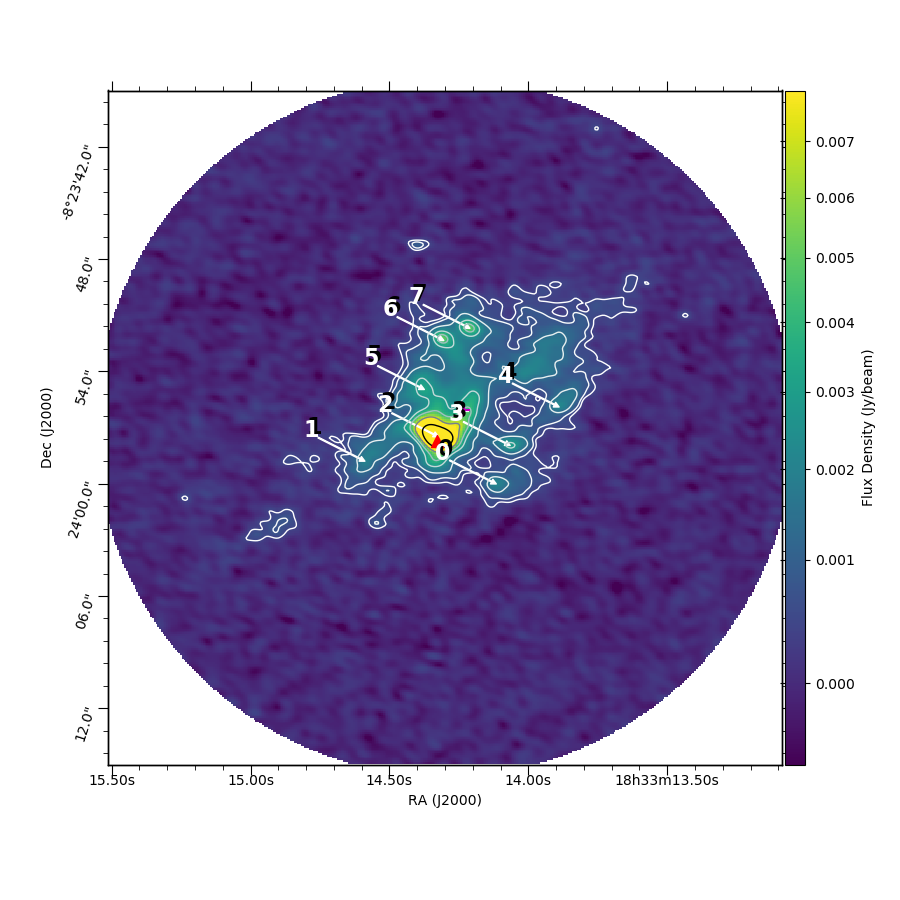}
        \caption{G023.3891$+$00.1851}
        \label{G023.3891$+$00.1851:fig}
    \end{subfigure}%
    ~ 
    \begin{subfigure}[t]{0.5\textwidth}
        \centering
	\includegraphics[scale=0.3]{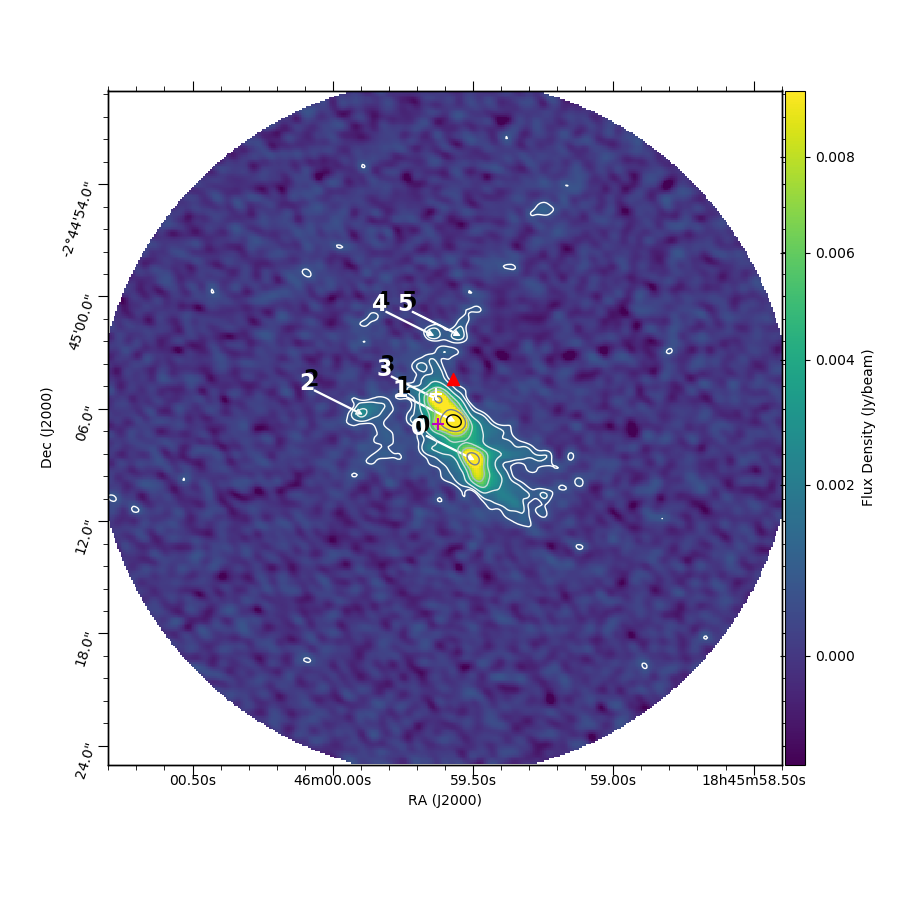}
        \caption{G029.8620$-$00.0444}
        \label{G029.8620$-$00.0444:fig}
    \end{subfigure}
    
    \begin{subfigure}[t]{0.5\textwidth}
        \centering
        \includegraphics[scale=0.3]{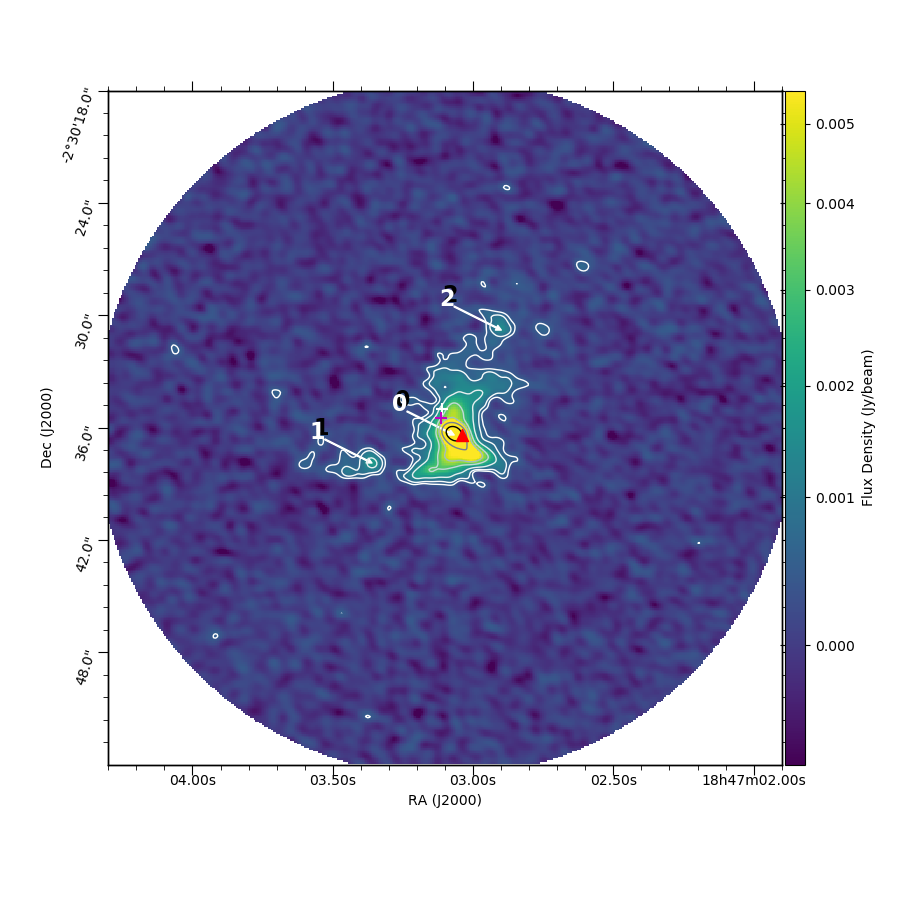}
        \caption{G030.1981$-$00.1691}
        \label{G030.1981-00.1691:fig}
    \end{subfigure}%
    ~ 
    \begin{subfigure}[t]{0.5\textwidth}
        \centering
	\includegraphics[scale=0.3]{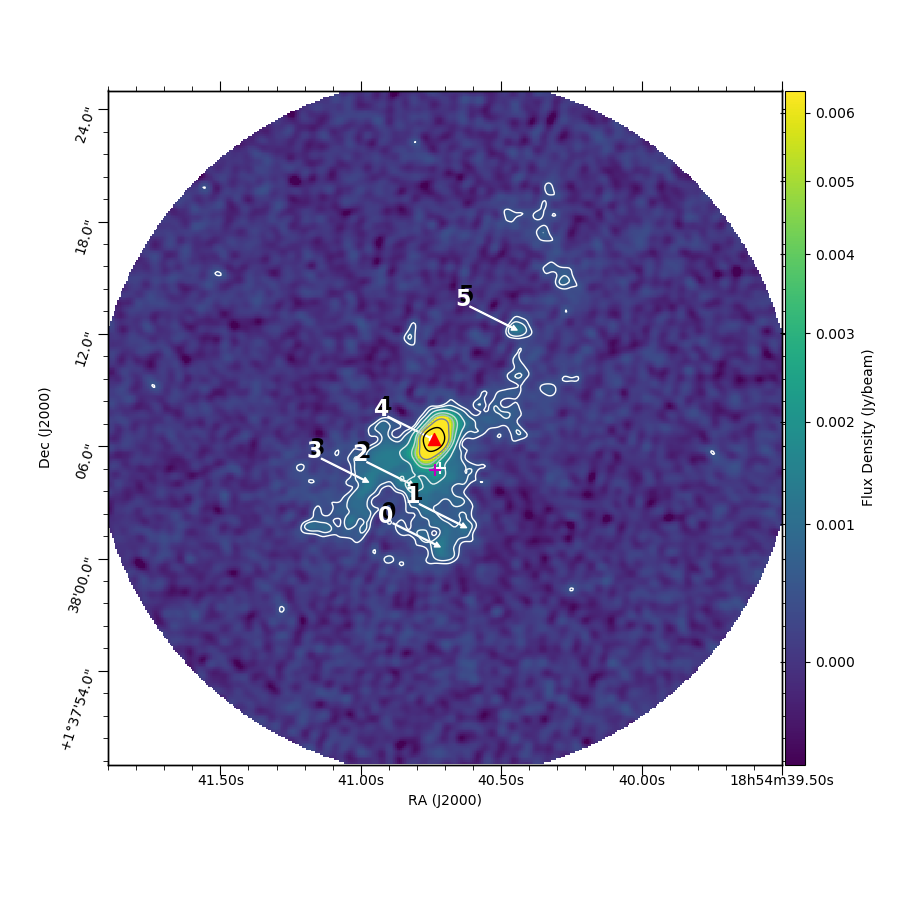}
        \caption{G034.7569$+$00.0247}
        \label{G034.7569+00p0247_sources:fig}
    \end{subfigure}
\caption{Catalogue of field images: Contours are at 3, 5, 10, 20, 30, 50 and 100$\times$ the fields rms-noise level, 0.17 and 0.46mJy respectively. The red triangle in (a) indicates the position of the 6.7~GHz methanol maser in that field with position from the MMB catalogues \citet{MMB006to020,MMB186to330,MMB345to006,MMB330to345,MMB020to060}. Numbers and arrows indicate the detected fragments in each field as per Table 4.}
\label{Images_of_all_fields:fig}
\end{figure*}

\renewcommand{\thefigure}{A\arabic{figure}}
\setcounter{figure}{1}
\begin{figure*}
	\centering
        \ContinuedFloat
	\begin{subfigure}[t]{0.5\textwidth}
		\centering
		\includegraphics[scale=0.3]{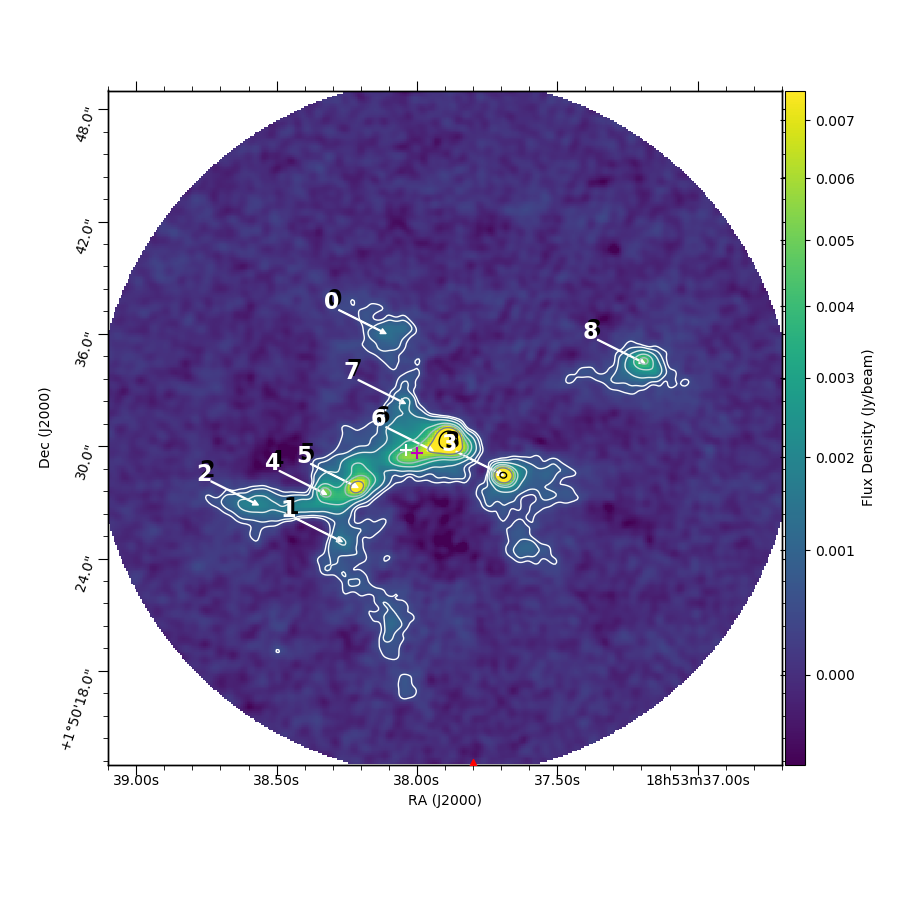}
		\caption{G034.8211$+$00.3519}
		\label{G034p8211+00p3519:fig}
	\end{subfigure}%
	~
	\begin{subfigure}[t]{0.5\textwidth}
		\centering
		\includegraphics[scale=0.3]{FIGURESV3/MAPS/G050p2213-00p6063_sources.png}
		\caption{G050.2213$-$00.6063}
		\label{G050p2213-00p6063:fig}
	\end{subfigure}%
	
	\begin{subfigure}[t]{0.5\textwidth}
		\centering
		\includegraphics[scale=0.3]{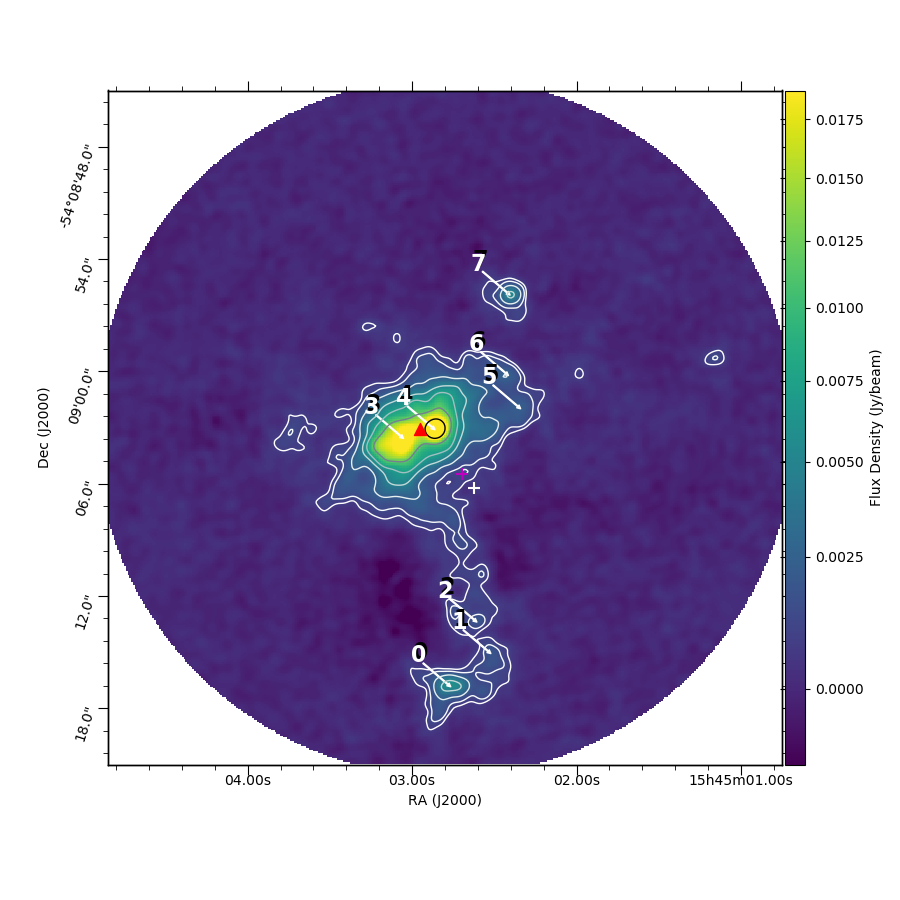}
		\caption{G326.6618$+$00.5207}
		\label{G326p6618+00p5207:fig}
	\end{subfigure}%
	~
	\begin{subfigure}[t]{0.5\textwidth}
		\centering
		\includegraphics[scale=0.3]{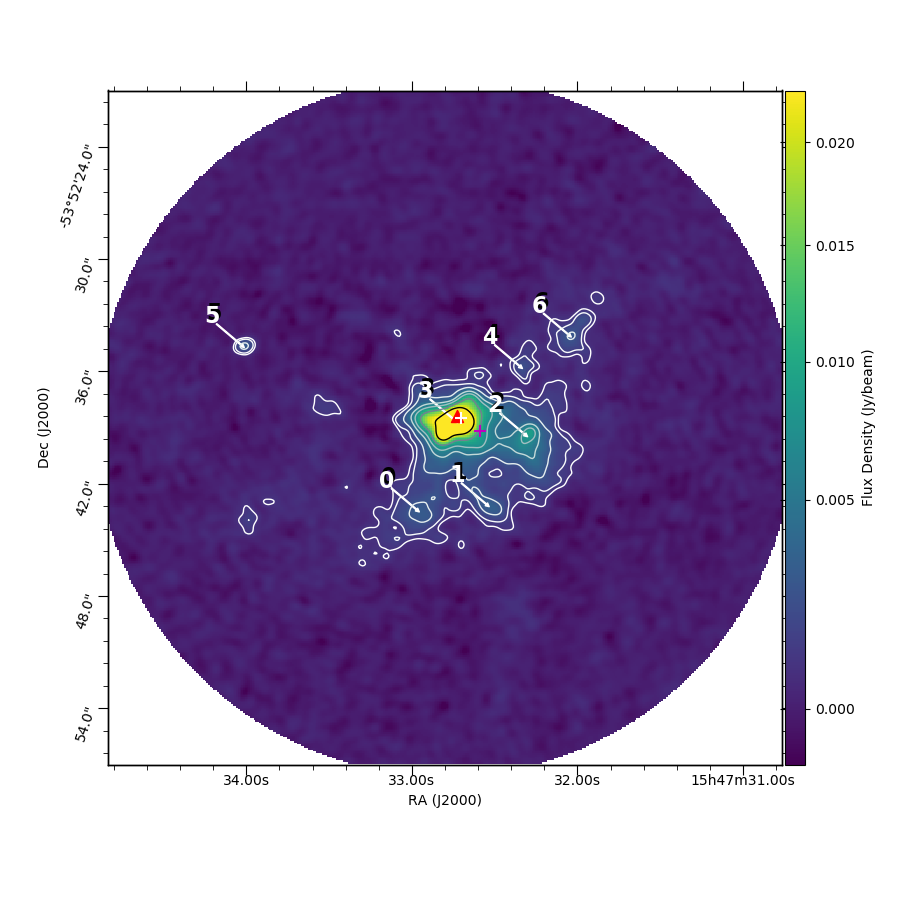}
		\caption{G327.1192$+$00.5103}
		\label{G327p1192+00p5103:fig}
	\end{subfigure}%
	
	\begin{subfigure}[t]{0.5\textwidth}
		\centering
		\includegraphics[scale=0.3]{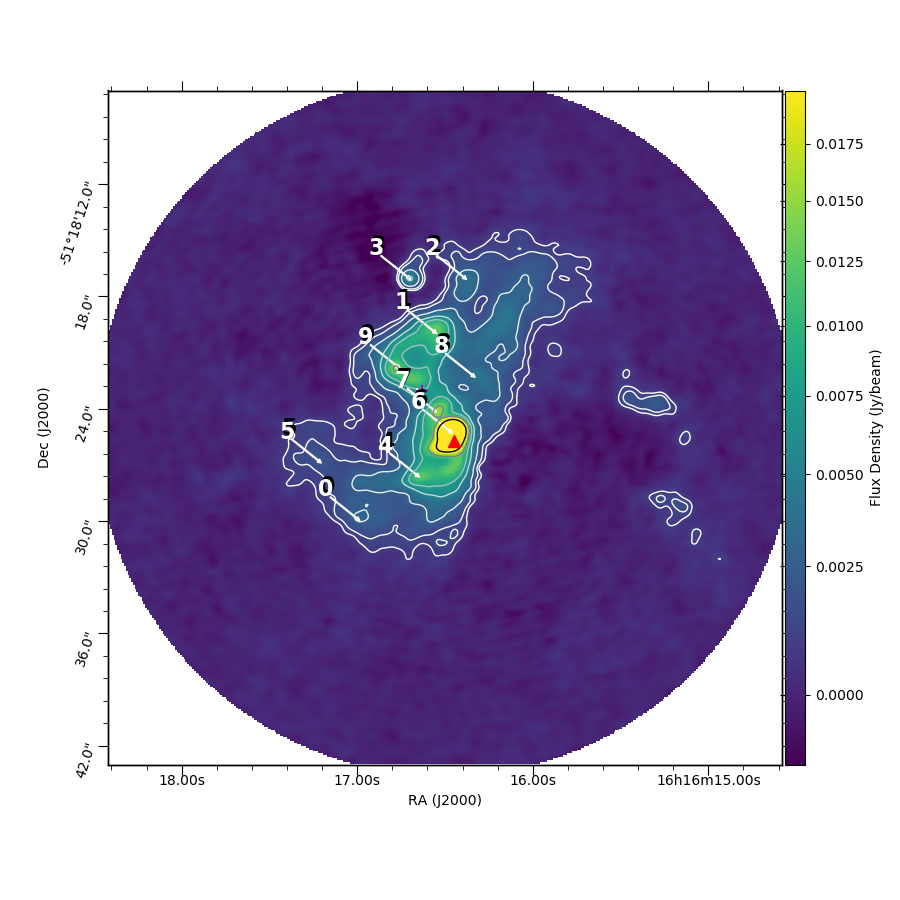}
		\caption{G332.0939$-$00.4206}
		\label{G332p0939-00p4206:fig}
	\end{subfigure}%
	~
	\begin{subfigure}[t]{0.5\textwidth}
		\centering
		\includegraphics[scale=0.3]{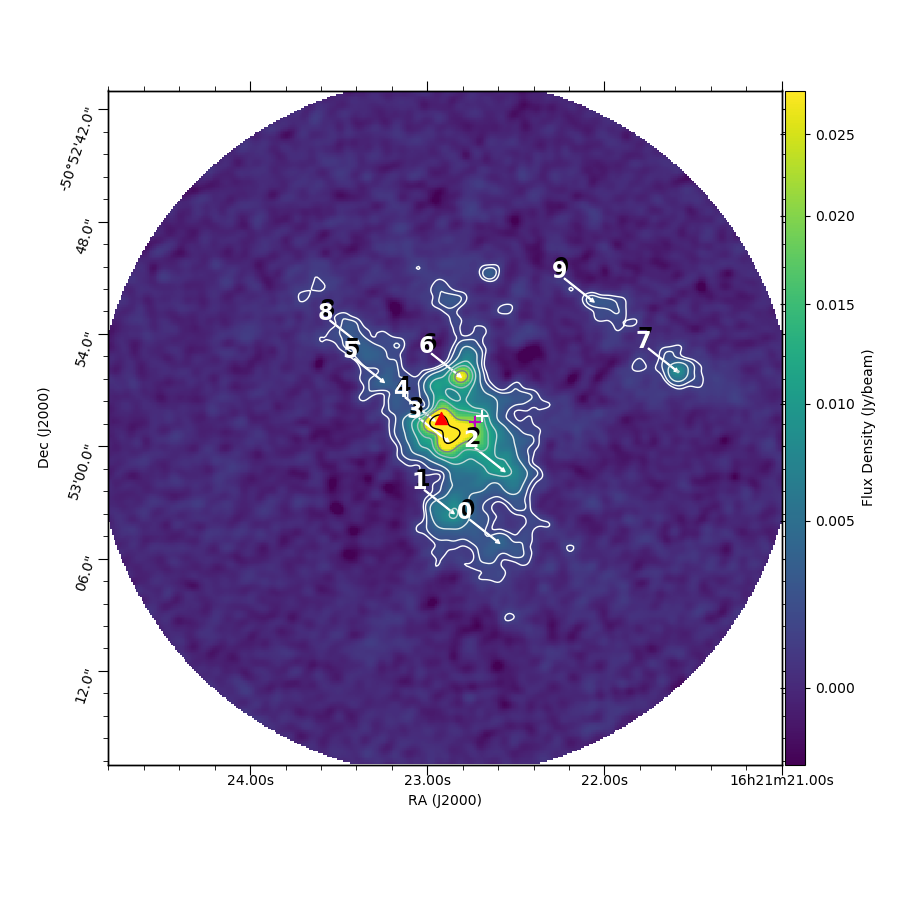}
		\caption{G332.9636$-$00.6800}
		\label{G332p9636-00p6800:fig}
	\end{subfigure}%
	
\caption{continued}
\label{Images_of_all_fields:fig}
\end{figure*}

\renewcommand{\thefigure}{A\arabic{figure}}
\setcounter{figure}{1}
\begin{figure*}
	\centering
        \ContinuedFloat
	\begin{subfigure}[t]{0.5\textwidth}
		\centering
		\includegraphics[scale=0.3]{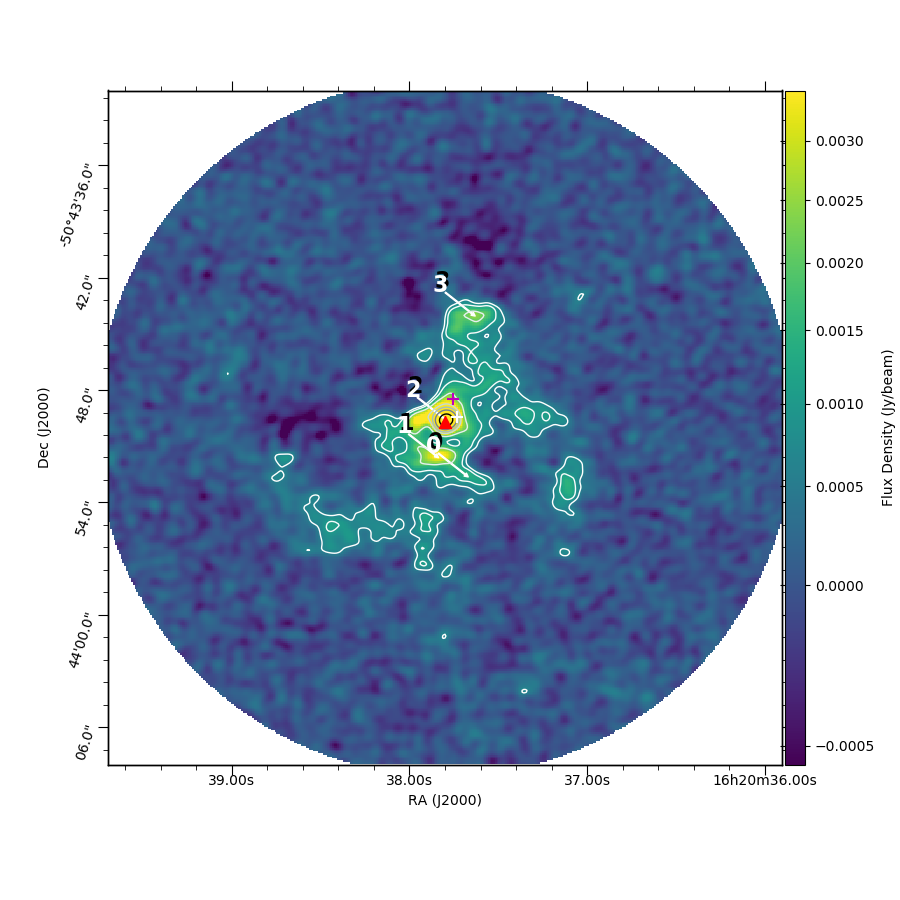}
		\caption{G332.9868$-$00.4871}
		\label{G332p9868-00p4871:fig}
	\end{subfigure}%
	~
	\begin{subfigure}[t]{0.5\textwidth}
		\centering
		\includegraphics[scale=0.3]{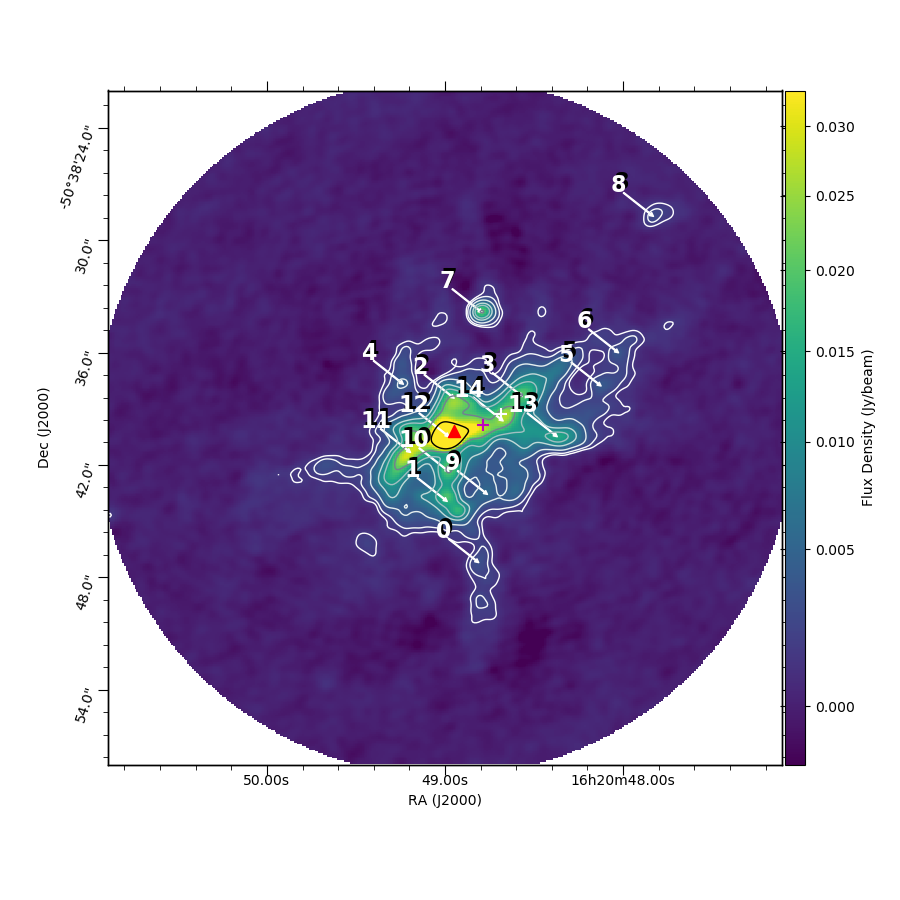}
		\caption{G333.0682$-$00.4461}
		\label{G333p0682-00p4461:fig}
	\end{subfigure}%
	
	\begin{subfigure}[t]{0.5\textwidth}
		\centering
		\includegraphics[scale=0.3]{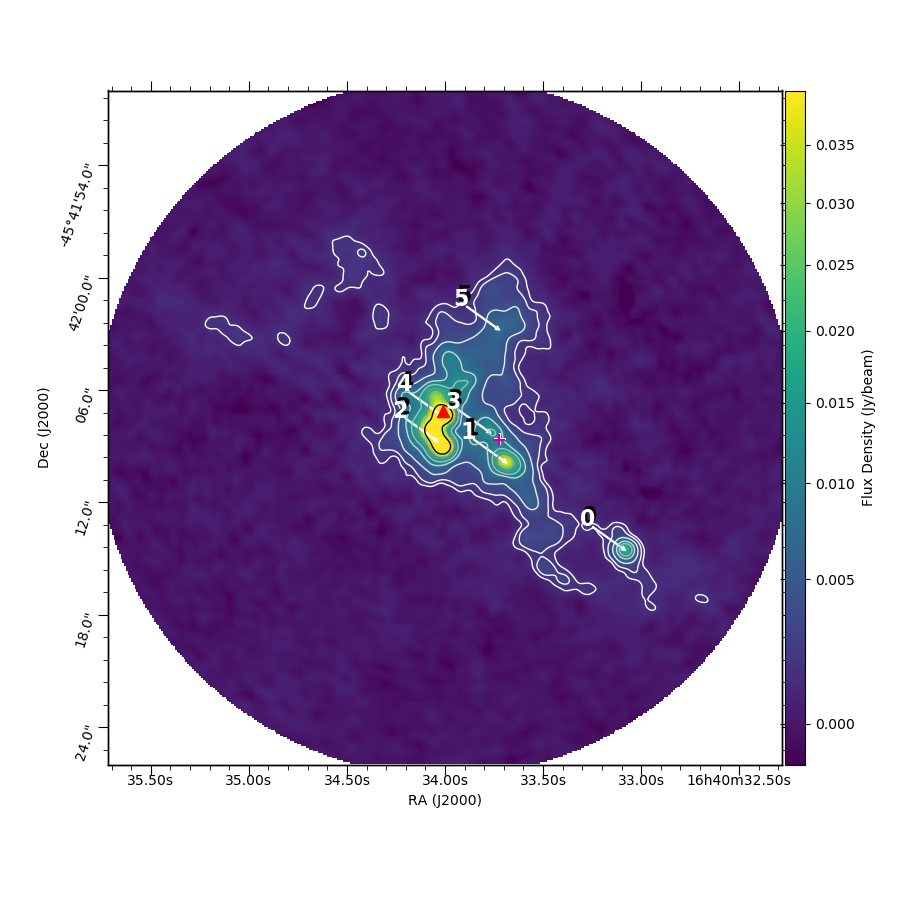}
		\caption{G338.9196$+$00.5495}
		\label{G338p9196+00p5495:fig}
	\end{subfigure}%
	~
	\begin{subfigure}[t]{0.5\textwidth}
		\centering
		\includegraphics[scale=0.3]{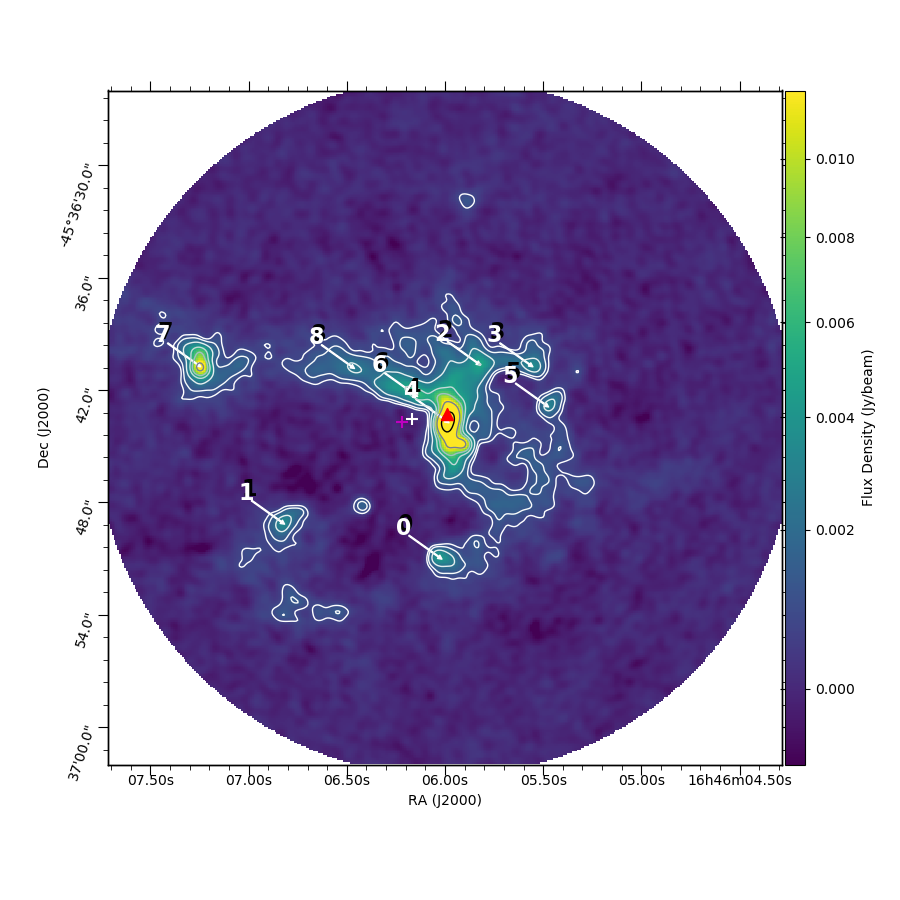}
		\caption{G339.6221$-$00.1209}
		\label{G339p6221-00p1209:fig}
	\end{subfigure}%
	
	\begin{subfigure}[t]{0.5\textwidth}
		\centering
		\includegraphics[scale=0.3]{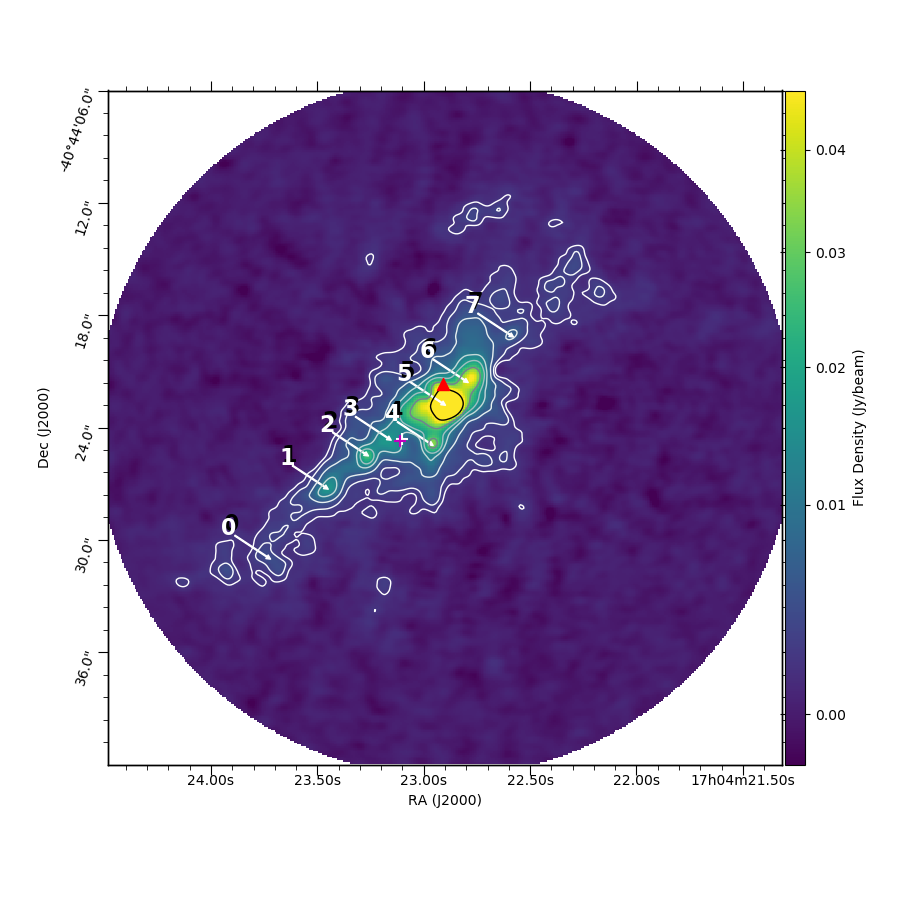}
		\caption{G345.5043$+$00.3480}
		\label{G345p5043+00p3480:fig}
	\end{subfigure}%
	~
	\begin{subfigure}[t]{0.5\textwidth}
		\centering
		\includegraphics[scale=0.3]{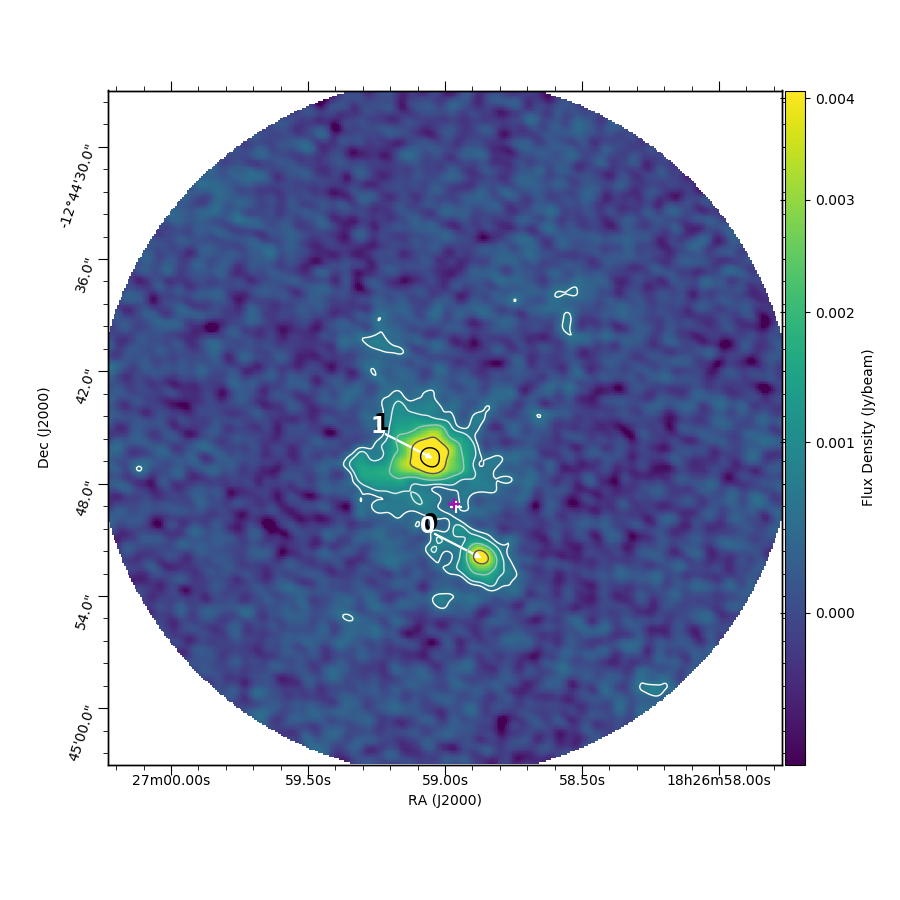}
		\caption{SDC18.816$-$0.447$\_$1}
		\label{SDC18p816-0p447_1:fig}
	\end{subfigure}%
\caption{continued.}
\label{Images_of_all_fields:fig}
\end{figure*}

\renewcommand{\thefigure}{A\arabic{figure}}
\setcounter{figure}{1}
\begin{figure*}
	\centering
        \ContinuedFloat
	\begin{subfigure}[t]{0.5\textwidth}
		\centering
		\includegraphics[scale=0.3]{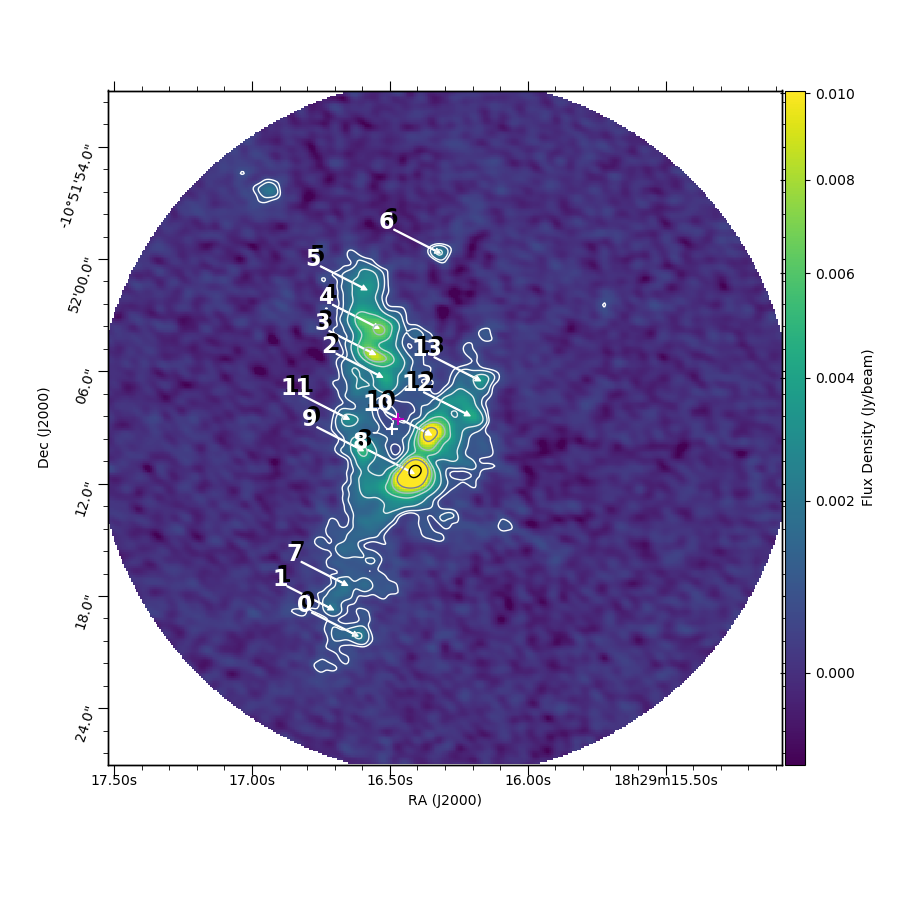}
		\caption{SDC20.775$-$0.076\_1}
		\label{SDC20p775-0p076_1:fig}
	\end{subfigure}%
	~
	\begin{subfigure}[t]{0.5\textwidth}
		\centering
		\includegraphics[scale=0.3]{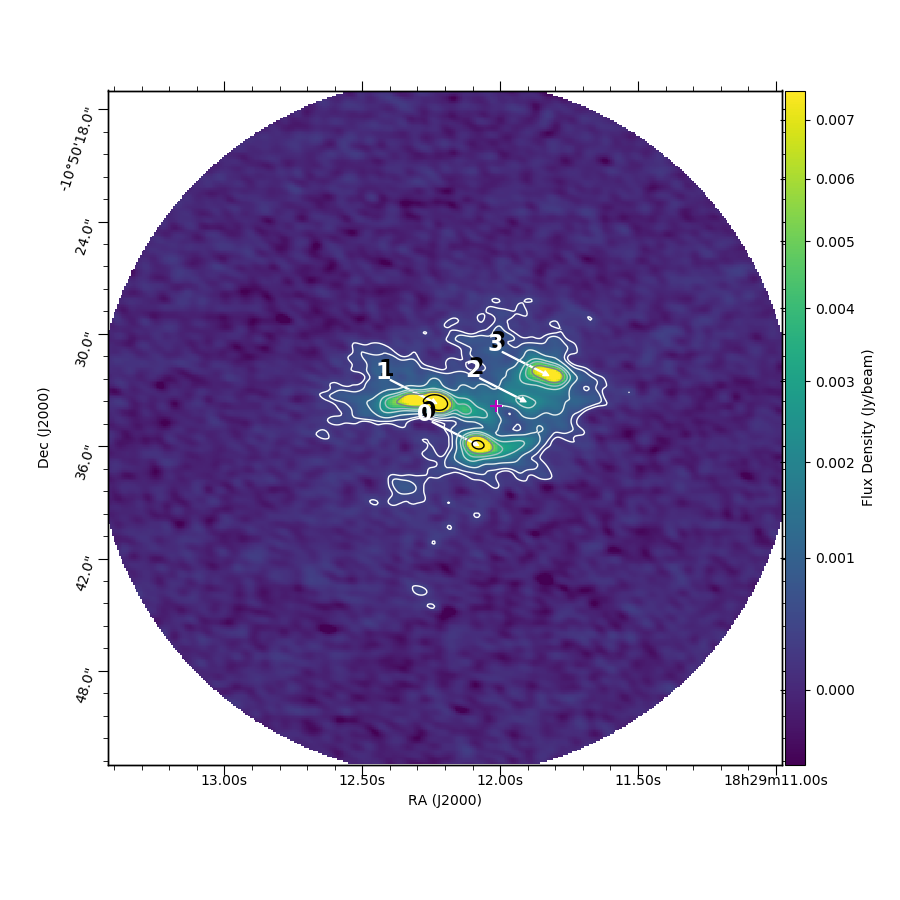}
		\caption{SDC20.775$-$0.076\_3}
		\label{SDC20p775-0p076_3:fig}
	\end{subfigure}%
	
	\begin{subfigure}[t]{0.5\textwidth}
		\centering
		\includegraphics[scale=0.3]{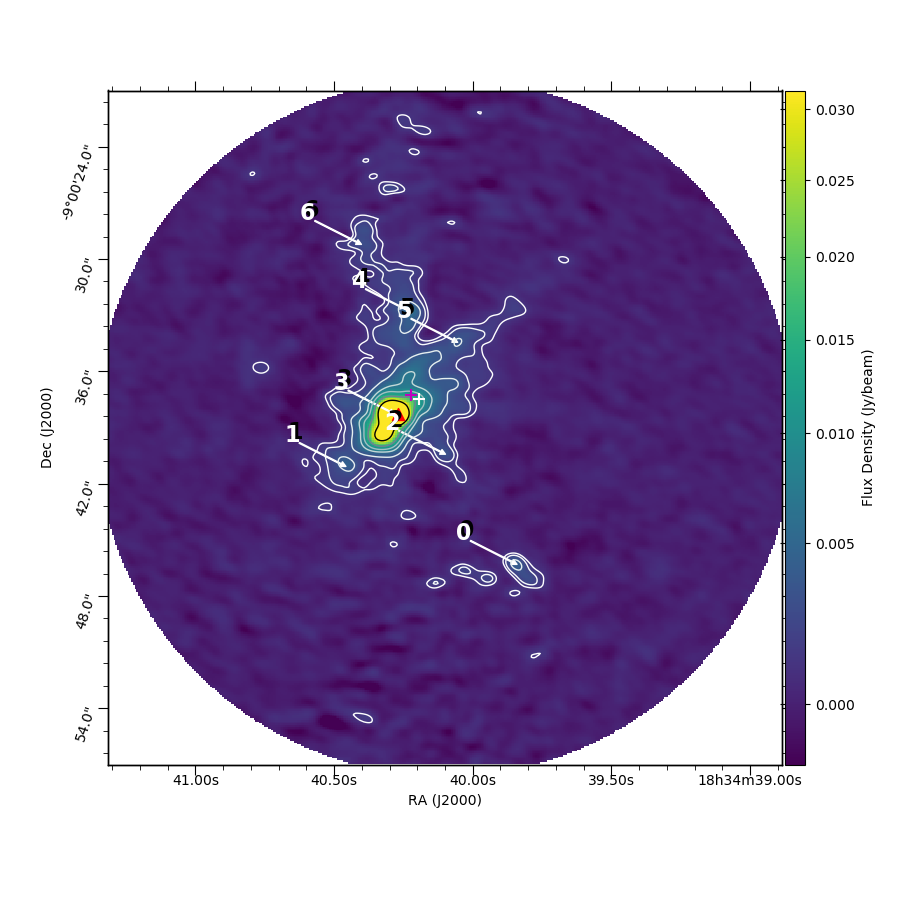}
		\caption{SDC22.985$-$0.412\_1}
		\label{SDC22p985-0p412_1:fig}
	\end{subfigure}%
	~
	\begin{subfigure}[t]{0.5\textwidth}
		\centering
		\includegraphics[scale=0.3]{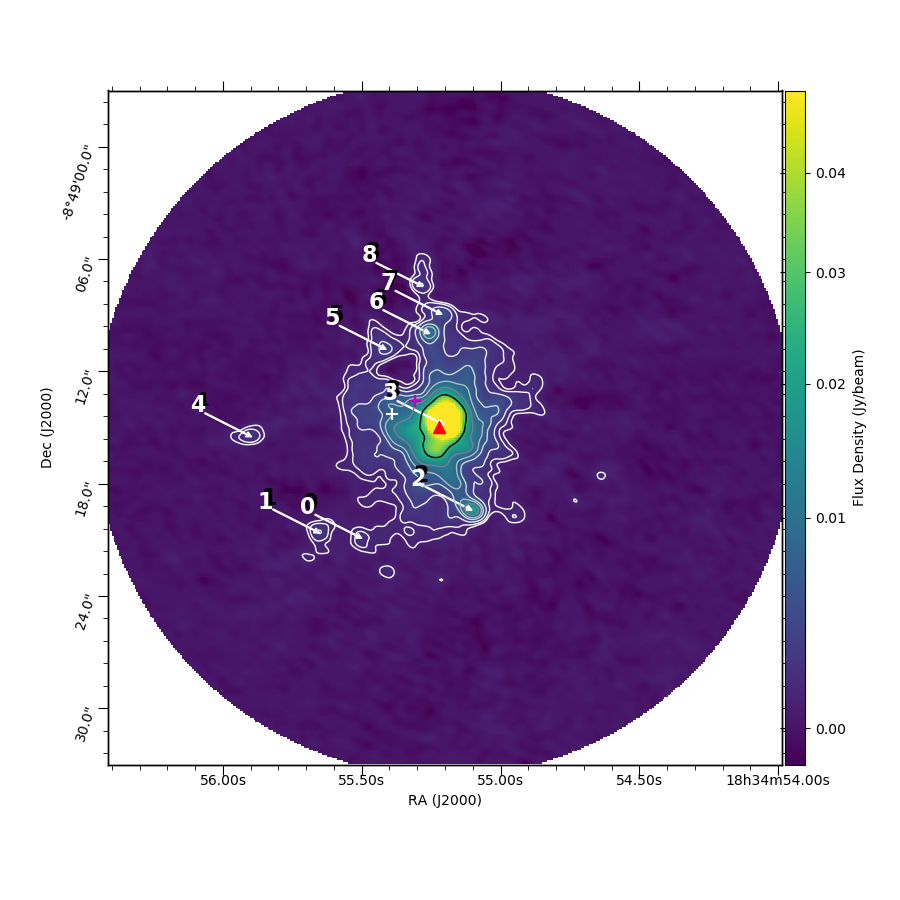}
		\caption{SDC23.21$-$0.371\_1}
		\label{SDC23p21-0p371_1:fig}
	\end{subfigure}%
	
	\begin{subfigure}[t]{0.5\textwidth}
		\centering
		\includegraphics[scale=0.3]{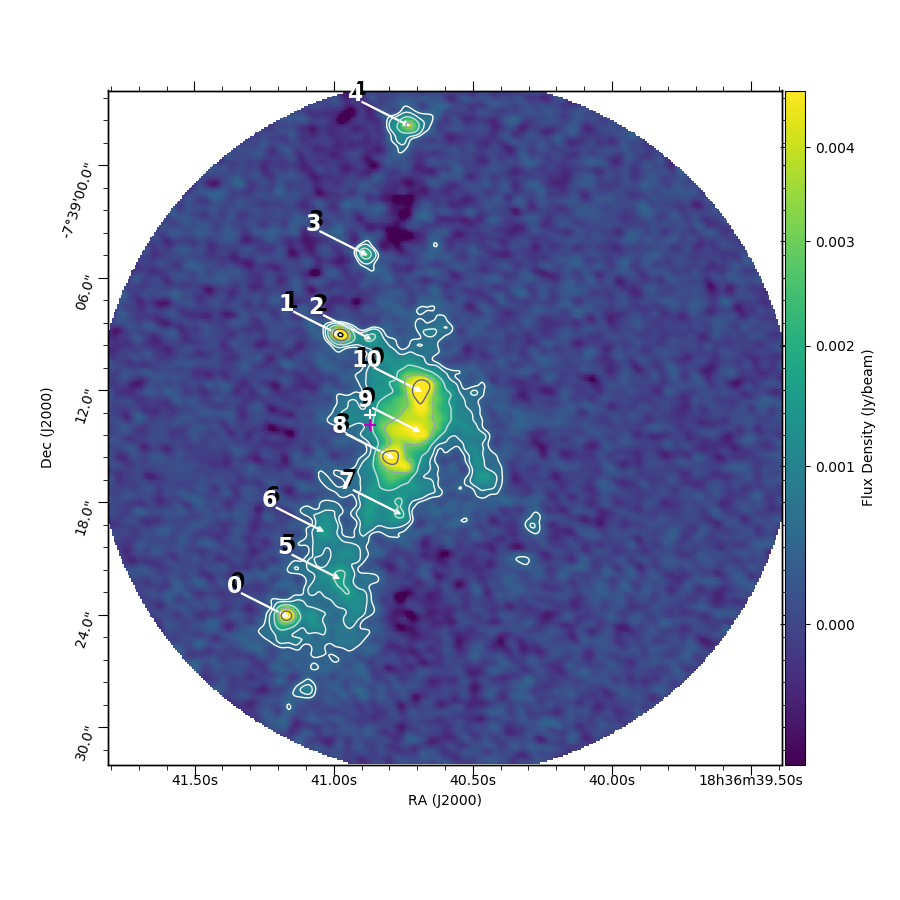}
		\caption{SDC24.381$-$0.21\_3}
		\label{SDC24p381-0p21_3:fig}
	\end{subfigure}%
	~
	\begin{subfigure}[t]{0.5\textwidth}
		\centering
		\includegraphics[scale=0.3]{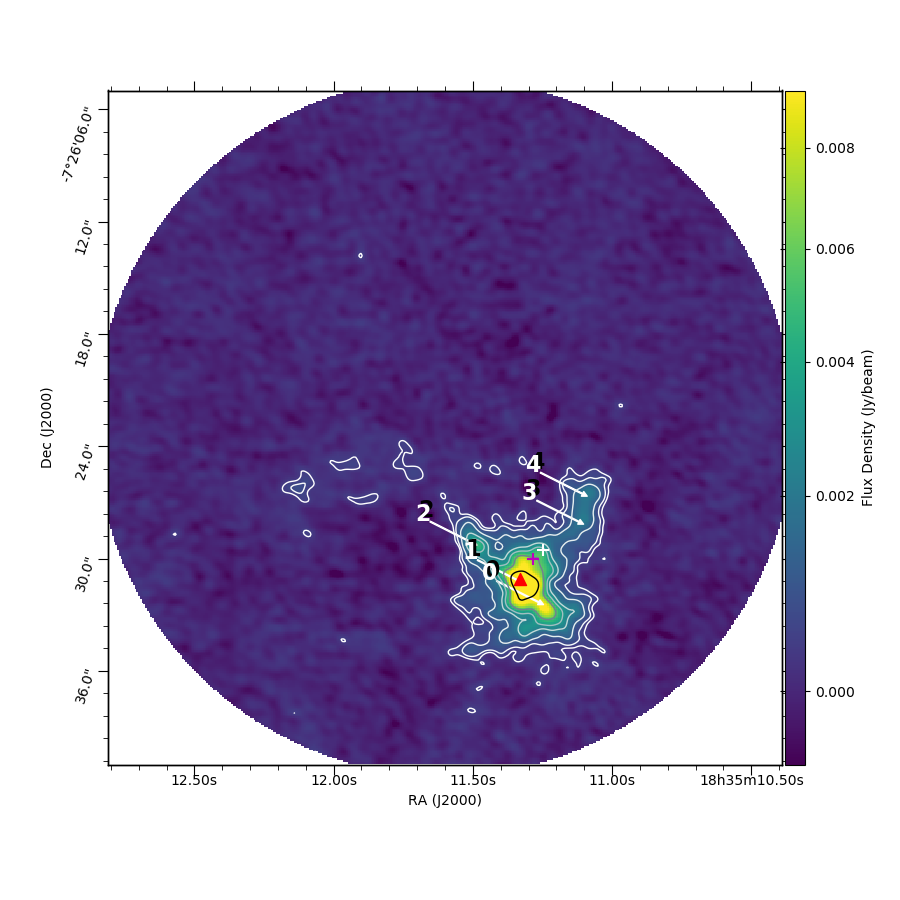}
		\caption{SDC24.462$+$0.219\_2}
		\label{SDC24p462+0p219_2:fig}
	\end{subfigure}%

\caption{continued}
\label{Images_of_all_fields:fig}
\end{figure*}
\renewcommand{\thefigure}{A\arabic{figure}}
\setcounter{figure}{1}

\begin{figure*}
	\centering
        \ContinuedFloat
	\begin{subfigure}[t]{0.5\textwidth}
		\centering
		\includegraphics[scale=0.3]{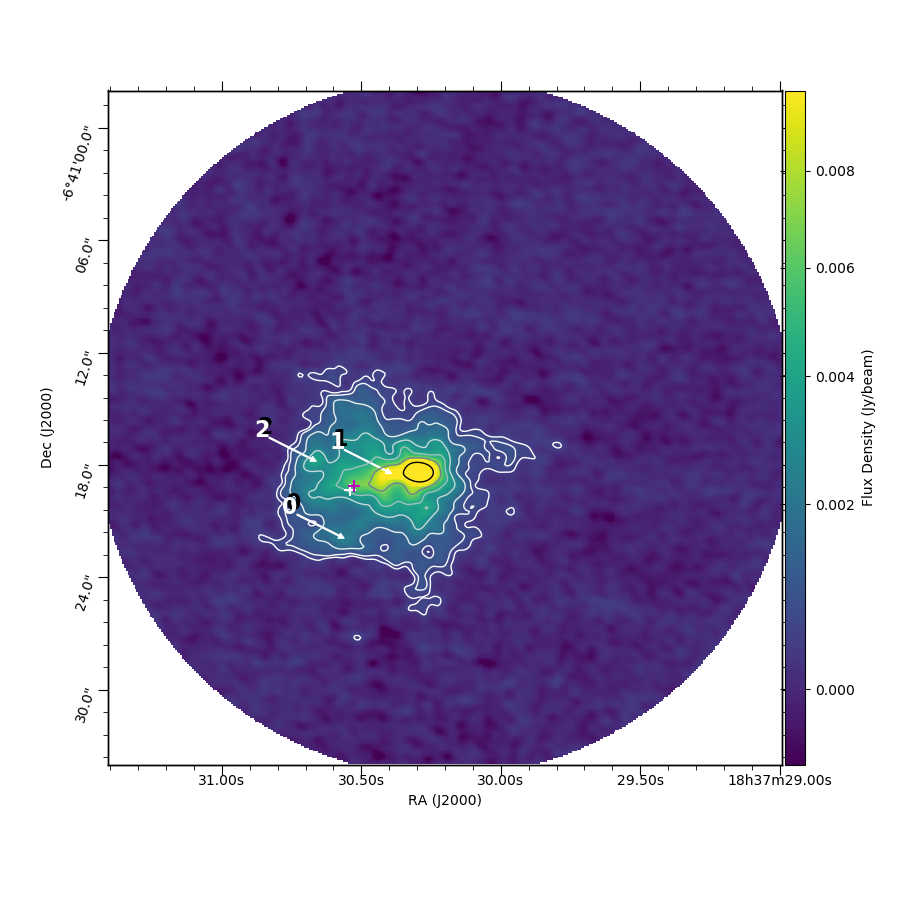}
		\caption{SDC25.426$-$0.175\_6}
		\label{SDC25p426-0p175_6:fig}
	\end{subfigure}%
	~
	\begin{subfigure}[t]{0.5\textwidth}
		\centering
		\includegraphics[scale=0.3]{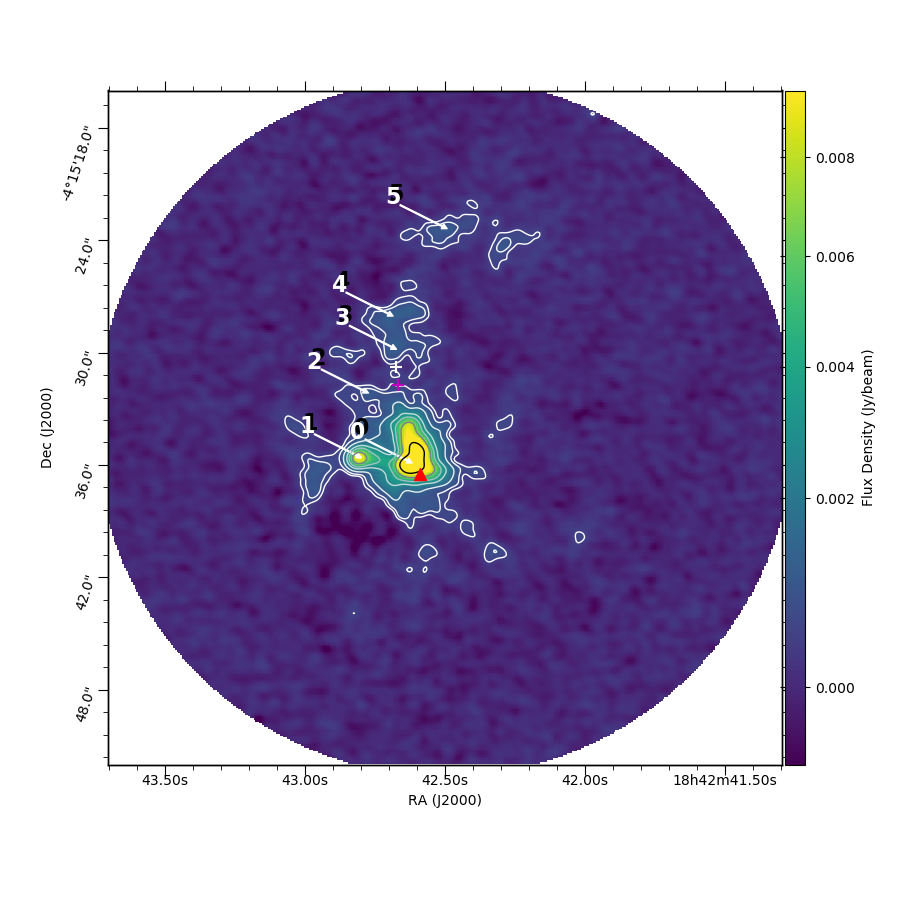}
		\caption{SDC28.147$-$0.006\_1}
		\label{SDC28p147-0p006_1:fig}
	\end{subfigure}%
	
	\begin{subfigure}[t]{0.5\textwidth}
		\centering
		\includegraphics[scale=0.3]{FIGURESV3/MAPS/SDC28p277-0p352_1_sources.png}
		\caption{SDC28.277$-$0.352\_1}
		\label{SDC28p277-0p352_1:fig}
	\end{subfigure}%
	~
	\begin{subfigure}[t]{0.5\textwidth}
		\centering
		\includegraphics[scale=0.3]{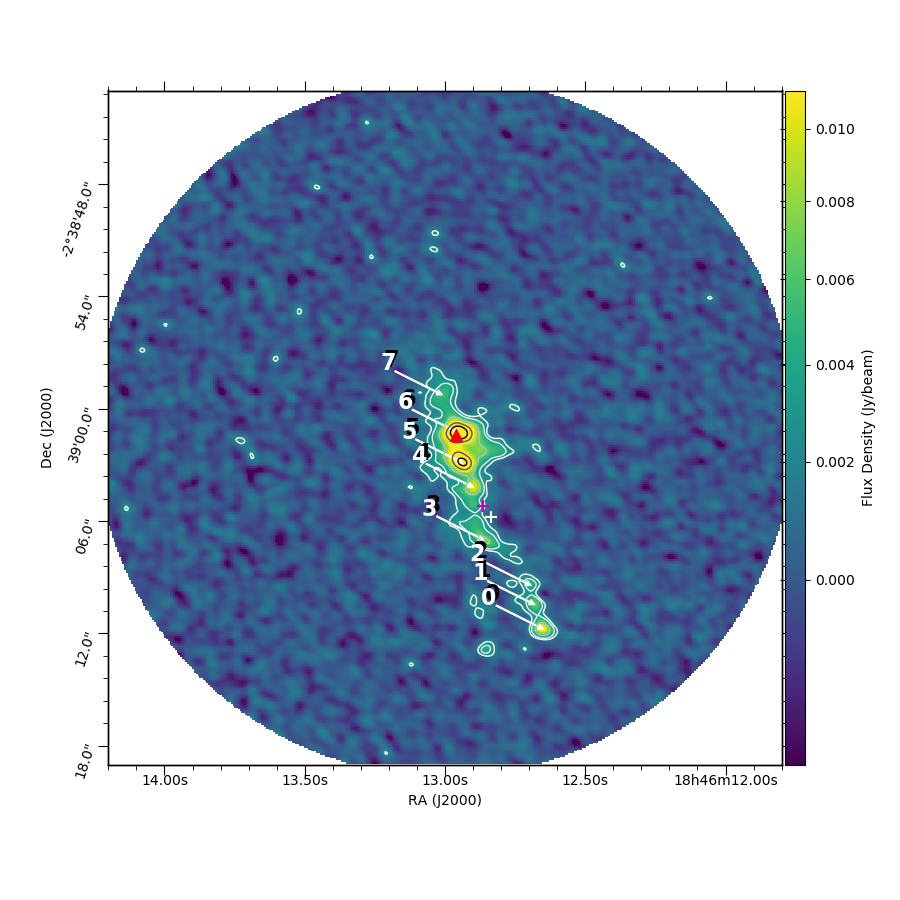}
		\caption{SDC29.844$-$0.009\_4}
		\label{SDC29p844-0p009_4:fig}
	\end{subfigure}%
	
	\begin{subfigure}[t]{0.5\textwidth}
		\centering
		\includegraphics[scale=0.3]{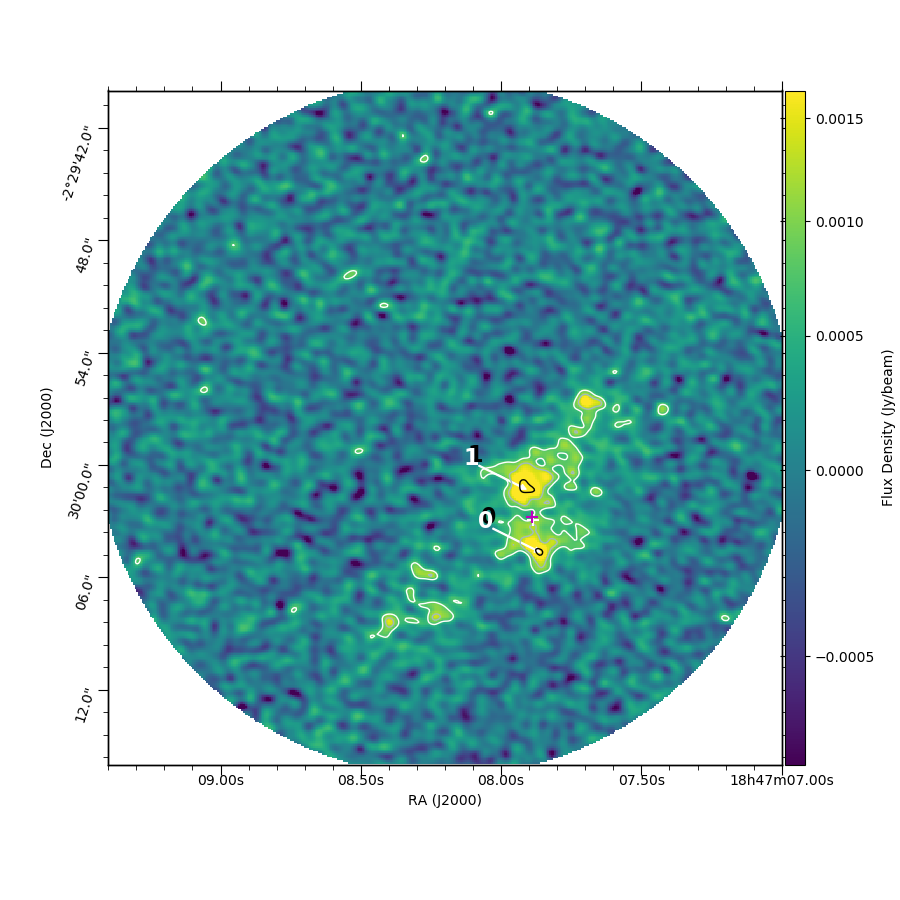}
		\caption{SDC30.172$-$0.157\_2}
		\label{SDC30p172-0p157_2:fig}
	\end{subfigure}%
	~
	\begin{subfigure}[t]{0.5\textwidth}
		\centering
		\includegraphics[scale=0.3]{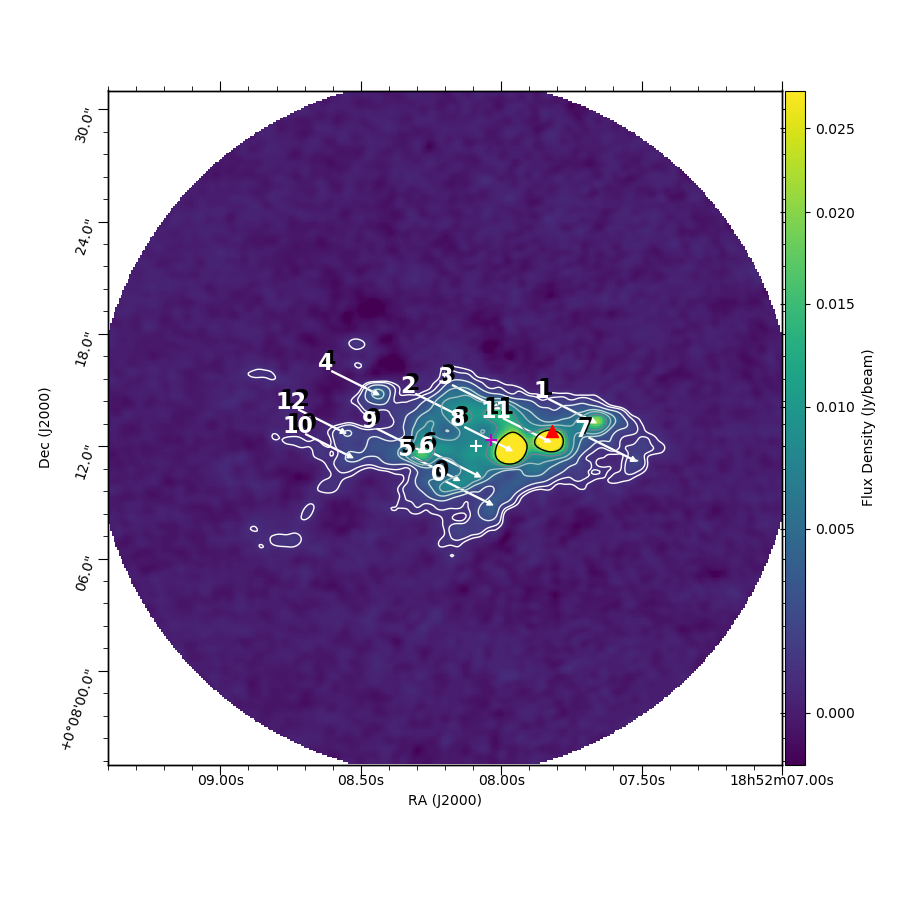}
		\caption{SDC33.107$-$0.065\_2}
		\label{SDC33p107-0p065_2:fig}
	\end{subfigure}%
\caption{continued.}
\label{Images_of_all_fields:fig}
\end{figure*}

\renewcommand{\thefigure}{A\arabic{figure}}
\setcounter{figure}{1}
\begin{figure*}
	\centering
        \ContinuedFloat
	\begin{subfigure}[t]{0.5\textwidth}
		\centering
		\includegraphics[scale=0.3]{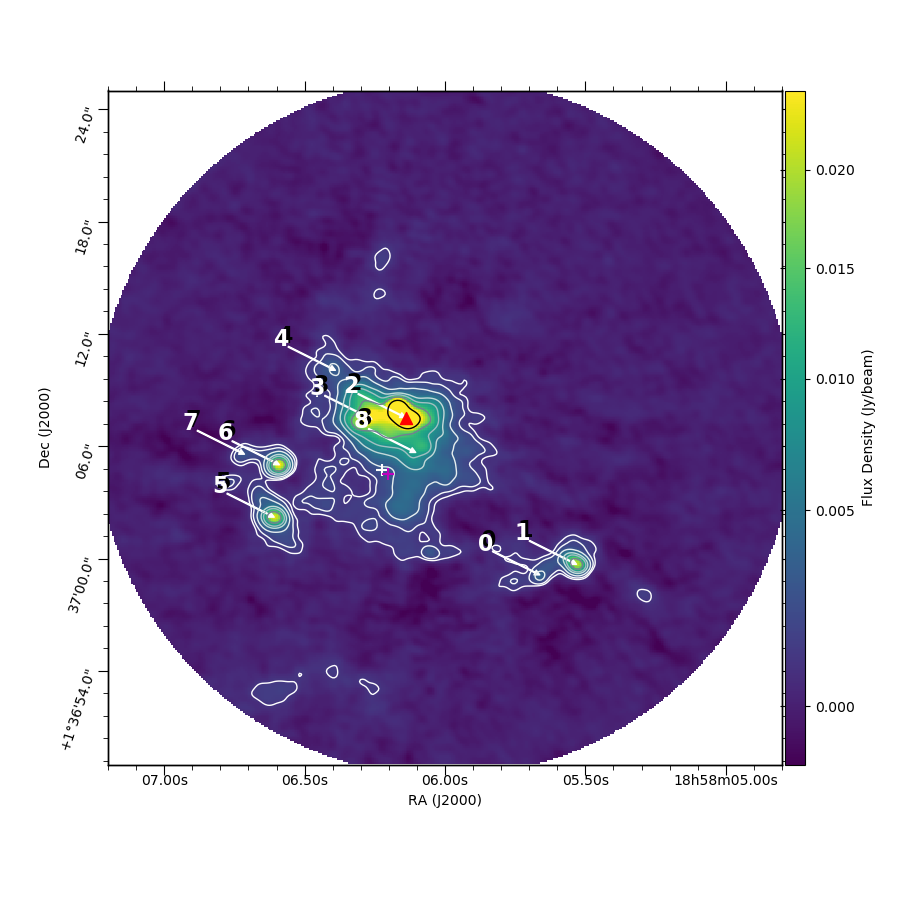}
		\caption{SDC35.063$-$0.726\_1}
		\label{SDC35p063-0p726_1:fig}
	\end{subfigure}%
	~
	\begin{subfigure}[t]{0.5\textwidth}
		\centering
		\includegraphics[scale=0.3]{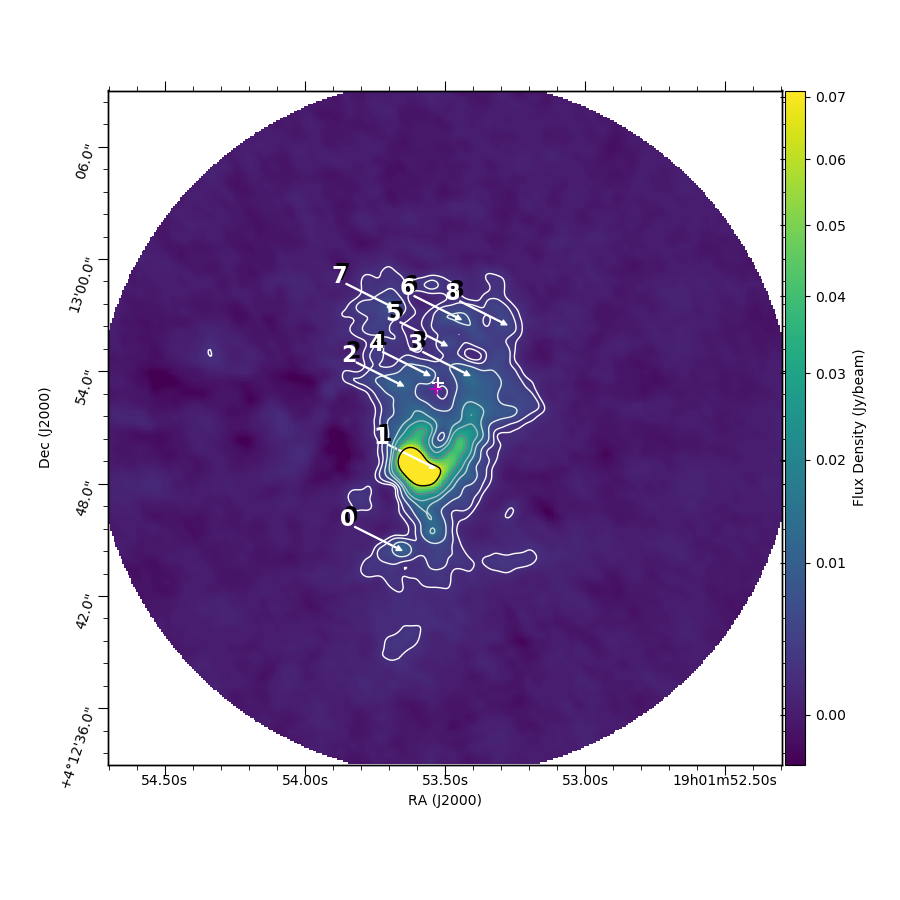}
		\caption{SDC37.846$-$0.392\_1}
		\label{SDC37p846-0p392_1:fig}
	\end{subfigure}%
	
	\begin{subfigure}[t]{0.5\textwidth}
		\centering
		\includegraphics[scale=0.3]{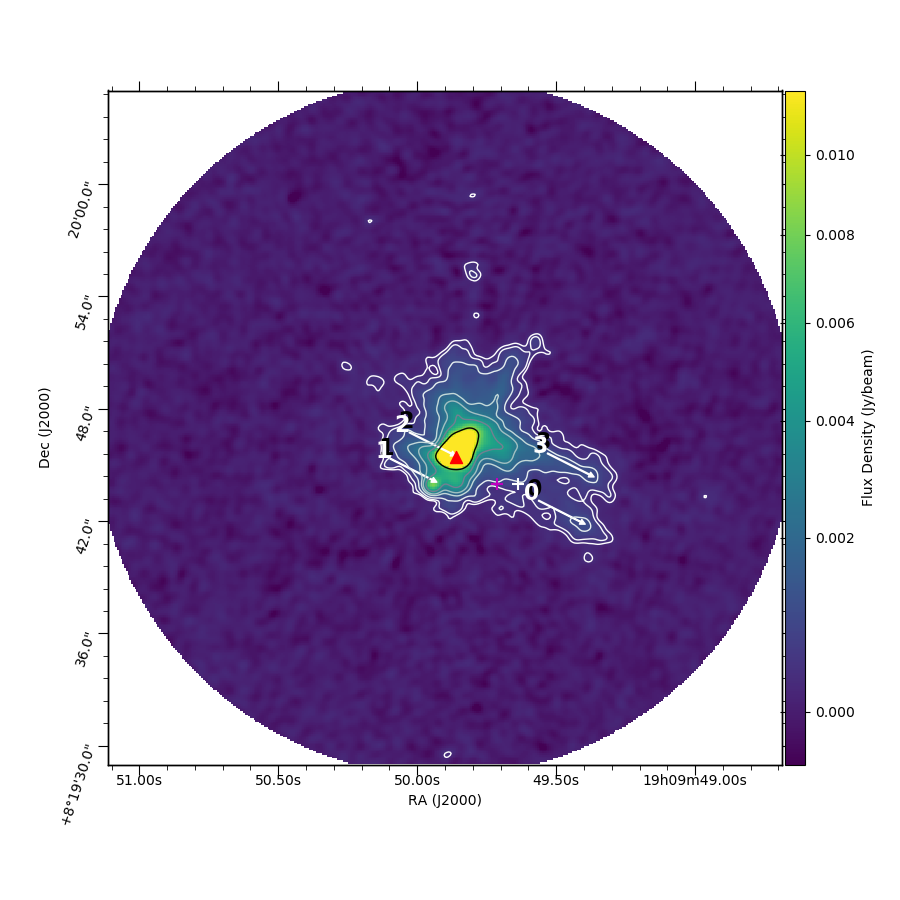}
		\caption{SDC42.401$-$0.309\_2}
		\label{SDC42p401-0p309_2:fig}
	\end{subfigure}%
	~
	\begin{subfigure}[t]{0.5\textwidth}
		\centering
		\includegraphics[scale=0.3]{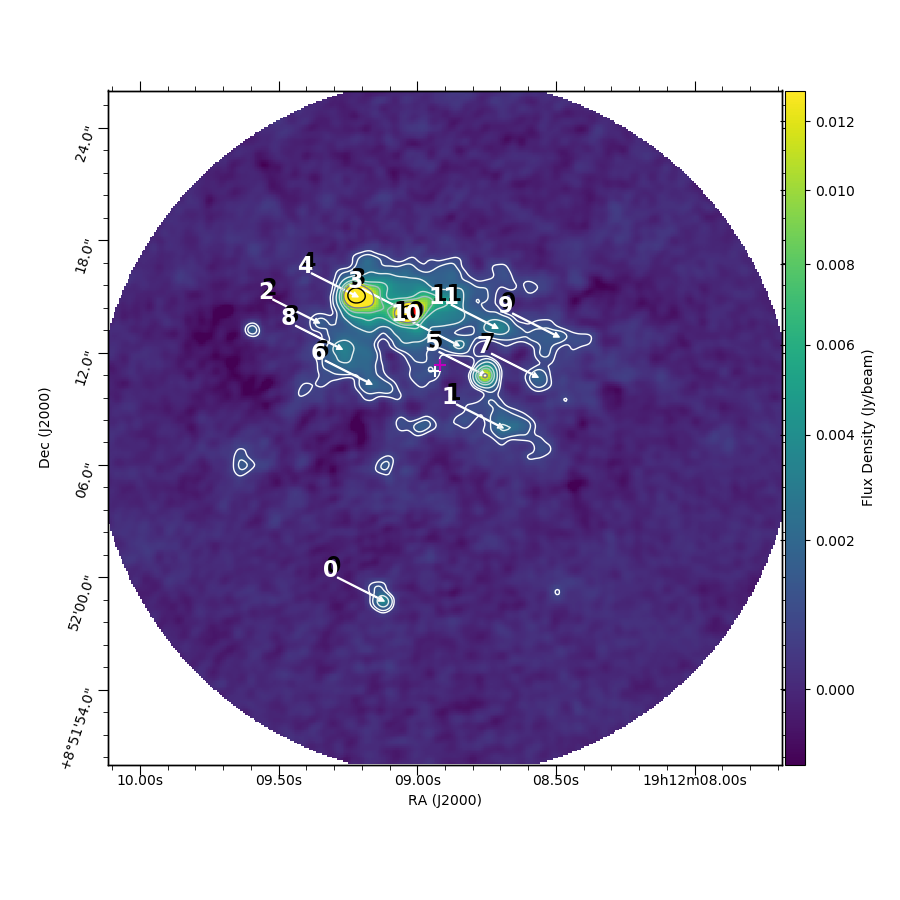}
		\caption{SDC43.186$-$0.549\_2}
		\label{SDC43p186-0p549_2:fig}
	\end{subfigure}%
	
	\begin{subfigure}[t]{0.5\textwidth}
		\centering
		\includegraphics[scale=0.3]{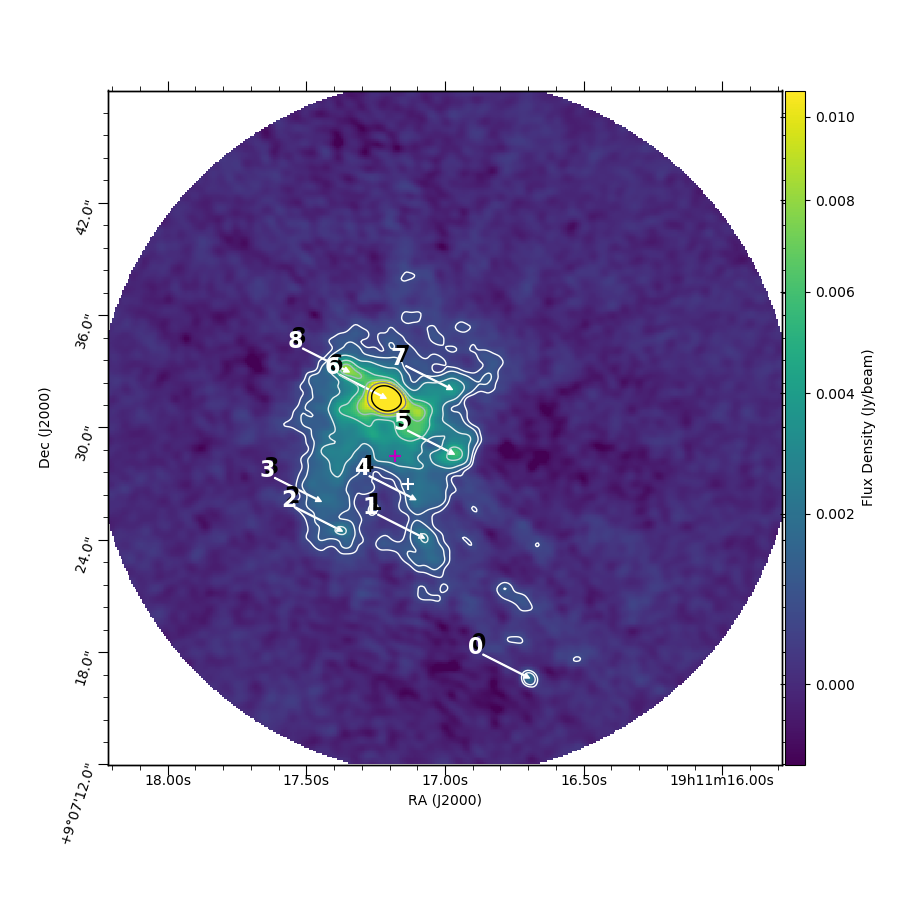}
		\caption{SDC43.311$-$0.21\_1}
		\label{SDC43p311-0p21_1:fig}
	\end{subfigure}%
	~
	\begin{subfigure}[t]{0.5\textwidth}
		\centering
		\includegraphics[scale=0.3]{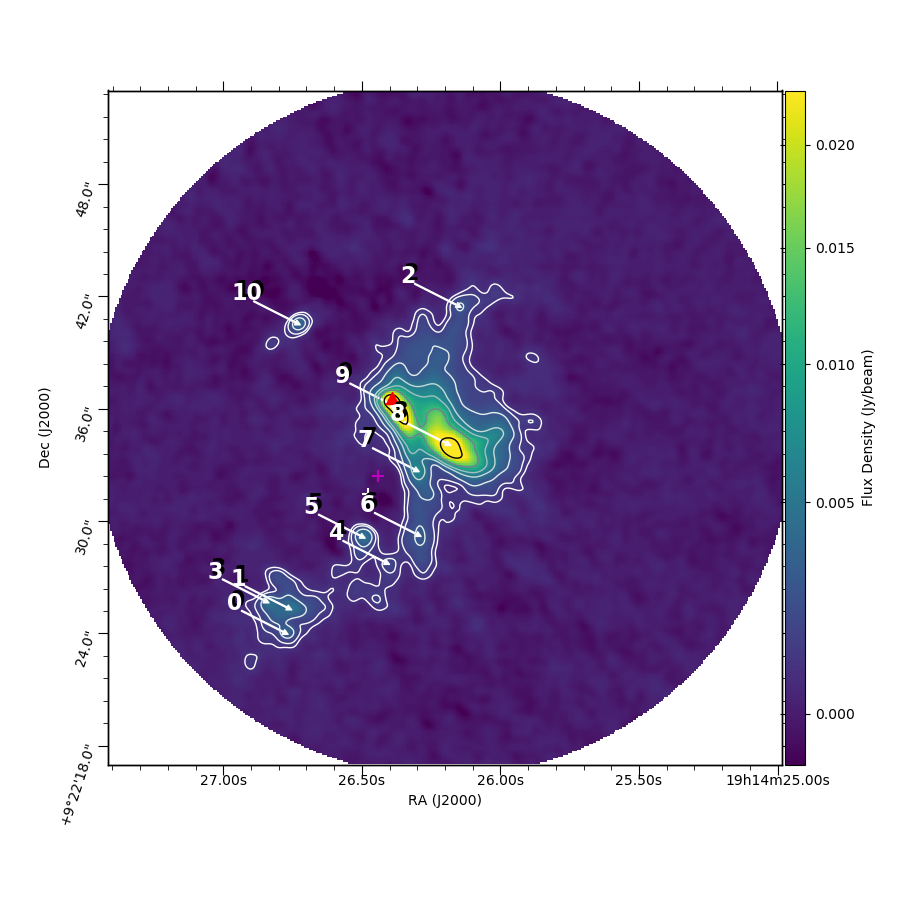}
		\caption{SDC43.877$-$0.755\_1}
		\label{SDC43p877-0p755_1:fig}
	\end{subfigure}%
\caption{continued.}
\label{Images_of_all_fields:fig}
\end{figure*}

\renewcommand{\thefigure}{A\arabic{figure}}
\setcounter{figure}{1}
\begin{figure*}
	\centering
        \ContinuedFloat
	\begin{subfigure}[t]{0.5\textwidth}
		\centering
		\includegraphics[scale=0.3]{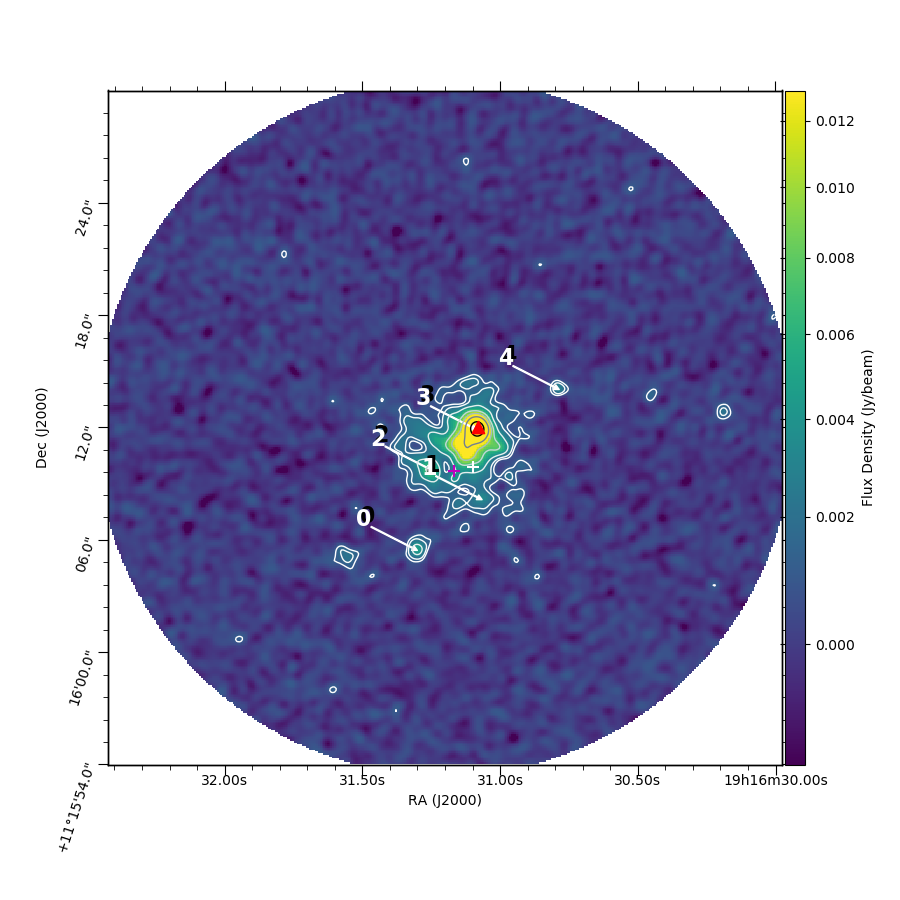}
		\caption{SDC45.787$-$0.335\_1}
		\label{SDC45p787-0p335_1:fig}
	\end{subfigure}%
	~
	\begin{subfigure}[t]{0.5\textwidth}
		\centering
		\includegraphics[scale=0.3]{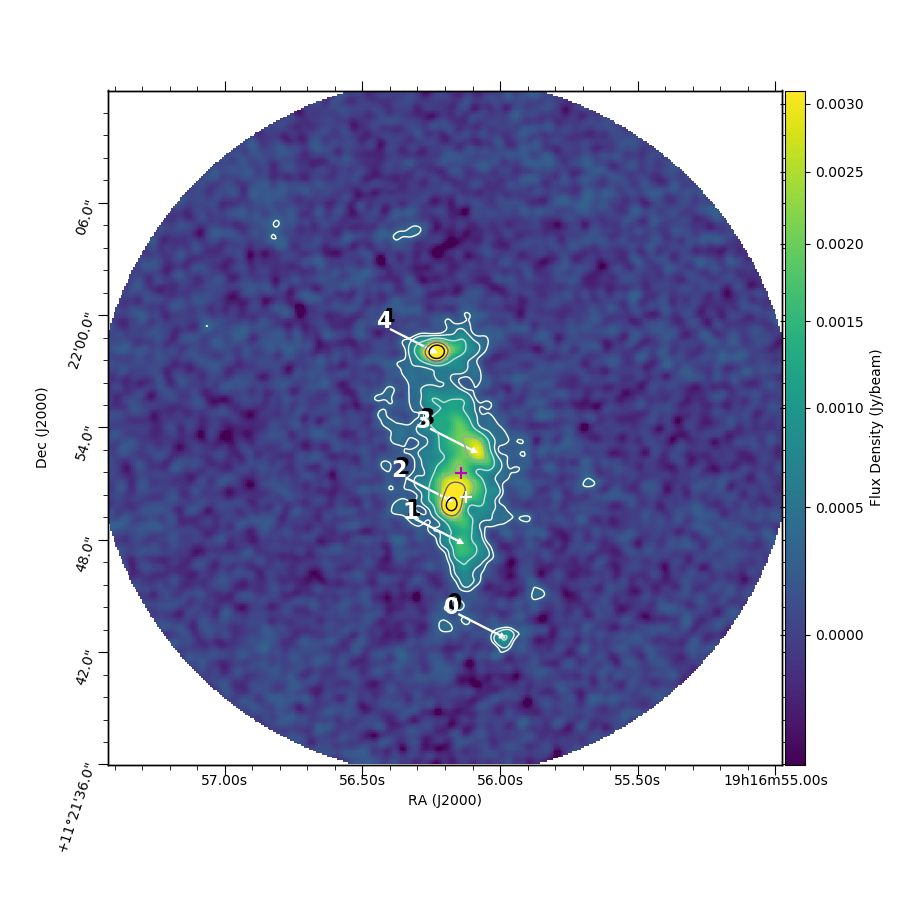}
		\caption{SDC45.927$-$0.375\_2}
		\label{SDC45p927-0p375_2:fig}
	\end{subfigure}%
\caption{continued.}
\label{Images_of_all_fields:fig}
\end{figure*}

Figure \ref{Images_of_all_fields:fig} provides continuum maps of all TEMPO fields.

\section{Using the \textit{Q}-parameter for small source counts}
\label{Qpara:appdx}
\renewcommand{\thefigure}{B\arabic{figure}}

\setcounter{figure}{-1}
\begin{figure*}
    \centering
    \begin{subfigure}[t]{0.5\textwidth}
        \centering
        \includegraphics[scale=0.38]{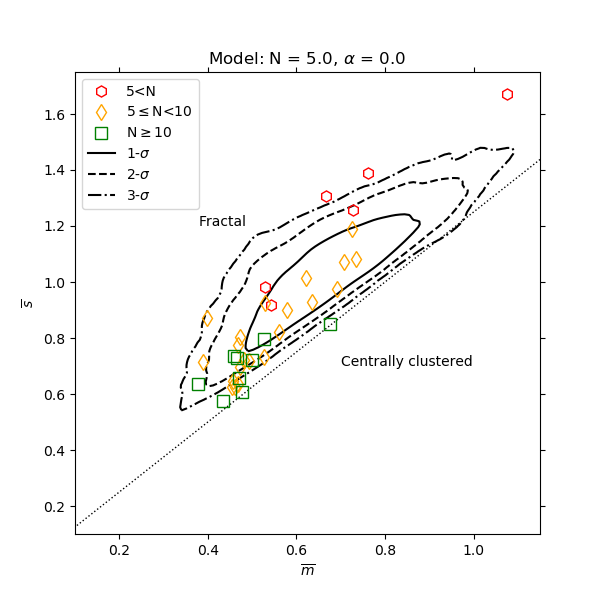}
        \caption{$\alpha = 0.0$, $N = 5$}
        \label{A00_N05:fig}
    \end{subfigure}%
    ~ 
    \begin{subfigure}[t]{0.5\textwidth}
        \centering
	\includegraphics[scale=0.38]{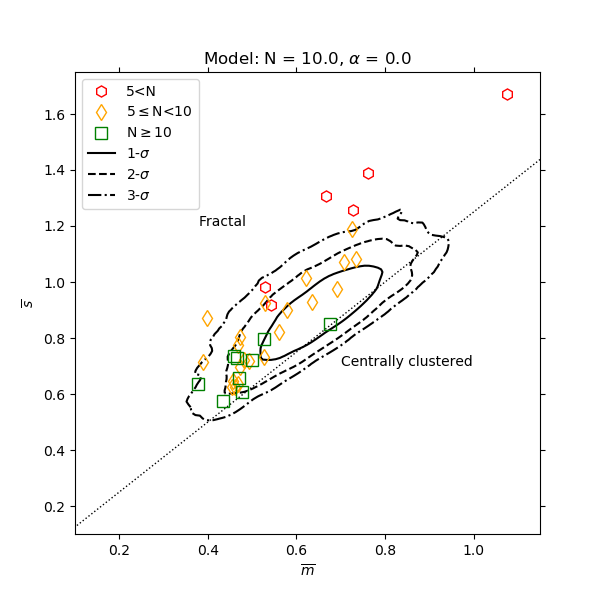}
        \caption{$\alpha = 0.0$, $N = 10$}
        \label{A00_N10:fig}
    \end{subfigure}
    
    \begin{subfigure}[t]{0.5\textwidth}
        \centering
        \includegraphics[scale=0.38]{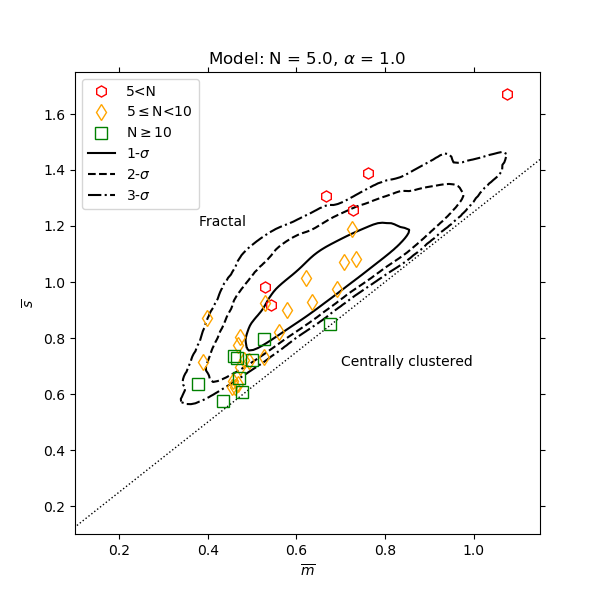}
        \caption{$\alpha = 1.0$, $N = 5$}
        \label{A01_N05:fig}
    \end{subfigure}%
    ~ 
    \begin{subfigure}[t]{0.5\textwidth}
        \centering
	\includegraphics[scale=0.38]{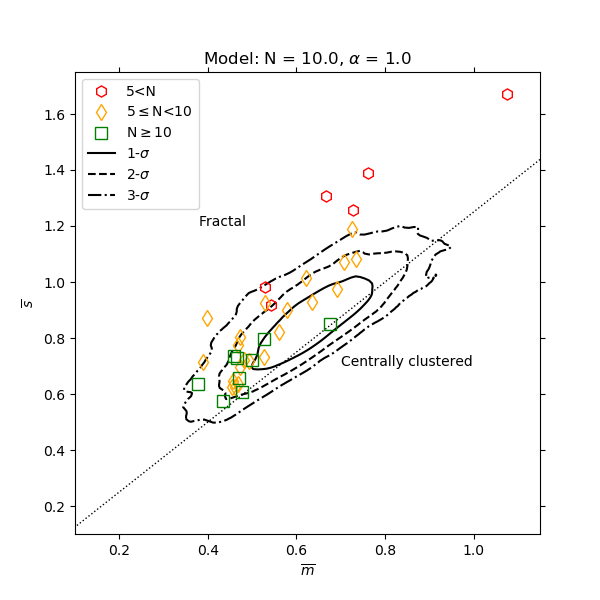}
        \caption{$\alpha = 1.0$, $N = 10$}
        \label{A01_N10:fig}
    \end{subfigure}
    
    \begin{subfigure}[t]{0.5\textwidth}
        \centering
        \includegraphics[scale=0.38]{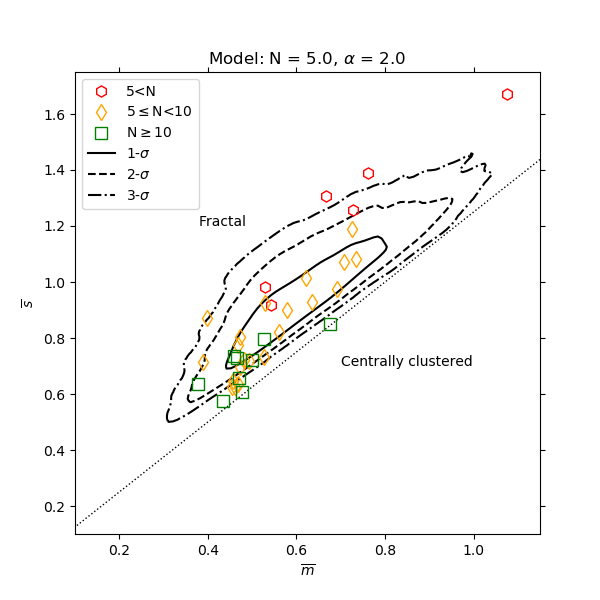}
        \caption{$\alpha = 2.0$, $N = 5$}
        \label{A02_N05:fig}
    \end{subfigure}%
    ~ 
    \begin{subfigure}[t]{0.5\textwidth}
        \centering
	\includegraphics[scale=0.38]{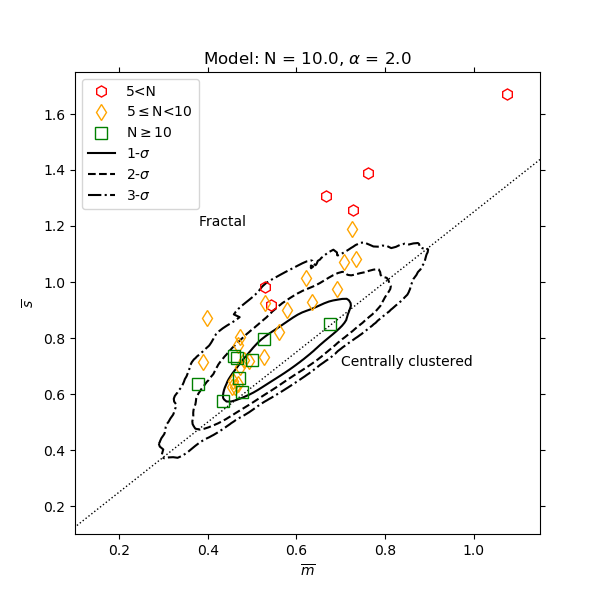}
        \caption{$\alpha = 2.0$, $N = 10$}
        \label{A02_N10:fig}
    \end{subfigure}
    
     \begin{subfigure}[t]{0.5\textwidth}
        \centering
        \includegraphics[scale=0.38]{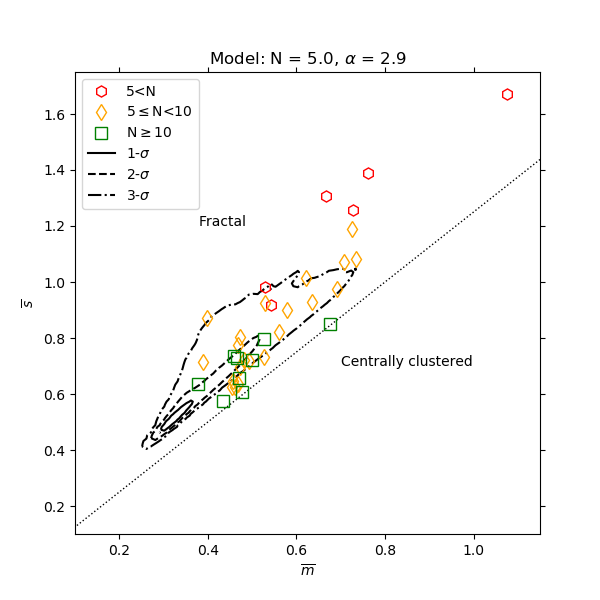}
        \caption{$\alpha = 2.9$, $N = 5$}
        \label{A03_N05:fig}
    \end{subfigure}%
    ~ 
    \begin{subfigure}[t]{0.5\textwidth}
        \centering
	\includegraphics[scale=0.38]{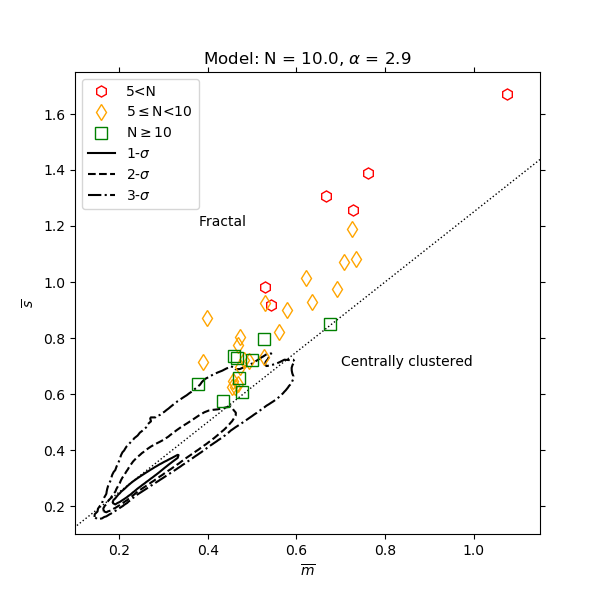}
        \caption{$\alpha = 2.9$, $N = 10$}
        \label{A03_N10:fig}
    \end{subfigure}
\caption{$\overline{m}$ and $\overline{s}$ parameter space for radial $n \propto r^{-\alpha}$ source distribution profiles. \textit{Right:} $N = 5$ models. \textit{Right:} $N = 10$ models.}
\label{Q_param_space_cc:fig}
\end{figure*}

\renewcommand{\thefigure}{B\arabic{figure}}
\setcounter{figure}{0}
\begin{figure*}
    \centering
    \begin{subfigure}[t]{0.5\textwidth}
        \centering
        \includegraphics[scale=0.38]{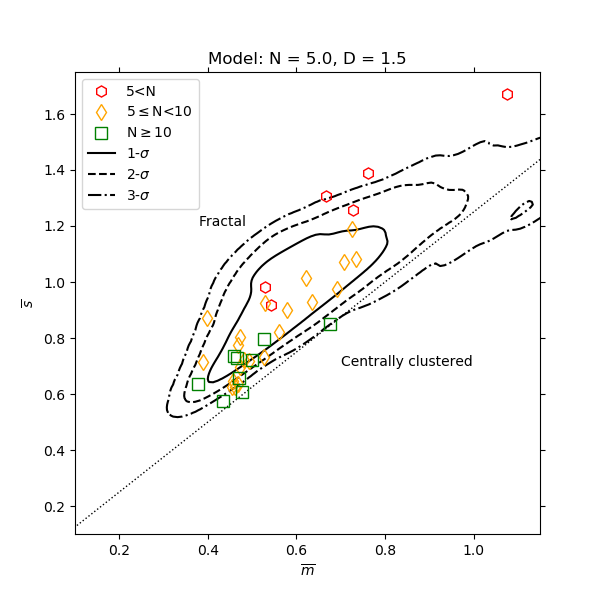}
        \caption{$D = 1.5$, $N = 5$}
        \label{D01p5_N05:fig}
    \end{subfigure}%
    ~ 
    \begin{subfigure}[t]{0.5\textwidth}
        \centering
	\includegraphics[scale=0.38]{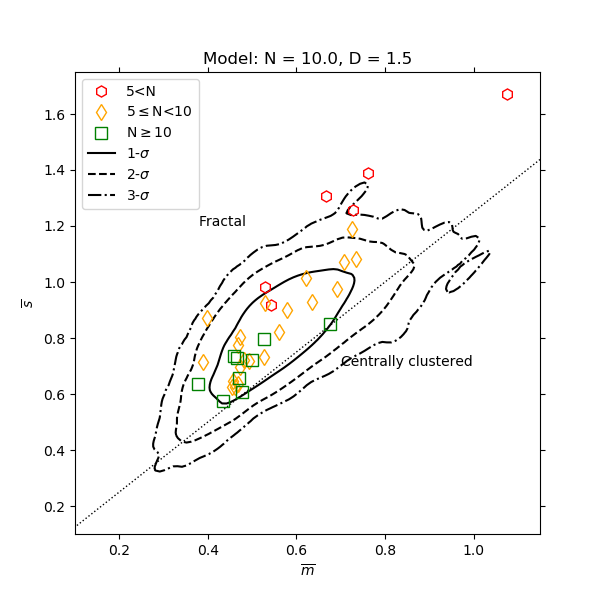}
        \caption{$D = 1.5$, $N = 10$}
        \label{D01p5_N10:fig}
    \end{subfigure}
    
    \begin{subfigure}[t]{0.5\textwidth}
        \centering
        \includegraphics[scale=0.38]{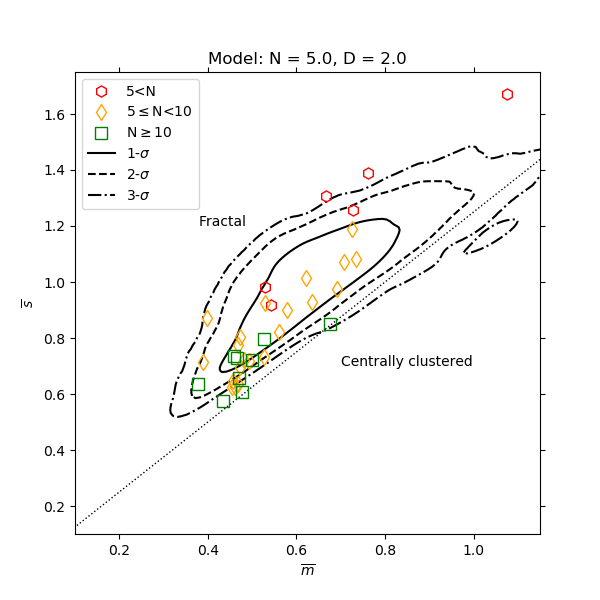}
        \caption{$D = 2.0$, $N = 5$}
        \label{D02p0_N05:fig}
    \end{subfigure}%
    ~ 
    \begin{subfigure}[t]{0.5\textwidth}
        \centering
	\includegraphics[scale=0.38]{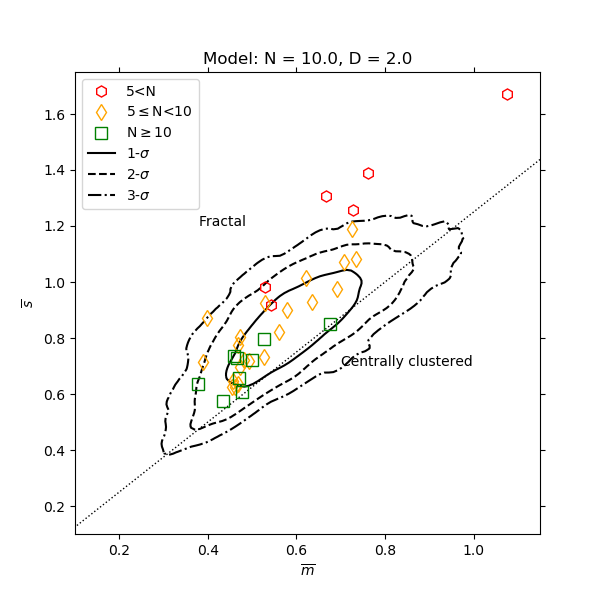}
        \caption{$D = 2.0$, $N = 10$}
        \label{D02p0_N10:fig}
    \end{subfigure}
    
    \begin{subfigure}[t]{0.5\textwidth}
        \centering
        \includegraphics[scale=0.38]{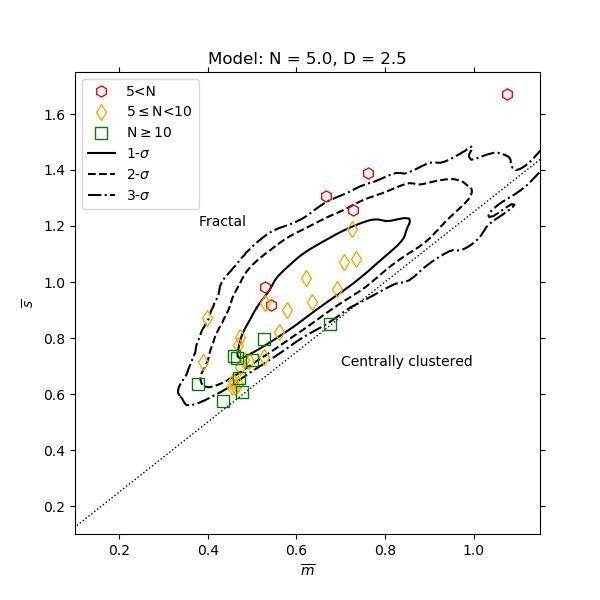}
        \caption{$D = 2.5$, $N = 5$}
        \label{D02p5_N05:fig}
    \end{subfigure}%
    ~ 
    \begin{subfigure}[t]{0.5\textwidth}
        \centering
	\includegraphics[scale=0.38]{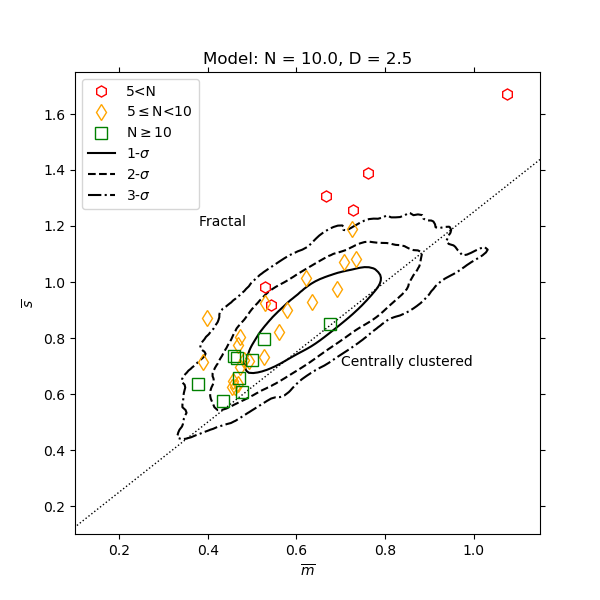}
        \caption{$D = 2.5$, $N = 10$}
        \label{D02p5_N10:fig}
    \end{subfigure}
    
     \begin{subfigure}[t]{0.5\textwidth}
        \centering
        \includegraphics[scale=0.38]{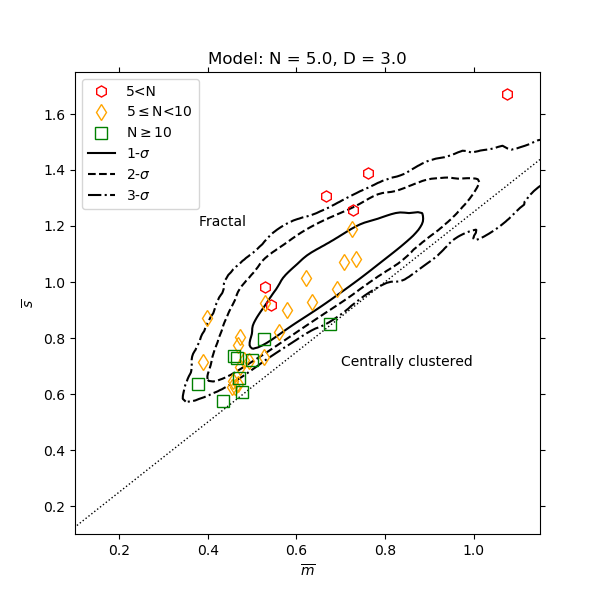}
        \caption{$D = 3.0$, $N = 5$}
        \label{D03p0_N05:fig}
    \end{subfigure}%
    ~ 
    \begin{subfigure}[t]{0.5\textwidth}
        \centering
	\includegraphics[scale=0.38]{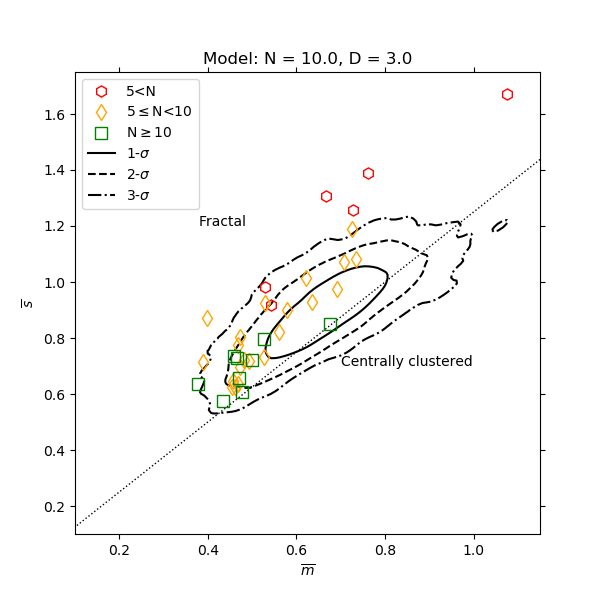}
        \caption{$D = 3.0$, $N = 10$}
        \label{D03p0_N10:fig}
    \end{subfigure}
    
\caption{$\overline{m}$ and $\overline{s}$ parameter space for fractal source distribution profiles. \textit{Right:} $N = 5$ models. \textit{Right:} $N = 10$ models.}
\label{Q_param_space_fr:fig}
\end{figure*}%
\begin{table*}
\caption[]{Calculated mean and standard deviations of values $Q$, $\overline{m}$, $\overline{s}$, for each run of 10,000 realisations of the 16 different model source distributions. Column 1 gives the number of sources, columns 2 and 3 denote the radial profile power,$\alpha$, and fractal dimension $D$ respectively. Columns 4-9 give, pairwise, the mean and standard deviations of values $\overline{m}$, $\overline{s}$ and $Q$. The last four rows give the combined values for the real observed fields, grouped by the number of source in each field.}
\begin{center}
\small
\begin{tabular}{c c c c c c c c c }
\hline
\hline
N & $\alpha$ & D &  mean($\overline{m}$) & std($\overline{m}$) & mean($\overline{s}$) & std($\overline{s}$)& mean($Q$) & std($Q$)\\
\hline
5.0 & 0.0 & - & 0.67 & 0.11 & 1.03 & 0.14 & 0.65 & 0.06 \\
5.0 & 1.0 & - & 0.66 & 0.11 & 1.02 & 0.14 & 0.65 & 0.05 \\
5.0 & 2.0 & - & 0.62 & 0.11 & 0.96 & 0.14 & 0.64 & 0.05 \\
5.0 & 2.9 & - & 0.39 & 0.09 & 0.63 & 0.14 & 0.63 & 0.03 \\
\hline
10.0 & 0.0 & - & 0.65 & 0.09 & 0.9 & 0.1 & 0.72 & 0.06 \\
10.0 & 1.0 & - & 0.63 & 0.08 & 0.86 & 0.1 & 0.74 & 0.06 \\
10.0 & 2.0 & - & 0.58 & 0.09 & 0.77 & 0.11 & 0.75 & 0.06 \\
10.0 & 2.9 & - & 0.3 & 0.08 & 0.37 & 0.11 & 0.83 & 0.08 \\
\hline
5.0 & - & 1.5 & 0.62 & 0.13 & 0.99 & 0.16 & 0.63 & 0.07 \\
5.0 & - & 2.0 & 0.64 & 0.12 & 1.01 & 0.15 & 0.64 & 0.06 \\
5.0 & - & 2.5 & 0.66 & 0.12 & 1.02 & 0.14 & 0.64 & 0.06 \\
5.0 & - & 3.0 & 0.68 & 0.12 & 1.04 & 0.14 & 0.65 & 0.06 \\
\hline
10.0 & - & 1.5 & 0.57 & 0.1 & 0.83 & 0.14 & 0.69 & 0.09 \\
10.0 & - & 2.0 & 0.6 & 0.1 & 0.85 & 0.13 & 0.71 & 0.07 \\
10.0 & - & 2.5 & 0.63 & 0.09 & 0.88 & 0.11 & 0.73 & 0.07 \\
10.0 & - & 3.0 & 0.66 & 0.09 & 0.9 & 0.1 & 0.74 & 0.06 \\
\hline
All data & & & 0.57 & 0.14 & 0.93 & 0.35 & 0.64 & 0.09 \\
$<$ 5 & & & 0.74 & 0.16 & 1.44 & 0.39 & 0.53 & 0.08 \\
5 $\leq N < 10$ & & & 0.54 & 0.1 & 0.83 & 0.16 & 0.65 & 0.07 \\
$> 10$ & & & 0.49 & 0.08 & 0.7 & 0.09 & 0.7 & 0.07 \\
\hline
\end{tabular}
\end{center}
\label{Q_para_test:tab}
\end{table*}


Introduced by \citet{CartwrightWhitworth04} the $Q$-parameter has been shown to be a useful diagnostic of stellar distributions within star clusters, making it possible to distinguish between stellar clusters with centrally concentrated, radial distributions and those displaying fractal distributions, \citep[e.g.][]{Maschberger10, Hunter14, Parker18}. The $Q$ value for a cluster, as given in Equation \ref{Qpara:eqn},  is the ratio of the mean edge length of the cluster's minimum spanning tree (MST), $\overline{m}$, and the cluster correlation length (the mean projected source separation within the cluster), $\overline{s}$. These two values are normalised values of measured quantities with the normalisation accounting for the cluster radial size and the number of sources present. The $Q$ value is given as:

\begin{equation}
Q = \frac{\overline{m}}{\overline{s}} =  \frac{X/\left(\frac{\sqrt{(N_{tot}\pi R_{clust}^2)}}{N_{tot}-1}\right)}{Y/R_{clust}}
\label{Qpara:eqn}
\end{equation}
\noindent where $X$ and $Y$ give the un-normalised measured values in arcseconds (for $\overline{m}$ and $\overline{s}$ respectively).  Their respective denominators give the normalisation factors, with $N_{tot}$ the total number of source in the cluster and $R_{clust}$ the distance from the arithmetic average source position to the most distant source in the cluster. 

\begin{figure}
\centering
\includegraphics[scale=0.58]{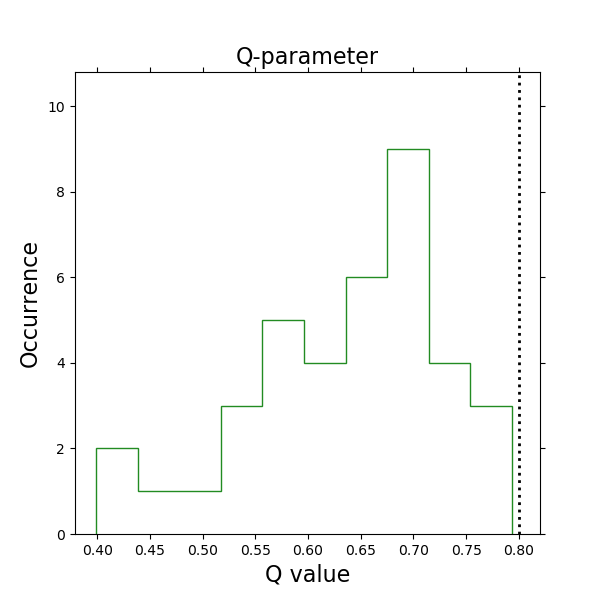}
\caption{Calculated $Q$-value for all 38 fields in the TEMPO sample, all fields have values consistent with a fractal source distribution, $Q$<0.8, however this may not be the case, see text. The dashed vertical line denotes a values of $Q$=0.8, representing a uniform source distribution.}
\label{Q_hist:fig}
\end{figure}

From \citet{CartwrightWhitworth04} values of $Q\simeq$0.8 indicate a uniform density. For centrally concentrated clusters $Q > 0.8$ and for fractal distributions $Q < 0.8$. The $Q$-parameter analysis was applied to each TEMPO field, finding all sources return a value of $Q$ less than 0.8, this can be seen in Figure \ref{Q_hist:fig}. 

These results would indicate that all TEMPO fields exhibit a fractal, rather than centrally concentrated population distribution. Whilst this appears to agree with the distribution seen in Figure \ref{norm_Radial_profile:fig} it is noted that the original and the majority of subsequent works which have used the $Q$-parameter have been applied to clusters with populations of several tens to hundreds. The TEMPO sample has comparatively small numbers of sources per field (between 2 and 15, c.f. Figure \ref{Sources_per_field:fig}). As such an investigation was undertaken to assess the applicability of the $Q$-parameter for small sample sizes. 


\subsection{Errors on \textit{Q},  $\overline{m}$ and $\overline{s}$  for small numbers }

First, to assess the validity of the calculated $Q$ values and the $\overline{m}$, $\overline{s}$ they are derived from, when used for small population clusters the $Q$-parameter analysis was repeated for each field in turn but iteratively excluding a single source from that field. For example in a field with 4 sources we repeat the $Q$ analysis with sources [2,3,4], [1,3,4], [1,2,4] and [1,2,3]. We denote the values created by this method with a subscript `N-1', e.g. $Q_{N-1}$. We note that this inspection fails for fields with only two or three fragments are detected, as dropping a source gives a point or single straight line in the MST and the $Q$-parameter analysis cannot be conducted with meaningful results.

From each iteration the $Q_{N-1}$, $\overline{m}_{N-1}$ and $\overline{s}_{N-1}$ values we recovered to inspect their maximum and minimum values and standard deviation in comparison to the value when all sources are included. Figures \ref{Q_errors:fig} to \ref{sbar_errors:fig} give a histogram of the fractional error ($\sigma_{x}/x$) distribution for $Q_{N-1}$, $\overline{m}_{N-1}$ and $\overline{s}_{N-1}$, respectively, on the left hand side. The right-hand plot of each of these figures shows the mean value per field (be it $Q_{N-1}$, $\overline{m}_{N-1}$ or $\overline{s}_{N-1}$) divided by the true value (e.g. $mean(x)_{N-1}$/$x$) plotted against the true value on the $x$-axis to inspect any trend in scatter and mean offset from the true value of each parameter. The error bars in each of these plots shows the distance to the maximum and minimum ${x_{N-1}}$ value from each iterated run, normalised as the mean was.

\renewcommand{\thefigure}{B\arabic{figure}}
\setcounter{figure}{2}
\begin{figure*}
    \centering
    \begin{subfigure}[t]{0.5\textwidth}
        \centering
        \includegraphics[scale=0.44]{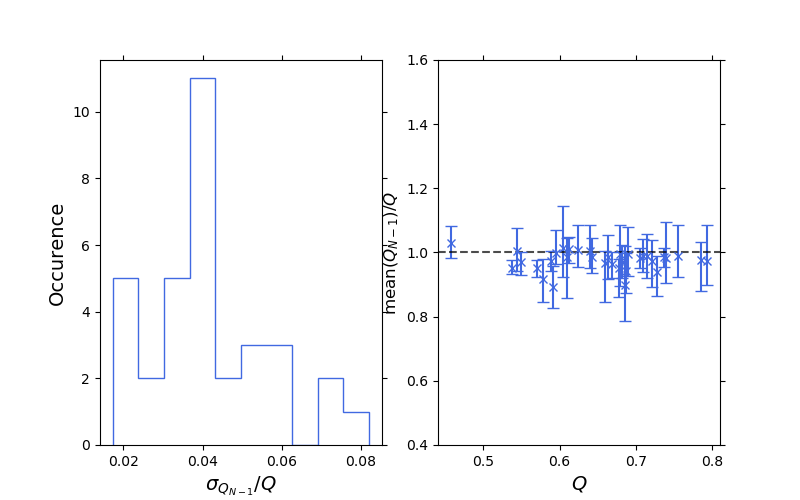}
        \caption{$Q$-error}
        \label{Q_errors:fig}
    \end{subfigure}%
    ~ 
    \begin{subfigure}[t]{0.5\textwidth}
        \centering
        \includegraphics[scale=0.44]{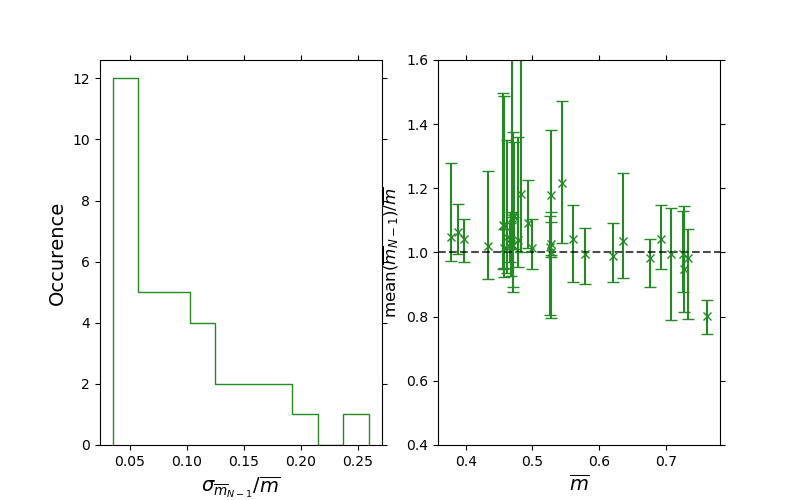}
        \caption{$\overline{m}$-error}
        \label{mbar_errors:fig}
    \end{subfigure}
    
    \begin{subfigure}[t]{0.5\textwidth}
        \centering
        \includegraphics[scale=0.44]{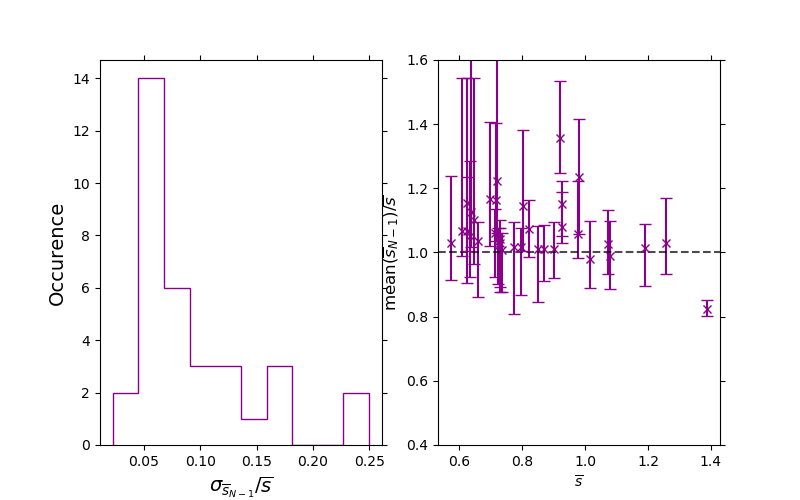}
        \caption{$\overline{s}$-error}
        \label{sbar_errors:fig}
    \end{subfigure}
    \caption{For each subfigure: \textit{Left:} Histogram of the fractional error ($\sigma_{x}/x$) distribution for $Q_{N-1}$, $\overline{m}_{N-1}$ and $\overline{s}_{N-1}$, respectively. \textit{Right:} Mean value per field  divided by the true value (e.g. mean$({x})_{N-1}$/$x$) plotted against true value, $x$, where $x$ is $Q$, $\overline{m}$ or $\overline{s}$. Error bars give the maximum and minimum ${x_{N-1}}$ value from each iterated run, normalised as per the mean.}
\end{figure*}%


Figures \ref{Q_errors:fig}, \ref{mbar_errors:fig} and \ref{sbar_errors:fig} indicate that the fractional uncertainties in each parameter are of the order $15\%$ and less (with few exceptions in the values of $\overline{m}$) and all below a $10\%$ in $Q$. The size of the scatter (length of error bars in the right-hand plots) appear uncorrelated with the true value plotted on the $x$-axis, with perhaps the exception of $\overline{s}$ at higher values. The explanation of this potential trend in $\overline{s}$ is that fields with fewer sources are those with smaller error-bar ranges and higher $\overline{s}$. This is a necessary outcome of the computation of $\overline{s}$ with very small numbers, as the correlation length will remain similar across iterations and the normalising cluster size $R_{clust}$ will also be relatively small (c.f. Equation \ref{Qpara:eqn}).

Given the uncorrelated scattering and small fractional uncertainties in our parameters, particularly in $Q$, very few TEMPO fields are likely to have $Q$ > 0.8 based on the detected sources therein, suggesting the calculated $Q$ values are indeed valid.

\subsection{$Q$-parameter for small numbers}

Next, the impact of the small number of sources per field on the effectiveness of the $Q$-parameter as a diagnostic tool for cluster structure was addressed. To achieve this a series of synthetic clusters containing either 5 or 10 sources were generated and their $Q$, $\overline{m}$  and $\overline{s}$ properties measured. 

The simulated clusters followed the prescription of \citet{CartwrightWhitworth04} for 3D spherical clusters with a volume density of $n \propto r^{-\alpha}$ and 3D fractal stellar clusters. We use $\alpha$ values of 0, 1, 2 and 2.9 and fractal dimension values, $D$ of 3.0, 2.5, 2.0 and 1.5, again as per \citet{CartwrightWhitworth04}. We note that for $D$ = 3.0 and $\alpha$ = 0 both cluster types give a uniform distribution.

For each $\alpha$ and $D$ value 10,000 realisations were generated for clusters of total population 5 and 10 sources (160,000 synthetic clusters in total).

Table \ref{Q_para_test:tab} gives the mean and standard deviations of the $Q$, $\overline{m}$ and $\overline{s}$ over the 10,000 realisations for each of the synthetic cluster type. The bottom four rows give the values for the real fields in our sample, from the whole sample and then sub-divided into fields with various source counts.

It is clear from these results that the use of $Q > 0.8 $ or $Q < 0.8$ to distinguish between clustered and fractal distributions is not possible for small populations. Only the $N = 10$ centrally clustered models show a mean $Q$ value $ > 0.8 $. For $N = 5$ models $Q$ is approximately equal within the one standard deviation irrespective of the cluster type used.

However, the $\overline{m}$ and $\overline{s}$ properties do show more significant variation across the models, particularly in the extreme cases. In Figures \ref{Q_param_space_cc:fig} and \ref{Q_param_space_fr:fig} we plot $\overline{m}$ and $\overline{s}$ for each of our sources overlaid on contours showing the 1-, 2- and 3-$\sigma$ level of those parameters over the 10,000 simulated cluster per model.

The figures demonstrate that the for small sample sizes the majority of clusters will show $Q$ values $<0.8$ and that either distribution type can display at least some clusters with $Q > 0.8 $. From this the use of the 0.8 value as a diagnostic between the two clustering types is clearly not viable. 

In relation to the targets the TEMPO sample, it can be seen in Figures \ref{Q_param_space_fr:fig} and \ref{Q_param_space_cc:fig} that all real TEMPO fields are inconsistent with a radial density profile of $\alpha$ = 2.9 at $N=10$ at the 1- or 2-$\sigma$ level. Similarly, relatively few TEMPO fields are consistent with a uniform density distribution ($\alpha$ = 0.0 and $D$ = 3.0) at the 1-$\sigma$ level. 

Both these findings agree with our interpretation of Figure 9 in \S4.1.2, that the TEMPO fields are densely clustered but do not display a simple radial power law.

\section{The use of interferometric visibility data to assess whether sources are centrally condensed}
\label{centCond:apdx}
\renewcommand{\thefigure}{C\arabic{figure}}
\setcounter{figure}{0}

Interferometric telescopes provide data of higher spatial resolution than is possible from single dish instruments. However, interferometers lack sensitivity to structure on large scales. The maximum recoverable spatial scale achievable for an interferometric array is set by its minimum baseline length (distance between the two closest antennas in the array). This limitation results in the `filtering out' of any emission from extended structures in the target field beyond this maximum recoverable scale. Such filtering can lead to images which appear to show several distinct clumpy regions, which are in reality only denser regions of a larger extended structure. 

In addition to this imaging limitation, and of significance to this work, in star forming regions there may exist dense clumpy regions which either do not yet contain a protostellar core or are simply transient phenomena which will never collapse to form a star.

In this appendix, we review the expected visibility properties for a set of simulated source models used to compare to the TEMPO source sample, discuss the simulated observations undertaken to define empirical relations between these simulated source model types and diagnostic visibility properties. Finally, discuss application of empirical relations of these diagnostic properties to the real TEMPO source data.

\subsection{Visibility Theory}

The cross correlation of signals for pairs of antennas in an interferometric array are used to measure complex visibilities, which are the Fourier transform counterpart of the sky brightness distribution being observed.

Complex visibilities are of the form:
\begin{equation}
V(u,v)=|V|e^{i\phi_{V}} = \int A(l,m)I(l,m) e^{-i2\pi(ul+vm)} \frac{{\rm{d}}l {\rm{d}} m}{\sqrt{1-l^2-m^2}}
\label{Vis:eqn}
\end{equation}

\noindent where $l$ and $m$ are the direction cosines of a vector \textit{\textbf{s}} from the phase centre of the observation. Interferometers measure one complex visibility per antenna pair per integration time interval. They are characterised by an amplitude, |V|, and phase, $\phi_V$. The amplitude relates directly to the flux density of the sky brightness distribution on the spatial scales observable by a given antenna pair and the phase relates to the distribution of emission on the sky.

\subsection{Simulated Source Models for comparison with TEMPO data}

To assess the nature of the sources in the TEMPO continuum source catalogue, an analysis of the Fourier/Visibility space properties of each object in the catalogue was undertaken. This analysis compared each detected TEMPO source to the properties of simulated source models for an unresolved point source (Point), a source with a Gaussian profile (Gaussian) and a source with a point source in a Gaussian envelope (Gaussian$+$Point), over a range of signal to noise ratios comparable to those seen in the TEMPO sample. The simulated source models used to compare to the TEMPO sources have the following properties in the visibility domain.

\subsubsection{Point source at the phase centre (Point)}
An unresolved point source observed with an interferometer has two identifying characteristics in visibility space. First, as the point source will be unresolved on all baselines of the interferometer the amplitude component of the complex visibilities measured will be the same on all baselines of the array. Second, for an unresolved point source all emission is localised at a single position on the sky, thus when the point source is at the phase centre of the observation the phases of the observed complex visibilities are zero (as by definition $u$ and $v$ are zero at the phase centre). This is again true on all baselines. 

\subsubsection{Gaussian source at the phase centre (Gaussian)}
A Gaussian source has slightly more complex visibility characterisitcs. The visibility amplitudes will decrease as a function of the baseline length (and be at a maximum on the shortest baseline). Conceptually, it is perhaps easier to think about this feature in terms of the angular scales being probed. Shorter baselines are probing emission from larger spatial scales. As such, for a Gaussian source which is smaller than the angular scale measured by the shortest baseline the Gaussian appears unresolved and 100\% of its emission is being measured by that baseline. For increasing baseline length, the source begins to be resolved into smaller and smaller angular elements thus less emission is being recovered. For the phase properties, on shorter baselines, which provide measurements of angular scales greater than the extent of the Gaussian source, the phases appear point-like and are zero degrees. Beyond these baselines the phases become scattered away from zero.

\subsubsection{Point source within a Gaussian envelope at the phase centre (Gaussian+Point)}
Combines the behaviour of the above two types of simulated source. The visibility amplitudes will decrease as a function of baseline length for baselines where it is possible to recover the emission of the Gaussian envelope. Beyond this at longer baselines the amplitude will be offset from zero as the point source at the centre will provide a constant amplitude to all baselines. Similarly in phase, at short baselines where the Gaussian is unresolved the phases will cluster around zero, as per a point source at those resolutions. The embedded point source will lead to a clustering of phases around zero degrees on all baselines, a signature which will distinguish it from a purely Gaussian profile source.
\\
\begin{figure}
\begin{center}
\includegraphics[width=.42\textwidth]{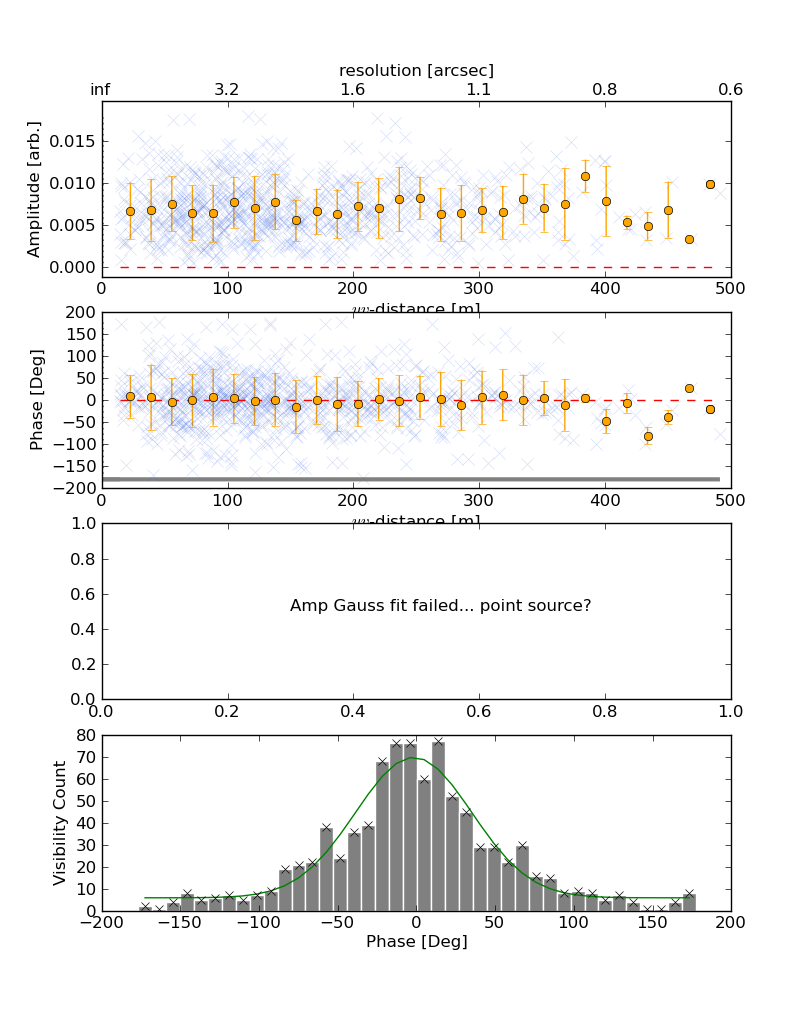}
\caption{Simulated point source model visibility properties. The model has a SNR of 50. From top to bottom the panels give: (\textit{Top panel}) The complex visibility amplitudes as a function of $uv$-distance, (\textit{second panel}) the complex visibility phases as a function of $uv$-distance. In the \textit{Top} and \textit{Second} panels the blue `x' are the visibilities values extracted from the simulated data and the orange circles averaged values in $uv$-distance bins. The errorbars associated with the binned data are $\pm$ 1 standard deviation. (\textit{Third panel}) Blank for this model, as the fitting of the $AG\_FWHM$ parameter defined in the text was unsuccessful. (\textit{Bottom panel}) A histogram of the unaveraged phase values. The green line describes a Gaussian fit to the data used to extract parameters the parameters for $ph2\_x0$ and $ph2\_FWHM$ as defined in the text.} 
\label{PointSourceModel:fig}
\end{center}
\end{figure}
\begin{figure}
\begin{center}
\includegraphics[width=.42\textwidth]{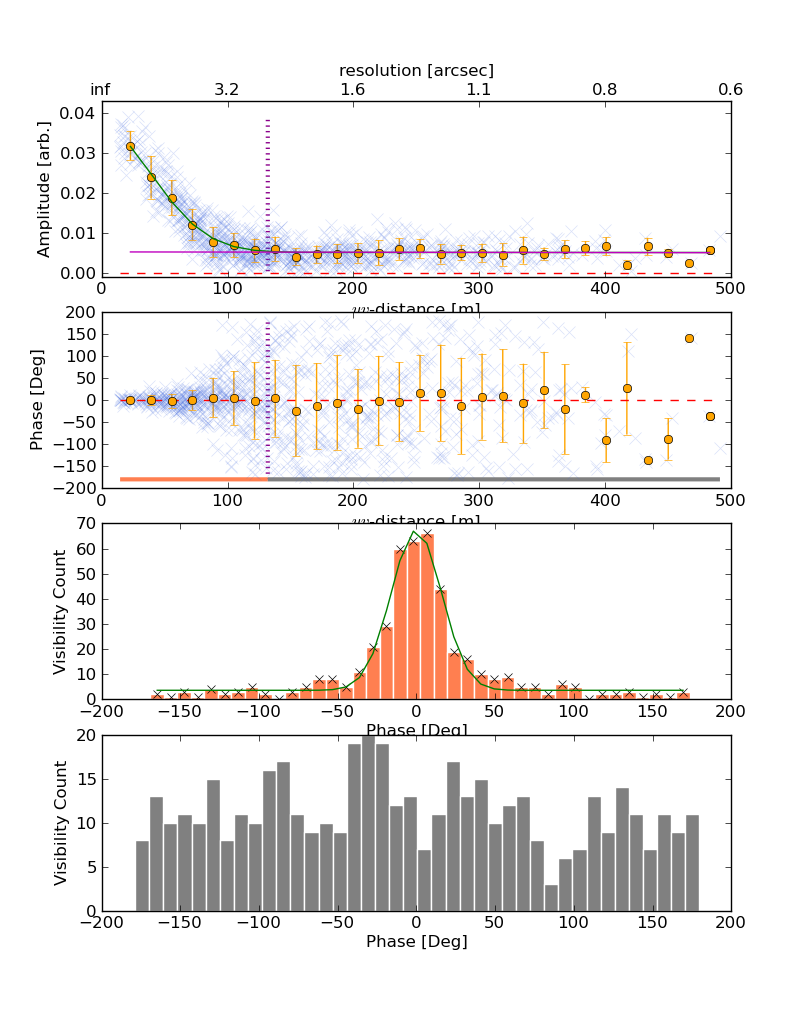}
\caption{Simulated Gaussian source model visibility properties. The model has a SNR of 50 and major/minor axes of 2\arcsec. From top to bottom the panels give: (\textit{Top panel}) and (\textit{second panel}) as per Figure \ref{PointSourceModel:fig}. In the \textit{top panel} the green line marks a Gaussian fit to the binned amplitude data. This is from the fit to measure the $AG\_FWHM$ parameter. The vertical dashed line gives the 3$\sigma$ limit of the Gaussian fit. In the \textit{second panel} the vertical dashed line gives the same 3$\sigma$ limit as in the \textit{top panel}. The coral and grey lines are used to denote colours used in plotting phase histograms in the \textit{third} and \textit{bottom} panels. (\textit{Third panel}) A histogram of the unaveraged phase values in the $\leq3\sigma$ $uv$-distance range (phase values with a $uv$-distance below the vertical dashed line in \textit{second} panel. The green line describes a Gaussian fit to the data used to extract the $ph1\_x0$ and $ph1\_FWHM$ parameters referred to in the text. (\textit{Bottom panel}) A histogram of the unaveraged phase values at $uv$-distances $>3\sigma$, the parameters for $ph2\_x0$ and $ph2\_FWHM$, are fit from data in this plot. In this specific case the fit is unsuccessful so now line is shown.} 
\label{GaussianSourceModel:fig}
\end{center}
\end{figure}

\begin{figure}
\begin{center}
\includegraphics[width=.42\textwidth]{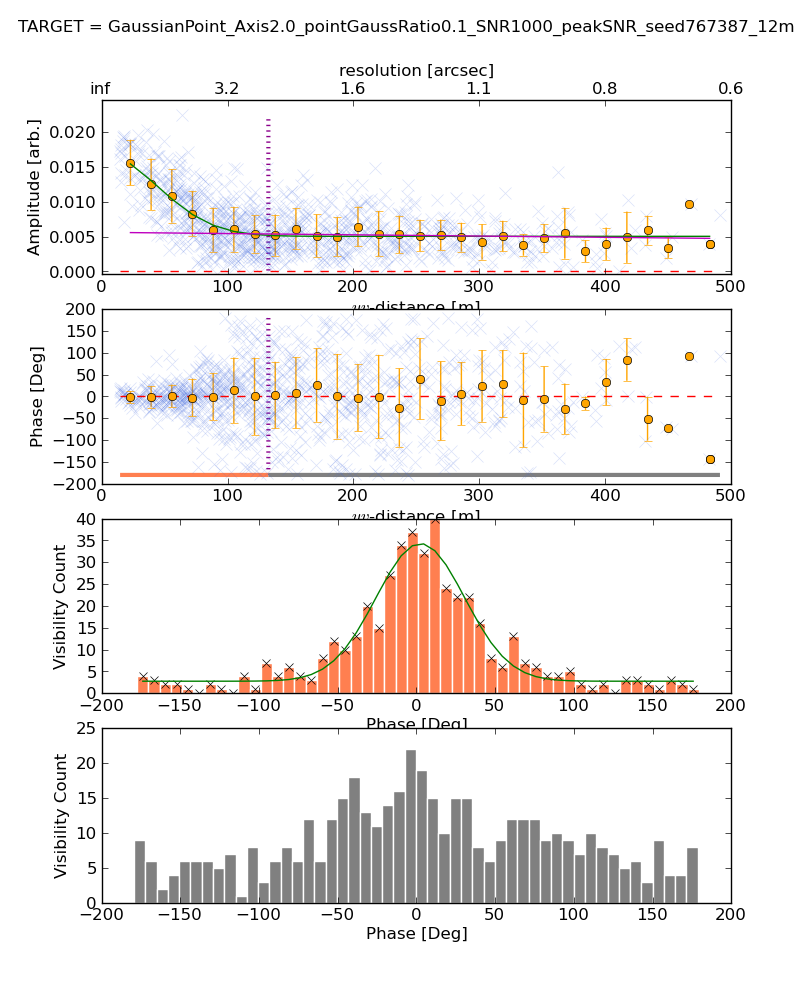}
\caption{Simulated Gaussian$+$ point source model visibility properties. The model has a SNR of 1000, a major/minor axes of 2\arcsec and a point source flux to Gaussian peak flux density of 0.1. From top to bottom the panels are as per Figure \ref{GaussianSourceModel:fig}.} 
\label{GaussianPOINTSourceModel:fig}
\end{center}
\end{figure}

Real observations have a noise per visibility which contributes to the recovered amplitudes and phases and thus will cause a deviation from the idealised properties described above. Figures \ref{PointSourceModel:fig}, \ref{GaussianSourceModel:fig} and \ref{GaussianPOINTSourceModel:fig} display in the upper two panels the described source properties as a function of $uv$-distance (equivalent to baseline length), with noise included. The noise in the amplitude and phase data impacts our ability to reliably compare the TEMPO sources to the idealised case, specifically for the low signal-to-noise sources. Steps to mitigate the impact of this are discussed in section \ref{SimSuite:sec}.

Another level of complexity to consider in real observations, is a field with more than one source of emission. Here emission from the sources which are not at the phase centre will cause deviation from the phase and amplitude properties the idealise cases described above. A step to mitigate the effects of this was used in this analysis and is described in \ref{prepTempo:sec}

\subsection{Generating Simulated Source Properties}
\label{SimSuite:sec}
The simulated observations were created using the CASA task \texttt{simobserve} and emulate closely the TEMPO observing characteristics \S\ref{obsprops:sec}. The simulated observing properties are as follows: An observing time of 300s, total BW of 1.875GHz (equivalent to 1 SPW), a simulated percipitble water vapour (PWV) of 1.796mm (a typical ALMA value for Band 6 observations) and the CASA ALMA array configuration file \texttt{alma.cycle6.3.cfg} was used, as it was closest to the true configuration used during TEMPO observations.

The simulated source models were generated using the CASA \texttt{componentlist} tools. Point sources were purely point like objects with a flux density set to give a desired SNR. For Gaussian models (Gaussian and Gaussian$+$Point) multiple size Gaussian sources models were simulated. Major axis values of 1\arcsec, 2\arcsec and 3\arcsec\ were used in combination with major to minor axes ratios of 1:1 (radially symmetric), 2:1 and 3:1 each with position angle set to 0 degrees in all cases. The major axis values chosen provide both marginally resolved (the 1\arcsec Gaussian) and fully resolved (2 and 3 \arcsec Gaussians) in a simulated TEMPO synthesised beam.

For Gaussian$+$Point sources models the peak flux density is given by the addition of the point-like object and the Gaussian object, with the peak flux set to provide the required SNR. For these models a Gaussian to point flux density ratio was set to provide simulated sources where the point source object had a flux density equal to the Gaussian profile, half that of the Gaussian and a tenth of the Gaussian envelope.

In each case the model source was placed at the phase centre of the observation. The models created such that the peak emission divided by the off source noise (measured in a simulated blank sky of the same observing properties) gave a desired SNR. The SNR values used were, 5, 10, 15, 20, 30, 40, 50, 60, 80, 100, 120, 150, 200, 300, 500, 750, 1000.

At each SNR value, 100 simulations per model type were conducted (giving 62,900 simulations in total) with a unique random seed defining the ‘phase screen’ used to apply the noise to the data, effectively changing the distribution of the thermal noise in the recovered map. 

\subsection{Measuring simulated source properties}
With the suite of simulated observations, the CASA task \texttt{plotms} (with its graphical user interface deactivated) is used to record to text file the simulated amplitudes and phases as a function of $uv$-distance (in units metres, though the choice of $x$-axis values is arbitrary) at the phase centre. Given the data size, the visibilities were averaged up in both time and frequency channel and only the `XX’ correlation was recorded to reduce the number of data points required for analysis. After this the following values are extracted from the simulated visibility properties:

\subsubsection{Amplitude vs $uv$-distance Gaussian profile (\textit{AG\_FWHM})}
Here the amplitude and phase data of the recorded visibilities are first further averaged into 30 $uv$-distance bins to again reduce the data volume. An attempt is then made to fit a Gaussian profile to the amplitude as a function of $uv$-distance. If this fitting succeeds then the full width half maximum (hereafter \textit{AG\_FWHM}) value of the Gaussian profile is recorded (in arcseconds, by conversion from metres to angular size at the observing frequency). If the fitting algorithm returns a FWHM greater than the maximum $uv$-distance or less than the minimum $uv$-distance the fit is rejected. In these cases, or if the fitting fails no \textit{AG\_FWHM} is recorded.

In cases where an \textit{AG\_FWHM} is recorded then the visibilities (no longer binned by uv-distance as above) are split between the `Unresolved domain' which includes visibilities with $uv$-distances from the minimum $uv$-distance up to 3 times the fitted Gaussian variance ($\sigma$)\footnote{related to \textit{AG\_FWHM} by $AG\_FWHM = 2\sqrt{2ln2}\sigma$} and the `Resolved domain', data $uv$-distances from 3$\sigma$ to the max $uv$-distance. Where no \textit{AG\_FWHM} is recorded all the visibilities are considered to be in the `Resolved domain'.

The upper panel of Figure \ref{GaussianSourceModel:fig} shows a successful fit for the $AG\_FWHM$ parameter.

\subsubsection{`Unresolved domain' visibility histogram centre and FWHM (\textit{ph1\_x0} and \textit{ph1\_FWHM})}
When an \textit{AG\_FWHM} is recorded this indicates that a Gaussian source is likely present in the model. On short baselines the (assumed) Gaussian will be marginally to completely unresolved and behave like a point source on these baselines, with the phases tightly clustered around zero. To measure this a Gaussian profile is fit to a histogram of the recorded phase values, in 50 bins. The centre of the Gaussian profile peak \textit{ph1\_x0} and its FWHM \textit{ph1\_FWHM} are recorded. In the case of a failure to fit a dummy value is recorded and ignored in further analysis.

The third panel from the top of Figure \ref{GaussianSourceModel:fig} shows a successful fit for the \textit{ph1\_x0} and \textit{ph1\_FWHM} parameters.

\subsubsection{`Resolved domain' visibility histogram centre and FWHM (\textit{ph2\_x0} and \textit{ph2\_FWHM})}
For visibilities in the `Resolved domain' a Gaussian profile is fit to a histogram of the recorded phase values, in 50 bins. The centre of the Gaussian profile peak \textit{ph2\_x0} and its FWHM  \textit{ph2\_FWHM} are recorded. In the case of a failure to fit a dummy value is recorded and ignored in further analysis.

The bottom panel of Figure \ref{PointSourceModel:fig} shows a successful fit for the \textit{ph2\_x0} and \textit{ph2\_FWHM} parameters.

\subsection{Defining empirical relations of SNR and recorded simulated source properties}
\label{emprels:sec}
The $x0$ and $FWHM$ values recovered from simulations were used to generate empirical bounding relations for these properties as a function of SNR. To do this, the average and standard deviation of both $x0$ and $FWHM$, for each model type (Point sources and Gaussian (including pure Gaussian and Gaussian+Point), were calculated. Upper and lower limit values were then set as the average value $\pm$ 3$\times$-the standard deviation at each SNR (across each model type). For  $x0$ values an inverse relation, and for $FWHM$ a power law relation between the data points and SNR were found to provide the best fits to the resultant profiles. For the lower $FWHM$ boundary the value at the maximum SNR in the simulated model suite was used as a fixed limit across all FWHMs as any value between this and the upper bound at any given SNR provides a realistic FWHM value.

\subsubsection{Point model parameter boundaries}

\begin{equation}
ph2\_x0_{upper} = \frac{319.21}{\rm{SNR}}+-0.17
\label{pointxoupperbound:eqn}
\end{equation}

\begin{equation}
ph2\_x0_{lower} = \frac{-301.91}{\rm{SNR}}+0.06
\label{pointxolowerbound:eqn}
\end{equation}

\begin{equation}
ph2\_FWHM_{upper} = 5067.0\rm{SNR}^{-0.96}
\label{pointfwhmupperbound:eqn}
\end{equation}
\begin{equation}
ph2\_FWHM_{lower} = 4169.8\rm{SNR}^{-0.98}
\label{pointfwhmlowerbound:eqn}
\end{equation}

\subsubsection{Gaussian model parameter boundaries}

\begin{equation}
ph1\_x0_{upper} = \frac{113.23}{\rm{SNR}}+3.8
\label{gaussianxoupperbound:eqn}
\end{equation}
\begin{equation}
ph1\_x0_{lower} = \frac{-100.12}{\rm{SNR}}+-4.1
\label{gaussianxolowerbound:eqn}
\end{equation}

\begin{equation}
ph1\_FWHM_{upper} = 534.6\rm{SNR}^{-0.34}
\label{gaussian1fwhmupperbound:eqn}
\end{equation}
\begin{equation}
ph1\_FWHM_{lower} = -572.3\rm{SNR}^{-0.75}
\label{gaussian1fwhmlowerbound:eqn}
\end{equation}

\begin{equation}
ph2\_x0_{upper} = \frac{36.83}{\rm{SNR}}+20.04
\label{gaussianxo2upperbound:eqn}
\end{equation}
\begin{equation}
ph2\_x0_{lower} = \frac{-52.24}{\rm{SNR}}+-18.8
\label{gaussianxo2lowerbound:eqn}
\end{equation}

\begin{equation}
ph2\_FWHM_{upper} = 722.9\rm{SNR}^{-0.3}
\label{gaussian2whmupperbound:eqn}
\end{equation}
\begin{equation}
ph2\_FWHM_{lower} = -396.5\rm{SNR}^{-0.27}
\label{gaussian2fwhmlowerbound:eqn}
\end{equation}

\subsection{Application to the TEMPO data}
With the simulated model boundaries in place a comparison to the real TEMPO data is then possible. Firstly, the visibility data for each TEMPO field was prepared to mitigate the effects of multiple sources in the same field.

\subsubsection{Preparing the data}
\label{prepTempo:sec}
To extract the visibility data for a specific TEMPO source in its host field the following steps were taken:
\begin{itemize}
    \item Using the position, major and minor axis, position angle and measured flux density for all sources in the current TEMPO field, \textit{excluding} the source under investigation, a CASA component list of Gaussian sources was generated.
    \item This component list was subtracted from the visibility data using the CASA task \texttt{uvsub}. This removes, or minimises, the effect of multiple sources in a given field adding extra 'noise' to the expected source properties.
    \item The phase centre of the field visibilities is shifted to the source position and the amplitude and phase data extracted by the same method as used for the simulated models.
\end{itemize}

In theory, the properties of the amplitudes and phases should then match those of the models in the case that the target has centrally condensed properties. In practice, there are some additional, unavoidable issues which much be considered. These are discussed in $\S$\ref{outstanding:sec}

\subsubsection{Assigning a star forming classification}

\begin{figure*}
\begin{center}
\includegraphics[width=.8\textwidth]{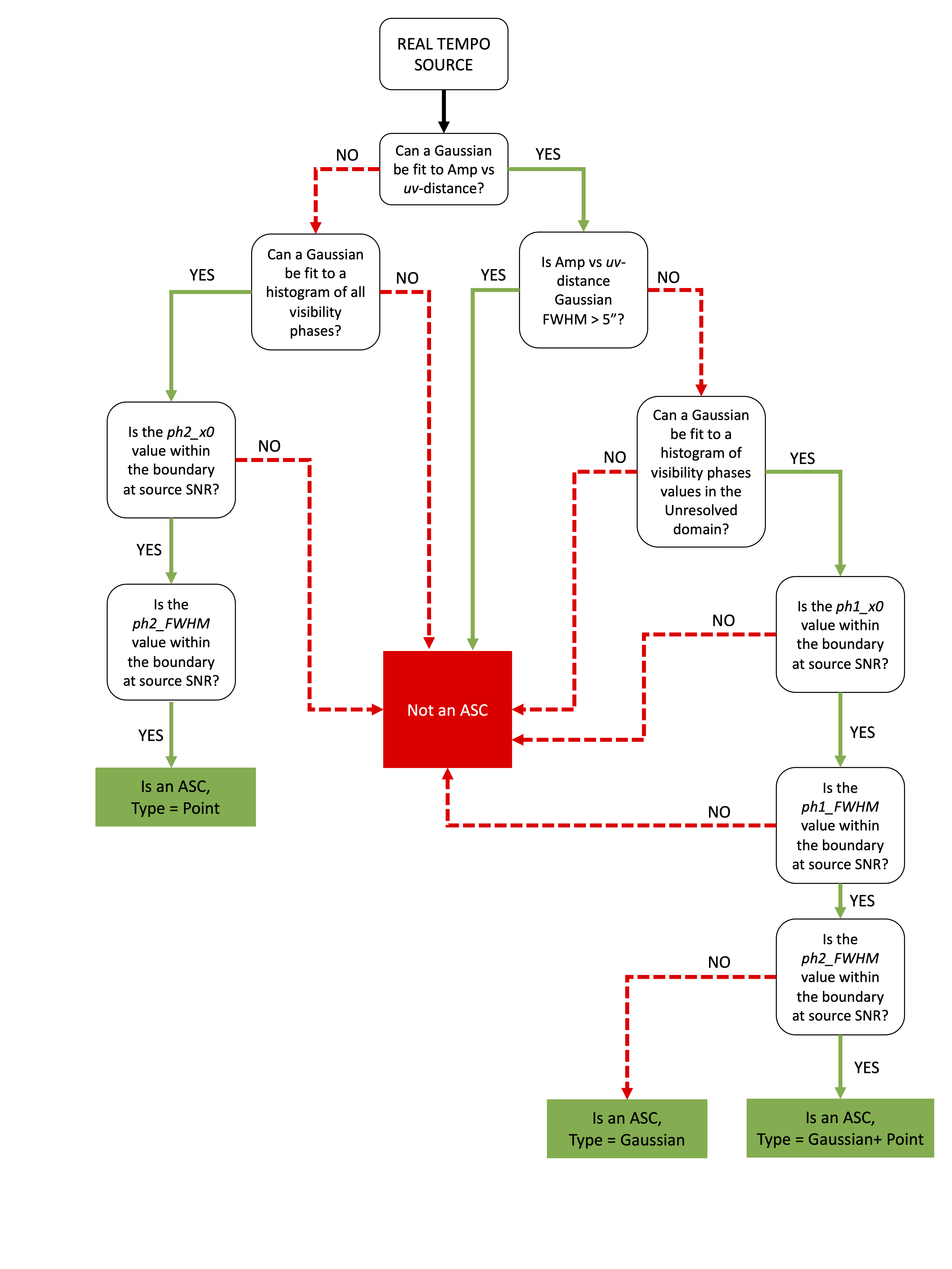}
\caption{Decision tree used to determine if fragments within the TEMPO sample are actively star-forming or not, and to specify the type if so. For each query a fragment either meets the criteria, and so follows the YES (solid green arrows) branch or fails to do so, and so follows the NO (red dashed arrows) branch until an end point is reached.} 
\label{ASCflow:fig}
\end{center}
\end{figure*}

The same method used to measure the simulated model properties was then applied to the real TEMPO sample giving, for each source, values (or null results) for the parameters \textit{AG\_FWHM}, \textit{ph1\_FWHM}, \textit{ph1\_x0}, \textit{ph2\_x0} and \textit{ph2\_FWHM}. With these values and the data boundaries from the empirical relations given in \S \ref{emprels:sec}, each TEMPO source was assessed at the recorded SNR in the combined continuum images following a series of descision steps, used to assign if a source was a Point, Gaussian, Gaussian$+$Point or None of the Above type source. For Point, Gaussian and Gaussian$+$Point types these are considered Actively Star-forming Candidates. Figure \ref{ASCflow:fig}, gives the decision tree used to determine if a TEMPO source is considered actively star-forming or not.

Given the spread in the empirically derived bounds used a lower SNR cut off was used so that only sources with SNR $\geq$ 30 were considered int the classification analysis.

\subsection{Limitations to this analysis}
\label{outstanding:sec}

Two limitations to this method when applied to real data exist. The first concerns the subtraction of field sources from the visibilities. The source properties (major and minor axes and position angle) used to generate the component list which is subtracted from the visibility data are based on those reported by the dendrogram analysis as discussed in the main text. In cases that assuming a Gaussian profile is a poor fit to the true source shape, for example if the source is structured or extended then subtracting a Gaussian will lead to residual emission structure in the remaining visibilities. When imaged such residual emission structure could exhibit features like negative holes with positive emission halos or arcs around them. Similarly, subtracting a Gaussian profile for a source which is a combination of unresolved components in the TEMPO data will leave residual structure in the visibility data.

Both effects will limit our ability to assign a star-forming status to some sources within the TEMPO sample. This effect is hard to mitigate, as correctly modelling the emission properties of discreet $>$200 sources in complex fields containing extended structures is both time and computationally expensive and beyond the scope of the work conducted here.

The second artefact which can present itself in this method is setting an incorrect position when shifting the phase centre of the visibilities. For a point source small positional offsets from the true source position results in the phase data showing an `arrow' or `$<$’-like profile. This indicates a delay-like behaviour caused by the offset between the phase centre and the true source position. The slope of the $<$ can be used to indicate how large this offset is, as the $uv$-distance in which the phase slope would take to trace a full 360 degrees gives you a baseline length. Convert that baseline length to an angular scale ($\lambda/b$) gives the magnitude of the offset. Unfortunately, to probe all positions at this magnitude offset is again beyond the scope of this analysis. 
\bsp	
\label{lastpage}
\end{document}